\documentclass[lettersize,journal]{IEEEtran}

\usepackage{graphicx}
\graphicspath{{figures/}{pictures/}{images/}{./}} 

\usepackage[hidelinks]{hyperref}
\usepackage{enumitem}
\usepackage{wrapfig}
\usepackage{multirow}
\usepackage{tabu}                      
\usepackage{booktabs}                  
\usepackage{soul}
\usepackage{algorithm}
\usepackage{algorithmicx}
\usepackage{algpseudocode}
\algdef{SE}[DOWHILE]{Do}{doWhile}{\algorithmicdo}[1]{\algorithmicwhile\ #1}
\usepackage{makecell}
\usepackage{amsmath}
\usepackage{relsize}

\newcommand{\eg}{e.g. }

\usepackage{stackengine}

\usepackage{color}
\usepackage{balance}

\definecolor{red}{rgb}{1,0,0}

\newcommand{\lk}[1]{{\color{black} #1}}

\definecolor{marked}{rgb}{0,0,0}
\newcommand{\revised}[1]{{\color{black}#1}}
\newcommand{\revision}[1]{{\color{black}#1}}
\newcommand{\revisionB}[1]{{\color{black}#1}}

\begin{document}

\title{Color-Name Aware Optimization to Enhance the Perception of Transparent Overlapped Charts}

\author{Kecheng Lu,
	Lihang Zhu,
        Yunhai Wang\IEEEauthorrefmark{1},
	Qiong Zeng,
	Weitao Song,
	and Khairi Reda

\IEEEcompsocitemizethanks{
	\IEEEcompsocthanksitem \IEEEauthorrefmark{1} Yunhai Wang is the corresponding author.
	\IEEEcompsocthanksitem Kecheng Lu \& Yunhai Wang are with the School of Information, Renmin University of China, China.	E-mail: lukecheng0407@gmail.com, wang.yh@ruc.edu.cn
	\IEEEcompsocthanksitem Lihang Zhu \& Qiong Zeng are with the School of Computer Science and Technology, Shandong University, China. E-mail: \{leon20192050, qiong.zn\}@gmail.com
	\IEEEcompsocthanksitem Weitao Song is with the School of Optics and Photonics, Zhengzhou Research Institute, Beijing Institute of Technology, China. E-mail: swt@bit.edu.cn
	\IEEEcompsocthanksitem Khairi Reda is with the Luddy School of Informatics \& Computing, Indiana University Indianapolis, USA. E-mail: redak@iu.edu
}

\thanks{Manuscript received August 9, 2024; revised XXXX XX, XXXX.}}

\markboth{IEEE Transactions on Visualization and Computer Graphics}%
{Lu \MakeLowercase{\textit{et al.}}: Color-Name Aware Optimization to Enhance the Perception of Transparent Overlapped Charts}

\maketitle
\begin{abstract}
Transparency is commonly utilized in visualizations to overlay color-coded histograms or sets, thereby facilitating the visual comparison of categorical data. However, these charts often suffer from significant overlap between objects, resulting in substantial color interactions. Existing color blending models struggle in these scenarios, frequently leading to ambiguous color mappings and the introduction of false colors. To address these challenges, we propose an automated approach for generating optimal color encodings to enhance the perception of translucent charts. Our method harnesses color nameability to maximize the association between composite colors and their respective class labels. We introduce a color-name aware (CNA) optimization framework that generates maximally coherent color assignments and transparency settings while ensuring perceptual discriminability for all segments in the visualization. We demonstrate the effectiveness of our technique through crowdsourced experiments with composite histograms, showing how our technique can significantly outperform both standard and visualization-specific color blending models. Furthermore, we illustrate how our approach can be generalized to other visualizations, including parallel coordinates and Venn diagrams. We provide an open-source implementation of our technique as a web-based tool.
\end{abstract}



\begin{figure*}[!ht]
	\centering
	\includegraphics[width=0.95\linewidth]{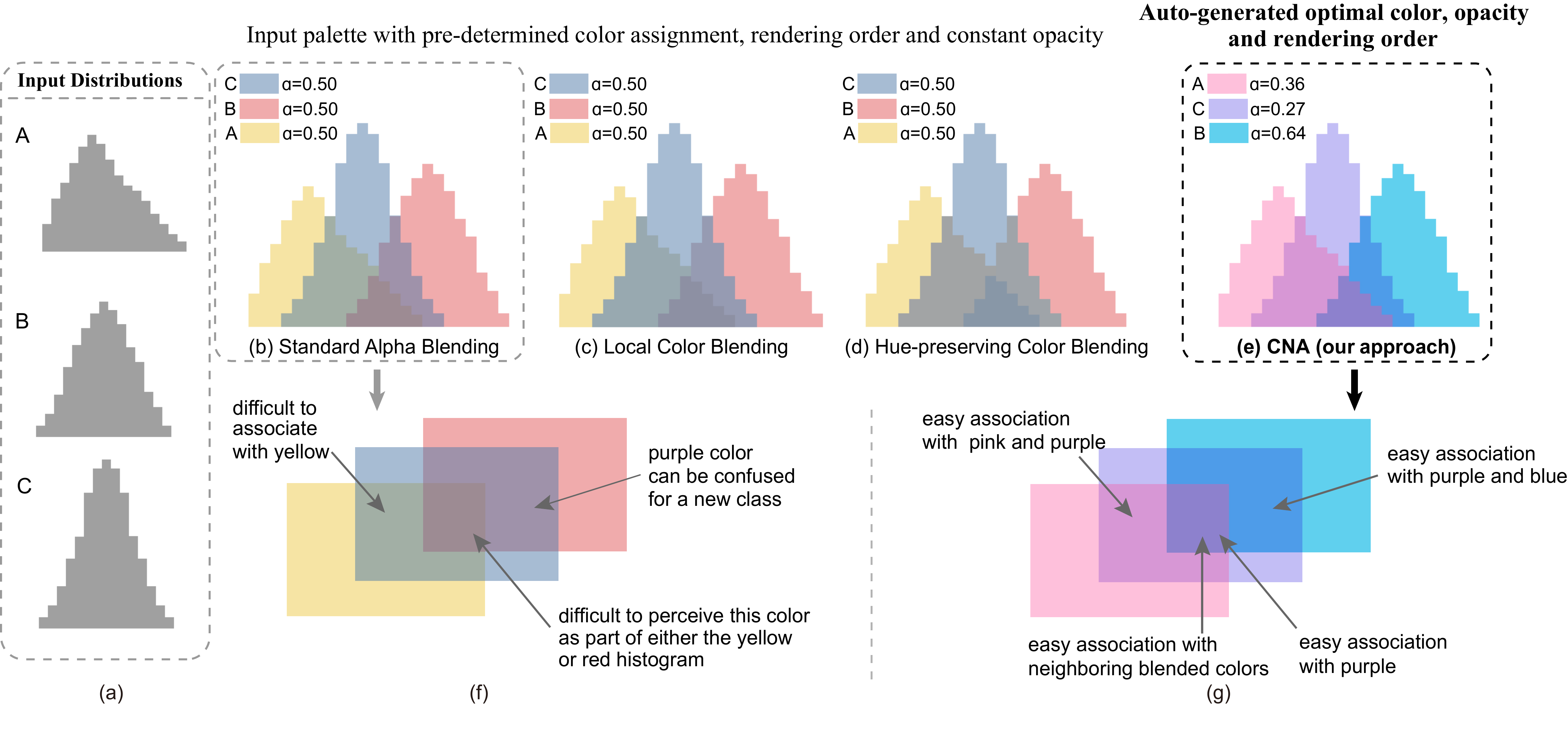}
	\caption{
 Results of applying our approach (right-most) to optimize three overlapping histograms.  (a) shows the input component distributions; (b, c, d) illustrates the results of applying three comparable benchmarks: standard alpha blending, local color blending ~\cite{wang2008color}, and hue-preserving color blending~\cite{chuang2009hue} using base colors from the Tableau-10 palette and a uniform opacity of $\alpha=0.5$. By comparison, our approach (e), \lk{which embodies a color-name aware optimization,} auto-generates optimal color, transparency and rendering order settings, ensuring discriminability for all segments while improving whole-from-parts perception. Our technique can also optimize other overlapped visualizations, including Venn diagrams (f vs. g).
 }
	\label{fig:teaser}
\end{figure*}

\section{Introduction}

\IEEEPARstart{O}{verlapping}, transparent visualizations are commonly used to show various data representations, from histograms to Venn diagrams to multi-variate parallel coordinates~\cite{mcdonnell2008illustrative}. These charts create strong cues of visual superposition~\cite{Gleicher11}, facilitating effective comparison of distributions and sets across categorical variables. Typically, each category is depicted in a distinct color, resulting in multiple color-coded objects that are then composited using alpha blending techniques~\cite{porter1984compositing}. Despite their conceptual simplicity, crafting perceptually accurate translucent visualizations is challenging, requiring careful assignment of colors and opacity levels. Transparency rendering, in particular, can lead to complex interactions between colors, resulting in the introduction of false categories, or to a reduction in the coherence of the composite segments to their original class color. \revision{Moreover, resolving composite colors becomes more challenging when the original class colors are taken from hues that lie further apart in the color wheel~\cite{gama2014studying}, particularly when blending results do not align with recognizable or expected hues.} For instance, consider the visualizations in \autoref{fig:teaser}-b, where the blended areas suggest the presence of a new, non-existent feature (a purple `histogram', in this case). Such artifacts can complicate the interpretation of the visualization.

To address these issues, prior work suggests the addition of visual cues to help disambiguate the visualization. For example, Wilke~\cite{wilke2019fundamentals} recommends adding a kernel density estimate to each histogram. However, this approach can result in visual clutter, with multiple marks and mark-types overlapping in a potentially small space. An alternative line of research has focused on developing enhanced color-blending models that aim to reduce transparency-induced ambiguities. For instance, the \emph{local color blending} model~\cite{wang2008color} works by desaturating `background' colors before blending to preserve depth information (\autoref{fig:teaser}-c). Similarly, the \emph{hue-preserving color blending} model~\cite{chuang2009hue,kuhne2012data} works by reducing non-dominant hues (\autoref{fig:teaser}-d). Both of these techniques work by adjusting the blending mechanisms for a predetermined set of colors, yet they do not offer a means for users to select or generate an appropriate color palette that is resilient to blending effects, leading to suboptimal visualization. For example, a hue-preserving composite with a standard color palette (e.g., Tableau-10) frequently results in neutral, low-saturation grays that are difficult to associate with their original class (see~\autoref{fig:teaser}-d). Additionally, these techniques rely on non-standard color blending models, often requiring the implementation of custom pixel-level compositing pipelines, which may limit their applicability. Consequently, there is a need for colorization techniques that are tailored specifically for transparency considerations, yet compatible with standard graphical plotting platforms.

Inspired by recent automatic colorization methods~\cite{Lu21,Lu23}, we hypothesize that the perception of overlapping visualizations can be significantly improved by \emph{directly} optimizing color assignment, opacity, and rendering order. Moreover, we hypothesize that such optimization can be achieved while utilizing a standard alpha blending model, which is widely supported in graphics and visualization toolkits. We present a novel color generation and optimization technique tailored to address the challenge of interpreting semi-transparent histograms and other categorical visualizations with overlapping elements. Our technique is grounded in the understanding that associating colors with their respective categories depends not only on color appearance but also on semantics, such as color names~\cite{wertheimer1938gestalt}.  Leveraging this insight, we generate color assignments for overlapping elements by maximizing name similarity \lk{between parts that make up the distribution}, thus facilitating effective whole-from-parts perception. Our optimization framework takes into account various data characteristics, including the arrangement of distributions and the extent of overlap, ensuring coherence within color classes while maintaining visibility for smaller segments. To account for the potential complexity of color interactions, we employ a custom simulated annealing algorithm~\cite{aarts1989stochastic} to explore the solution space and pinpoint an optimal configuration.  The resulting visualizations enable accurate inference of distribution shapes while still maintaining discriminability of smaller distributional features, even when they overlap (\autoref{fig:teaser}-e \& g). 

We conducted two crowdsourced studies to evaluate our approach\footnote{Experimental data and analysis code are provided as supplemental materials, and can be accessed at \url{https://osf.io/xevk9/?view_only=1b79aeefec774209ad60f1a5b0cceda2}}. In the first study, we compared our method against custom color blending models intended for enhancing translucent visualizations. 
We find that our approach yields the best performance for histogram analysis among the benchmarks. In a second study, we test whether our optimization is still effective when histograms are augmented with kernel-density estimates, finding that our technique can still improve interpretation over the baseline. We further demonstrate the generalizability of our approach beyond histograms with case studies, showing its utility for parallel coordinate plots and set visualizations. To help disseminate this method, we developed a web-based implementation of our technique as an interactive color-design tool.
To summarize, our main contributions are:

\vspace*{-1mm}
\begin{itemize}[noitemsep]
\setlength{\itemsep}{5pt}
  \item A novel \lk{{C}olor-{N}ame {A}ware (CNA)} optimization approach that automatically generates appropriate color palettes and assigns opacity values and rendering order for overlapping, semi-transparent histograms. 
    
   \item Empirical validation of our approach. We show that our method allows for accurate analysis of distributional features, yielding better judgments than existing approaches.

  \item An extension of our method to other multi-class visualizations with overlap, including parallel coordinates and Venn diagrams.

  \item Finally, we contribute an implementation of our method as an open-source tool, available at \url{https://anon-link.github.io/transparency/}.
\end{itemize}

\section {Related Work}
We survey previous work on utilizing transparency for visualization. We then discuss color optimization techniques broadly and address the perception of histograms and distribution visualizations.

\subsection{Optimizing Semi-transparent Visualizations}
Transparency plays an important role in computer graphics~\cite{foley1996computer} and visualization~\cite{weiskopf2007gpu} and is often exploited to reduce occlusion in 3D objects. We limit our discussion to techniques related to information visualization, and to color blending models, opacity optimizations, as well as the perception of 2D transparent visualizations. 

\vspace{2mm}
\noindent\textbf{Color Blending Models}.
The most common approach for simulating partial transparency is the standard alpha blending model using the Porter-Duff operator~\cite{porter1984compositing}. 
However, this method often produces false colors (e.g., blending red and blue to produce purple), which can confuse the viewer. To address this issue, several alternative blending models have been proposed. 
Wang et al.~\cite{wang2008color} found that blending opponent colors produces neutral (e.g., greys) as opposed to categorically distinct colors. They accordingly propose to alleviate false color using a local blending approach that reduces the saturation of the background color. This results in blended color appearing closer to the front (e.g., see~\autoref{fig:teaser}-c), thus improving depth perception. Although useful in 3D settings, this technique can impede the perception of the background objects. Furthermore, local blending cannot entirely eliminate false colors.
Chuang et al.~\cite{chuang2009hue} developed a hue-preserving blending model to address the issue of false colors. This approach works by modifying the hue component of the non-dominant color such that it is the opponent of the dominant hue. The results thus preserve the hue of the dominant color. However, as shown in~\autoref{fig:teaser}-d, this method often results in the complete loss of hue (except for the dominant color), which is a serious limitation for visualizations that rely extensively on color-encoding to communicate categories. 
Instead of heuristics-based optimization, 
Zhang et al.~\cite{kuhne2012data} propose a data-driven color blending model based on a lab study that specifically attempts to avoid false colors and preserve depth ordering, and demonstrate the effectiveness of their method with a use case based on interpreting parallel coordinates. Unfortunately, the collected data and model have not been released and thus we cannot reproduce their results. 
Almost all these methods aim to generate effective visualization for better preserving depth order,  whereas our goal is to help users perform distributional analyses like identifying and comparing the shapes of distributions. 

In lieu of compositing colors, Urness et al.~\cite{urness2003effectively} propose an alternative method to convey transparency via color weaving~\cite{urness2003effectively}, that interweaves individual colors of multiple variables to form a high-frequency texture. A user study by Hagh-Shenas et al.~\cite{hagh2007weaving} found that color weaving is more efficient than traditional color blending at conveying the values of individual data distributions in a multivariate visualization, but the performance degrades as the number of overlapping classes increases. Luboschik et al.~\cite{luboschik2010new} extended this technique for representing overlapping regions in scatterplots. Although promising, the use of color weaving is uncommon in practice and could lead to poor performance for some comparison tasks. Furthermore, this technique is highly sensitive to choices of textural and statistical parameters~\cite{albers2014task}.


\vspace{2mm}
\noindent\textbf{Opacity Optimization}.
Specifying proper opacity values can be an effective way to reduce the occlusion. A few opacity optimization techniques have been developed for a variety of 2D/3D objects, including lines~\cite{gunther2013opacity,lu2021curve}, surfaces~\cite{gunther2014opacity}, and volume~\cite{wang2011efficient}. To address over-plotting in parallel coordinates, Johansson et al.~\cite{johansson2005revealing} and Zhou et al.~\cite{zhou2008visual} introduce tools that assign opacity based on local data density. However, these tools still require considerable manual effort to optimize the visualization. Based on a crowdsourced study, Matejka et al.~\cite{matejka2015dynamic} proposed a user-driven opacity scaling model for scatterplots that automatically assigns an appropriate transparency for each data point. Micallef et al.~\cite{micallef2017towards} and  Quadri et al.~\cite{quadri2022automatic} extended this effort by considering additional visual variables alongside mark opacity to better support certain data analysis tasks. All of these optimization techniques do not consider color blending artifacts caused by composite colors, which is a central focus of our work. 

\vspace{2mm}
\noindent\textbf{Transparency Perception}.
Transparency can serve as an important `channel' in visualization, with several models having been developed to measure and quantify the effect of this cue.  For example, Metelli's episcotister model~\cite{metelli1974perception} is one of the earlier models to quantify the perception of transparency. 
The X-junction model~\cite{adelson1990ordinal} utilizes luminance to predict depth perception. These models can be used to optimize transparency configurations for volume visualizations~\cite{zheng2012perceptually}. \lk{Beck et al.~\cite{beck1984perception} test the Metelli model through multiple experiments, and report that the degree of perceived transparency varied linearly with lightness, not the reflectance. Similarly, Singh et al.~\cite{singh2002toward} conduct further experiments toward building a perceptually based theory of transparency that addresses the limitations of the episcotister model. Readers are referred to a recent survey ~\cite{gigilashvili2021translucency} for a more detailed review of the translucency perception.}
Gama and Gonçalves~\cite{gama2014guidelines} investigate people's ability to resolve the original colors in blended composites. Their findings indicate that participants perform poorly when resolving opponent hues that are composited together, or when identifying source colors from three (as opposed to two) colors. Notably, blending performed in the CIE LCh model space resulted in higher accuracy compared to other color spaces. Our research similarly seeks to create translucent charts that enhance the `provenance' of composite colors for categorical visualization, which is important for distribution analysis, among other types of data.

\subsection{Color Optimization}
Once a source palette is chosen (e.g., from ColorBrewer~\cite{harrower2003colorbrewer}), an important step in categorical visualization design entails assigning colors from the palette to different categories. A few optimization techniques have been proposed to streamline this process, often with the goal of maximizing semantic association between colors and the concept they represent. For example, 
Lin et al.~\cite{lin2013selecting} propose semantically resonant color assignment for categorical data based on a semantic affinity score calculated from a set of representative images for each category.
Setlur and Stone~\cite{setlur2016linguistic} refine this model and introduce a semantic co-occurrence measure that leverages color-name frequencies. 
\lk{
El-Assady et al.~\cite{el2022semantic} developed a pipeline for mining semantic color associations from the literature. Schloss et al. developed a framework to facilitate colormap interpretation by ensuring a level of `semantic discriminability' for the represented categorical variables~\cite {schloss2018color,schloss2020semantic,mukherjee2021context}. However, some categorical visualizations might not contain clear semantic associations. Our technique instead relies on optimizing color name consistency to provide additional cues for discriminating displays with overlapping and potentially crowded elements.}  The impact of color names has been emphasized in recent studies, with results suggesting an important role for nameability in colormap interpretation~\cite{reda2020rainbows,reda2021color,reda2022rainbow}. Wang et al.~\cite{Wang2018} propose a scheme aimed at maximizing the discriminability of multi-class scatterplots. This approach considers the spatial relationships between classes along with color contrast against the background. Subsequently, Lu et al.~\cite{Lu21} introduced Paletailor, a framework that combines palette generation and color assignment into a unified optimization process, showing its effectiveness for various types of charts. Most recently, Languenou~\cite{languenou2023contrast} extends the technique introduced by Wang et al.~\cite{Wang2018} to deal with area-based visualizations like streamgraphs and chord diagrams. All of these techniques, however, are tailored to opaque visualizations, making them less suited for application translucent visualizations.

\subsection{Perception of Data Distributions in Histograms}
Histograms are commonly used to visualize univariate distributions, supporting several distribution analysis tasks~\cite{blumenschein2020v}. However, understanding such charts is not easy and is influenced by different factors. 
Lem et al.~\cite{lem2013misinterpretation} and Kaplan~\cite{kaplan2014investigating} show that students have
misinterpretations in identifying frequencies of specific data values within histograms even when they have the requisite knowledge and adequate time. 
Correll et al.~\cite{correll2018looks} found that the number of histogram bins heavily impacts the tasks of detecting missing values and outliers. Sahann et al. evaluated the influence of the bin number, finding that accuracy stabilizes around 20 bins and does not improve by adding additional bins~\cite{sahann2021histogram}. 

To simultaneously compare multiple distributions, overlapping histograms are a widely used method~\cite{wilke2019fundamentals}, even if they frequently suffer from false colors introduced by color blending. The most prevalent way to alleviate this problem is to overlay kernel density curves to accentuate the underlying. Blumenschein et al.~\cite{blumenschein2020v} introduced v-plots that combine mirrored bar charts and violin-style plots, which are designed to aid the comparison of distributions at different levels. However, these plots are limited to comparing two distributions only at a time. By contrast, translucent color-coded histograms can theoretically support a larger number of distributions within the same chart. In this paper, we attempt to enhance the perceptibility of overlapping histograms (and other types of visualizations employing transparency) by optimizing color and opacity settings, as well as the rendering order.

\section{Method}
Given a set of $m$ distributions $\mathbf{X}=\{1,\cdots, m\}$ over a specific background color $c_b$,  our goal is to 
find a proper color $c_i$, an opacity value $\mathbf{\alpha_i}$ and the rendering order $o_i$ for each distribution. 
Our method can be combined with any color blending method. However, because we intended for our method to work with the standard Porter-Duff alpha-blending model~\cite{porter1984compositing}, we employ the latter for all illustrations and evaluations in this paper, without loss of generality. Specifically, when blending two colors $c_i$ and $c_j$ with opacity values $\alpha_i$ and $\alpha_j$, respectively, the color compositing operator generates a blended color $c'$ and a corresponding opacity $\alpha'$ using linear interpolation:
\begin{equation}
\begin{split}
&\alpha' = \alpha_i + \alpha_j(1-\alpha_i) \\ 
& c' = \frac{\alpha_i c_i + \alpha_j c_j(1-\alpha_i)}{\alpha'} \nonumber
\end{split}
\end{equation}
We illustrate our method with overlapping histograms, as they represent one of the most common chart types employing transparencies.
We assume that each distribution has at least one non-overlapping section, and apply the above operator to render $m$ distributions, yielding $n$ (where $n>m$) parts and $n-m$ overlapping regions where the resulting color is a blend of multiple colors.
To convey the membership of each overlapping region to its respective parent distribution(s) and to meet the needs of categorical data visualizations~\cite{Wang2018,Lu21}, we propose three design requirements:
\begin{enumerate}[label=(\roman*),nosep]
\item \textbf{DR1:} a blended color should have a strong association with at least one (and ideally all) of the classes it is a member of;
\item \textbf{DR2:} a color should not have a strong association with another color if they do not share membership in at least one class; and
\item \textbf{DR3:} all colors, including colors resulting from the blending process, should be sufficiently discriminable. Smaller segments, in particular, should have high contrast against surrounding colors.
\end{enumerate}

To meet DR1 and DR2, we reinforce the association between color and category by leveraging color nameability~\cite{heer2012color,reda2021color}.
Specifically, we ensure that blended segments in the visualization retain a high name similarity with their original class color, while minimizing name similarity among segments belonging to different categories. Simultaneously, we optimize the discriminability of all color segments to meet DR3.

\subsection{Objective Function}
\begin{figure}[!ht]
	\centering
	\includegraphics[width=0.95\linewidth]{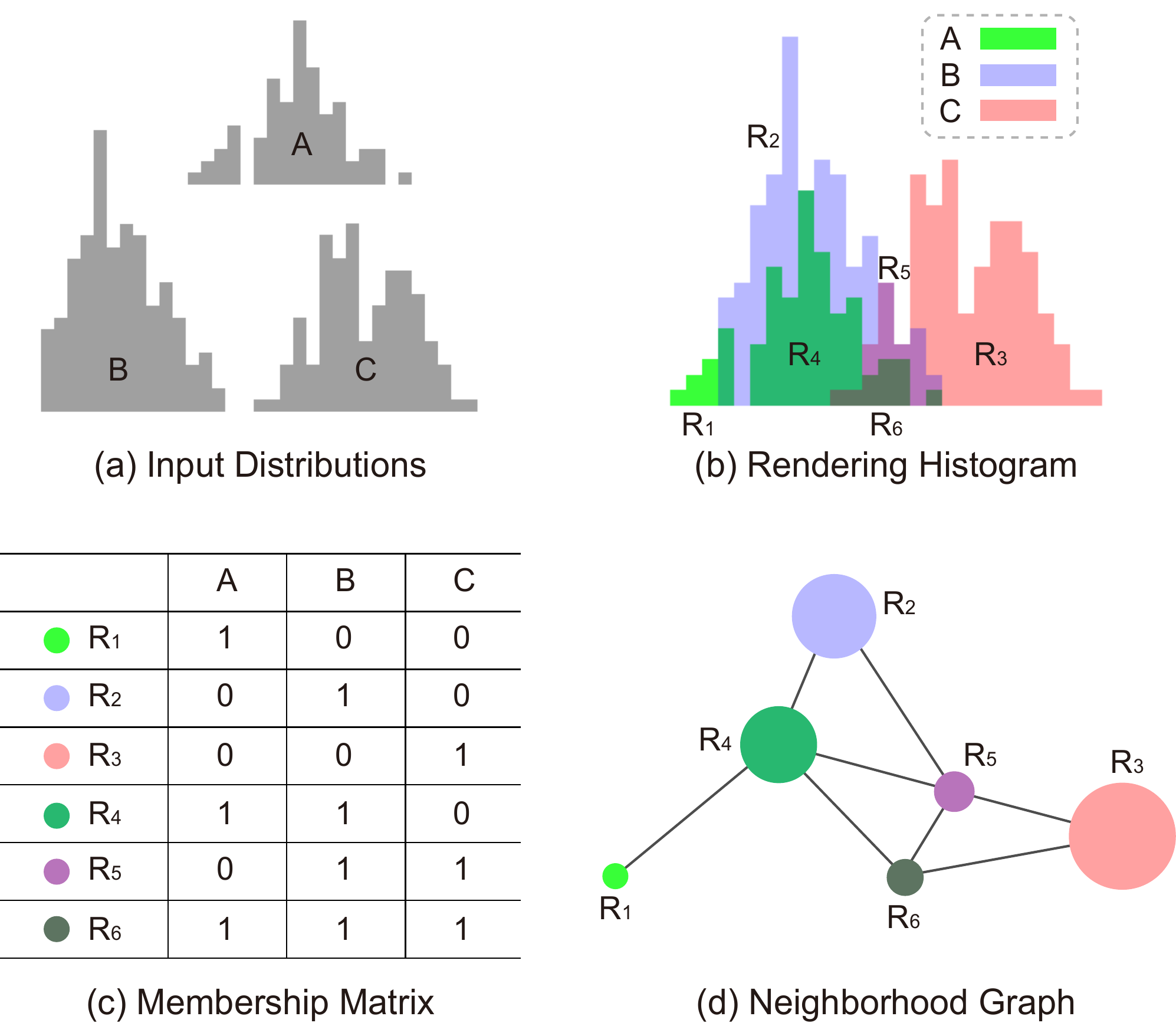}
	\caption{Preprocessing of the objective function. \lk{The input distributions are shown in (a). Rendering these three histograms (b) results in six regions $\{R_1, ..., R_6\}$, with some regions representing an intersection of two or more histograms. The binary membership matrix (c) lists the membership for each region, with $\mathbf{M}_{i,j}$ indicating whether region $i$ belongs to histogram (i.e., class) $j$. For example, the region $R_4$ belongs to the 1st class and 2nd class, which are classes A and B. The neighborhood graph (d) indicates region adjacency, with node size corresponding to the size of the region.}}
	\label{fig:structure}
	\vspace{-4mm}
\end{figure}

Based on the three design requirements, we formulate a search for an optimal color palette \revision{$\mathbf{P}= \{c_1, \cdots, c_m\}$}, a corresponding opacity set $\mathbf{A}= \{\alpha_1, \cdots, \alpha_m\}$, and a rendering order $\mathbf{O}= <o_1, \cdots, o_m>$ which collectively optimize the perception of the visualization. This optimization admits the following {objective function} $E(\mathbf{P}, \mathbf{A}, \mathbf{O})$:
\begin{equation}
\begin{split}
E(\mathbf{P},\mathbf{A}, \mathbf{O})~ &= 
   ~\omega_1 E_{WA}(\mathbf{P},\mathbf{A}, \mathbf{O}) \\
 &-~\omega_2 E_{BD}(\mathbf{P},\mathbf{A}, \mathbf{O}) \\[5pt]
 &+~\omega_3 E_{CS}(\mathbf{P},\mathbf{A}, \mathbf{O}) 
\end{split}
\label{eq:objectivefunction}
\end{equation}
Where the weights $\omega_i \in [0, 1]$ dictate the relative importance of the terms (all set to 1 by default).  
The three terms represent, respectively, Within-class Association (WA), Between-class Disassociation (BD), and Color Separability (CS). To facilitate the computation of these three terms efficiently, we pre-construct a $n \times m$ membership matrix $\mathbf{M}$, whereas $\mathbf{M}_{i,j}$
indicates whether the $i$th region $R_i$ of a histogram is part of the $j$th class. \autoref{fig:structure}-c illustrates how this membership matrix relates to the input distributions with an example. Here, $\mathbf{M}_{1,2}=0$ indicates that the region $R_1$ does not belong to the second histogram (i.e., class B) where \revision{$\mathbf{M}_{1,1}=1$ indicates that $R_1$ is part of the first histogram (class A).} 
\revision{The resulting colors of the $n$ regions are denoted by $\mathbf{C}= \{\mathbf{c}_1, \cdots, \mathbf{c}_m, \cdots, \mathbf{c}_n\}$, where the first $m$ regions, $\mathbf{c}_{i}$ ($i=1,\cdots, m$), are a mixture of the primary class colors defined by $\mathbf{P}$ and the background color $c_b$. These colors correspond to non-overlapping regions. The remaining colors, $c_j$ ( $n \geq j>m$), are composite colors that result from blending two or more colors from the palette $\mathbf{P}$. The opacity for each of the $m$ classes is determined by $\mathbf{A}=\{a_1, \cdots, a_m\}$, and the rendering order is determined by the sequence $\mathbf{O}=<o_1, \cdots, o_m>$. Note that our optimization is applied only to $\mathbf{P}$, $\mathbf{A}$, and $\mathbf{O}$. However, the full set of colors $\mathbf{C}$ is updated at each iteration to account for the blending results with the background and between overlapping histogram segments.}
We now describe each of the three optimization terms in detail.

\vspace{2mm}
\noindent\textbf{Within-Class Association}.
For each overlapping region, we define its degree of association with its `parent' classes based on color name similarity (recall that a region may belong to more than one class). We then maximize the name similarity between the blended color and its parent colors (\textbf{DR1}) in the objective function~\autoref{eq:objectivefunction}.
Accordingly, we can define within-class association degree as:
\begin{align}\label{eq:relateness}
   E_{WA}(\mathbf{P},\mathbf{A},\mathbf{O}) &= 
\frac{ \sum_{i \le m < j \le n} \Psi(i)^\frac{1}{2} {W_{i,j} } S(\mathbf{c}_i, \mathbf{c}_j)} {\sum_{i \le m < j \le n} W_{i,j}}  \nonumber \\
     &+ \min_{W_{i,j}>0}  S(\mathbf{c}_i, \mathbf{c}_j)
\end{align}

\revision{ The first term captures the \emph{average} name similarity ($S$) between a histogram $i$'s base class color and the colors of each of its constituent regions $j$. The second term finds the \emph{minimum} similarity within these pairings. The rationale for including the latter is to force the optimization to maximize both the mean and the lowest within-class name similarity, ensuring no histogram suffers from poor cohesion. Specifically,
$W_{i,j}$ represents the number of classes to which both region $i$ and region $j$ belong, which 
\revision{equals the dot product of the rows $i$ and $j$ of membership matrix $\mathbf{M}$.}
For example, $W_{1,5}=0$ indicates that the two regions ${R_1}$ and ${R_5}$, corresponding to the row vectors $(1,0,0)$ and $(0,1,1)$ in Figure~\ref{fig:structure}, are not members of a shared class, thus do not contribute to the final within-class association degree. By contrast, $W_{1,4}=1$, forcing a higher color similarity for $R_1$ and $R_4$.} 

$S(\mathbf{c}_i, \mathbf{c}_j)$  is the cosine-based name similarity between colors $c_i$ and $c_j$ proposed by Heer and Stone~\cite{heer2012color}:
\begin{equation}\label{eq:namesimilarity}
 S(\mathbf{c}_i, \mathbf{c}_j) = cos(\mathbf{T}_{\mathbf{c}_i}, \mathbf{T}_{\mathbf{c}_j})=\frac{\mathbf{T}_{\mathbf{c}_i} \mathbf{T}_{\mathbf{c}_j}}{||\mathbf{T}_{\mathbf{c}_i}|| ||\mathbf{T}_{\mathbf{c}_j}||},
\end{equation}
where  $\mathbf{T}$  is a color-term count matrix collected by an online study, and $\mathbf{T}_{\mathbf{c}_i}$ is the probability distribution of color names for a given color $\mathbf{c}_i$. A larger value of $ S(\mathbf{c}_i, \mathbf{c}_j) \in [0, 1]$  indicates the two colors are more likely to share the same name. 
We incorporate a class-specific weight  $\Psi(i)$ in ~\autoref{eq:relateness} based on two considerations: i) distributions that comprise a larger number of overlapping segments should be given a higher optimization priority; and ii) when the ratio of the overlapping to non-overlapping segments in one distribution is large, the corresponding class-weight is similarly increased. Accordingly, $\Psi(i)$ is defined as:
\begin{equation}\label{eq:overlapDegree}
 \Psi(i) = \frac{{\sum_{j=1}^n \mathbf{M}_{j,i}}}{\max_
{i \leq m }{\sum_{j=1}^n \mathbf{M}_{j,i}}}  
 \left(1- \frac{RS_i}
 {\sum_{j=1}^n{W_{i,j} RS_j} }\right),
\end{equation}
where $\sum_{j=1}^n \mathbf{M}_{j,i}$ is the number of \lk{regions that make up the $i$th class, computed by summing up column $i$ in the membership matrix.} \revision{
The size of the $i$-th region, denoted by $RS_i$, represents the fraction of pixel space occupied by the region relative to the entire visualization footprint}. Accordingly, $\sum_{j=1}^n{W_{i,j} RS_j} $ reflects the aggregate size of the $i$th distribution. 
In essence, the value of $\Psi(i)$ allows us to heuristically score the `difficulty' of perceiving distribution $i$, allowing the optimization to prioritize potentially problematic features.

\vspace{2mm}
\noindent\textbf{Between-Class Disassociation}.
While the within-class association term (\autoref{eq:relateness}) ensures that a single distribution with its overlapping segments is perceived as a coherent whole, we further need to maximize color dissimilarity between different distributions so that they can be perceived as separate entities. We achieve this by defining a between-class disassociation term. Specifically, this term assigns colors with larger color-name distances to all segments that do not belong to the same distributions:
\begin{equation}\label{eq:unrelateness}
\begin{split}
E_{BD}(\mathbf{P},\mathbf{A},\mathbf{O}) &= \frac{\sum_{i < j \le n } \delta(W_{i,j}) S(\mathbf{c}_i, \mathbf{c}_j)  }{\sum_{i < j \le n } \delta(W_{i,j})} \\
  &+ \max_{W_{i,j}=0 } S(\mathbf{c}_i, \mathbf{c}_j) ,
\end{split}
\end{equation}
Where $\delta(x)$ returns 1 if $x = 0$ and 0 for all other $x$ values. \revision{The first term computes the average color name similarity for pairs of regions that do not belong to the same class. For example, $\delta(W_{1,3})=1$ since $R_1$ and $R_3$ are not members of the same class, thus their similarity will be included in the average.}
\revisionB{By contrast, $\boldsymbol{R}_5$ and $\boldsymbol{R}_6$ share two classes, with $W_{5,6}=2$. Accordingly, $\delta(W_{5,6})=\delta(2)=0$ thus preventing these two regions from being penalized for having similarly sounding names.} \revision{The second term finds the highest name similarity among these pairings. Combining the two terms in the objective function minimizes both the mean and the highest similarity between disjoint histograms, thereby yielding distinct colors for regions that do not belong to the same class(es).}  In doing so, we ensure that \textbf{DR2} is met.

\vspace{2mm}
\noindent\textbf{Color Separability}.
\textbf{DR3} requires that all colors should be sufficiently discriminable. Smaller regions, in particular, should have high contrast against their surroundings so that they can be perceived sufficiently.
To incorporate this notion, we define a color separability term based on a region-based neighborhood graph (see \autoref{fig:structure}-d for illustration).
Since region size influences the perception of the color difference~\cite{szafir2017modeling} (specifically, smaller regions require larger differences), we operationalize the latter and define our separability term as:
\begin{equation}\label{eq:separability}
 E_{CS}(\mathbf{P},\mathbf{A},\mathbf{O}) = \min_{\forall i \leq n, \forall {R}_j \in \Omega(R_i)} D(\mathbf{c}_i, \mathbf{c}_j) \big(1+RS_i\big),
\end{equation}
where $\Omega(R_i)$ indicates all neighborhoods of the region $R_i$, \revised{
$D(\mathbf{c}_i, \mathbf{c}_j)$ represents the CIEDE2000 color distance ~\cite{sharma2005ciede2000} between two colors $\mathbf{c}_i$ and $\mathbf{c}_j$. In ~\autoref{eq:separability}, the color distance $D(\mathbf{c}_i, \mathbf{c}_j)$ is weighted by the area of the distribution $(1 + RS_i)$, as smaller areas make colors more difficult to distinguish. The additional constant ensures that the measure is not highly sensitive to very small $RS_i$ values.
This measure is computed for all neighboring pairs, with the overall performance determined by the worst separability, and thus, the minimum of all pairwise comparisons is used.}

Additionally, we introduce two hard constraints for discriminability: A \textbf{just-noticeable difference} constraint ensuring that all colors are perceivable:

\begin{equation}\label{eq:jnd}
 \forall i \leq n, j \leq n, i \neq j, D(\mathbf{c}_i, \mathbf{c}_j)>\eta ,
\end{equation}
where $\eta$ is a just noticeable difference (JND) threshold, which is 3 by default~\cite{yang2012color}. A second, \textbf{background contrast} constraint ensures that colors sufficiently stand out:

\begin{equation}\label{eq:bc}
 \forall i \leq n,  LD(\mathbf{c}_i, \mathbf{c}_b) \geq \sigma 
\end{equation}
where $LD(\mathbf{c}_i, \mathbf{c}_b)$ is the absolute luminance difference between the color $\mathbf{c}_i$ and background color $\mathbf{c}_b$ in the CIELAB space, and $\sigma$ is a threshold with the default value 5 units \revision{(approximately twice the empirical JND~\cite{jnd1994}), thus ensuring that the difference is perceptible to most people.}

\begin{algorithm}[!thb]
\caption{Customized Simulated Annealing}
\label{alg:sa}
\begin{algorithmic}[1]
\State randomly initializing color mapping $\mathbf{P}_0$ with $m$ classes, randomly initializing the set of opacity values $\mathbf{A}_0$, randomly initializing the rendering order $\mathbf{O}_0$
\State set initial temperature $T_0$, cooling coefficient $\gamma$
\State set the best color mapping $\mathbf{P}_{best}=\mathbf{P}_0$, opacity sets $\mathbf{A}_{best}=\mathbf{A}_0$ and rendering order $\mathbf{O}_{best}=\mathbf{O}_0$
\State $t=0$
\While{$T_t > T_{end}$}
\State set the current color mapping $\mathbf{P}=\mathbf{P}_t$, current set of opacity values $\mathbf{A}=\mathbf{A}_t$ and current rendering order $\mathbf{O}=\mathbf{O}_t$

\Do
\State p $\leftarrow random(0,1)$
\If{$p<\frac{1}{3}$} 
    \State randomly disturbs one color of $\mathbf{P}$
\ElsIf{$p<\frac{2}{3}$} 
    \State randomly changes one opacity value of $\mathbf{A}$
\Else 
    \State randomly exchanges two colors' rendering order of $\mathbf{O}$
\EndIf
\doWhile{$\exists D(\mathbf{c}_i, \mathbf{c}_j) < \eta $ \textbf{or} $\exists LD(\mathbf{c}_i, \mathbf{c}_b) < \sigma $}

\State $\Delta E=E(\mathbf{P}, \mathbf{A}, \mathbf{O})-E(\mathbf{P}_t, \mathbf{A}_t, \mathbf{O}_t)$
\If{$\Delta E>0$}
\State $\mathbf{P}_{t+1} \leftarrow \mathbf{P}$, $\mathbf{A}_{t+1} \leftarrow \mathbf{A}$, $\mathbf{O}_{t+1} \leftarrow \mathbf{O}$
\Else
\If{$random(0,1) < \exp(\Delta E /T_t)$}
\State $\mathbf{P}_{t+1} \leftarrow \mathbf{P}$, $\mathbf{A}_{t+1} \leftarrow \mathbf{A}$, $\mathbf{O}_{t+1} \leftarrow \mathbf{O}$
\EndIf
\EndIf

\State $T_{t+1} = \gamma T_t$, \; $t \leftarrow t+1$
\EndWhile
\end{algorithmic}
\end{algorithm}

\subsection{Optimization Strategy}
 
In addition to optimizing the color palette $\mathbf{P}$ and the opacity $\mathbf{A}$, we also optimize the rendering sequence $\mathbf{O}$ for the $m$ distributions. This is crucial as the outcome of alpha blending depends on the rendering sequence, which impacts the quality of the visualization. 
To efficiently solve this optimization, we developed a customized simulated annealing~\cite{aarts1989stochastic} algorithm (see Algorithm~\ref{alg:sa}). The algorithm starts at a high `temperature' that is gradually `cooled'. It begins with a random solution for each $(c_i, \alpha_i, o_i)$ and updates this solution at each iteration in two major steps: i) generating a new candidate solution from the current one, and ii) comparing the new candidate to the old solution and deciding whether to accept it. In line with prior work on simulated annealing for color selection, we choose a cooling coefficient of $\gamma = 0.99$, the starting temperature is $T=100,000$ and the ending is $0.001$~\cite{Lu21}. We next explain the above two steps in detail.

\vspace{2mm}
\noindent\textbf{Generating a new solution (lines 7-16):}
 At every iteration of the optimization, we generate a random solution that is one step removed from the current solution. This is done by randomly choosing from three possible perturbation mechanisms \lk{that are equally probable (lines 9, 11, and 13)}: i) slightly perturbing one color in the current palette $\mathbf{P}$ with a small RGB offset \lk{(in our implementation, we set this offset to 10 units for each component)}, ii) adjusting the opacity value $\mathbf{A}$ for one of the classes with a small offset \lk{(0.1 by default)}, and iii) swapping the rendering order $\mathbf{O}$ of two randomly selected colors. 
 The perturbation results in a new candidate solution that is then tested against the two hard constraints in~\autoref{eq:jnd} or~\autoref{eq:bc}. If the solution fails either of the two constraints, it is rejected, with a new candidate solution generated instead. \revision{We perform color perturbation in the RGB space to avoid the challenges associated with manipulations in the CIELAB space, where even small changes can yield colors outside the RGB gamut. This requires additional constraints, such as gamut clipping or mapping, introducing unnecessary complexities into the algorithm.}

\vspace{2mm}
\noindent\textbf{Accepting a new solution probabilistically (line 17-24):}
We score the new candidate solution $(\mathbf{P}_t, \mathbf{A}_t, \mathbf{O}_t)$ (line 17), accepting it if improves the score relative to the current solution. If, on the other hand, the candidate is objectively worse, we accept it with a probability of $e^{{\Delta E} / {T_t}}$, where $\Delta E$ is the difference in score between the new and the current solution and $T_t$ is the current temperature. Higher temperatures yield a higher probability of accepting a worse solution, which reduces the likelihood of encountering local minima earlier in the optimization.

\begin{figure}[t]
	\centering
	\includegraphics[width=0.98\linewidth]{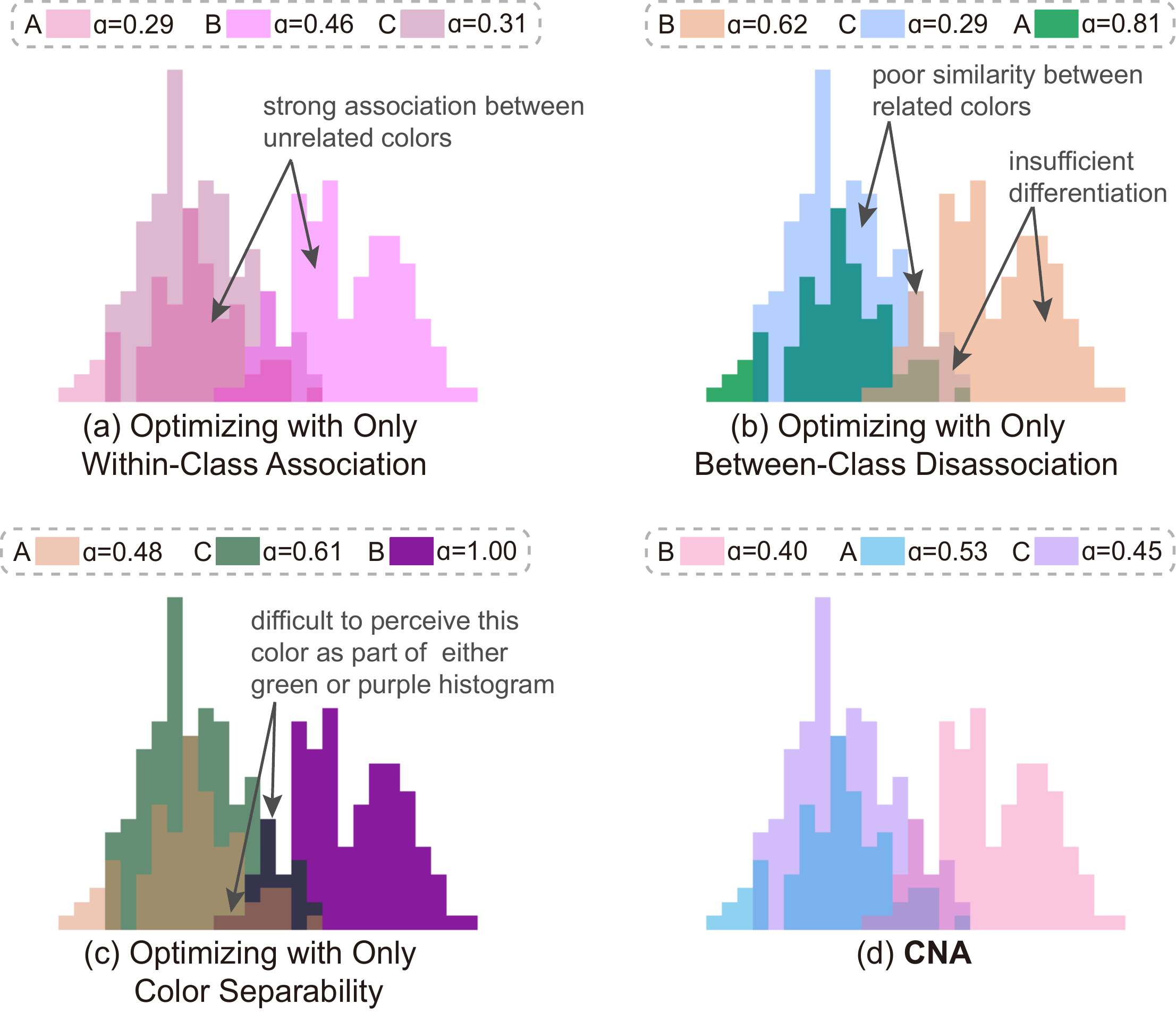}
	\caption{
 \lk{
 The influence of different weight settings ($\omega_1$, $\omega_2$, $\omega_3$) on the final colorization results, showing the contribution of each of the three objective-function terms individually. (a) illustrates the effects of optimizing within-class association only (1, 0, 0). (b) shows between-class discrimination only (0, 1, 0). (c) illustrates the effect of maximizing color separability (0, 0, 1). (d) illustrates the final result with all three terms equally weighted (1, 1, 1).} The legend also depicts the optimized rendering order (from top to bottom). \lk{The input distributions are the same as those shown in \autoref{fig:structure}-a.}
	}
	\label{fig:weightInfluence}
	\vspace{-4mm}
\end{figure}

\subsection{Parameter Analysis}

The three weights ($\omega_1, \omega_2, \omega_3$) in \autoref{eq:objectivefunction} exert significant influence on the final results, as shown in \autoref{fig:weightInfluence}.
When optimizing within-class association only, the overlapping regions within each distribution tend to be similar, facilitating easy grouping. Yet, establishing global membership across different distributions is still challenging. For instance, in \autoref{fig:weightInfluence}-a, similar colors (three pink shades) are assigned to different classes. By optimizing for between-class disassociation only (b), differentiation between classes becomes straightforward. Still, some distribution segments are rendered indistinguishable. 
The latter is helped by color separability, although this factor alone yields a low-quality solution  (c). Conversely, by integrating all three optimization terms (d), we obtain clearer class membership along with improved separability from both neighboring colors and the background. 
\revised{We assume that the three factors—within-class association, between-class disassociation, and color separability—are equally important. As a result, we assign equal weights to each factor and use this combination to generate all the results presented in this paper.}

\begin{figure}[t]
	\centering
	\includegraphics[width=0.98\linewidth]{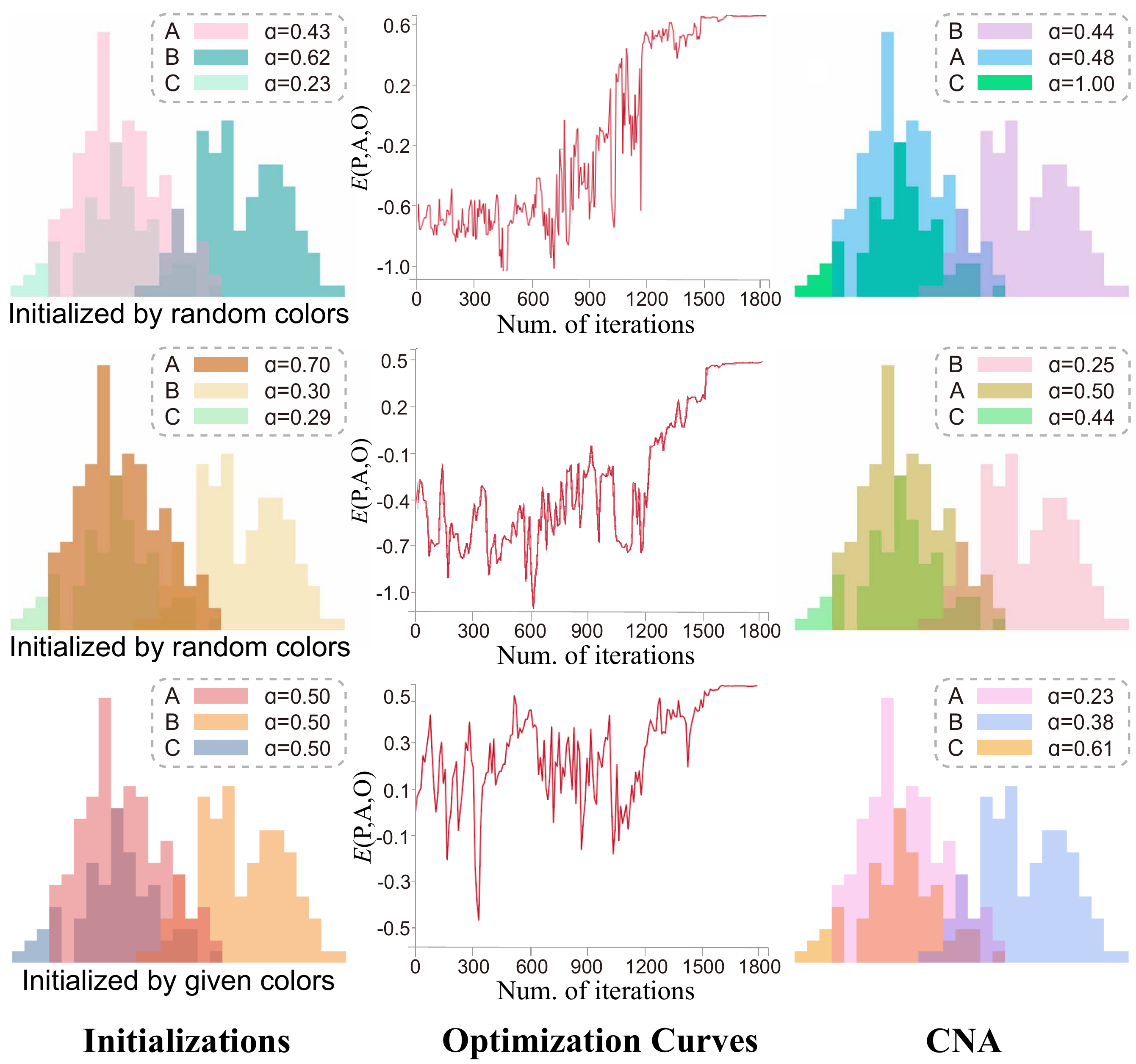}
	\caption{
 \revised{
  Different initializations (left) and the corresponding optimization results (right). The plots of E(P,A,O) versus the number of iterations illustrate the convergence of the proposed method (middle).
  }
	}
	\label{fig:initialization}
	\vspace{-4mm}
\end{figure}

\revised{
Although individual results may vary between runs of simulated annealing, our approach consistently converges to a robust and effective colorization solution. As shown in ~\autoref{fig:initialization}, even with different initializations, our method reliably produces satisfactory results. The convergence curves exhibit significant oscillations in the early stages, but the algorithm stabilizes at the end of the run.} \revision{Notably, when the initial colors are specified rather than randomized, as illustrated in the bottom row of ~\autoref{fig:initialization}, the method remains capable of identifying suitable results. However, we adopt random initialization by default as typically used in simulated annealing to encourage greater solution diversity.}

\newcommand{\standardBlending}{\emph{Standard}}
\newcommand{\localBlending}{\emph{Local}}
\newcommand{\hueBlending}{\emph{Hue-preserving}}
\newcommand{\ourBlending}{\emph{CNA}}
\newcommand{\standardOutline}{\emph{Standard with Curve}}
\newcommand{\ourOutline}{\emph{CNA with Curve}}

\section {Quantitative Evaluation}
\label{sec:evaluation}

\begin{figure}[t]
	\centering
	\includegraphics[width=0.95\linewidth]{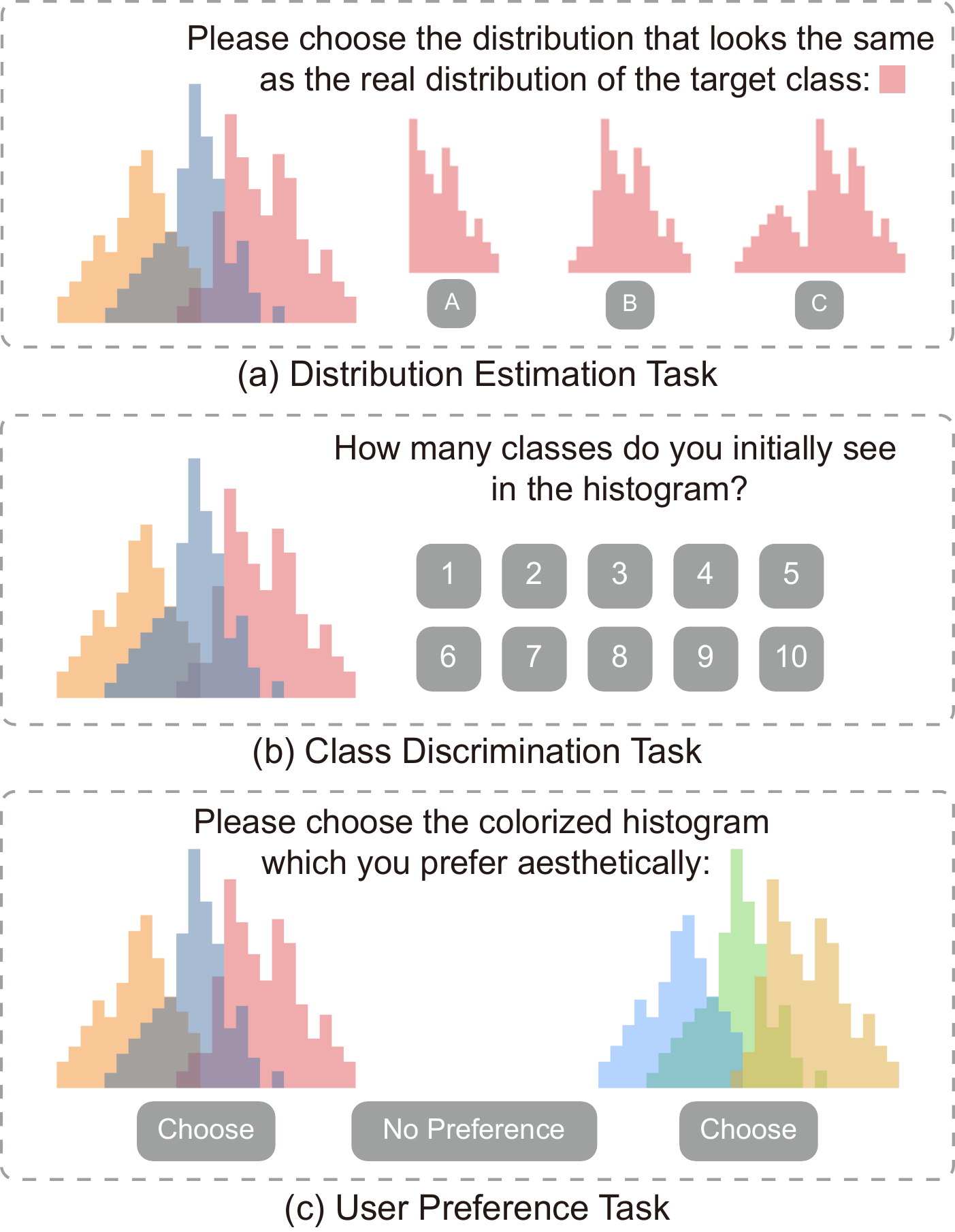}
	\caption{Illustration for the three tasks used in our experiments: (a) \emph{distribution estimation}, (b) \emph{class discrimination}, and (c) \emph{user preference}. \lk{The correct answer for the distribution estimation task is option B, while the correct number of classes in the discrimination task is three classes.}
	}
	\label{fig:tasks}
	\vspace{-4mm}
\end{figure}

We present a quantitative evaluation of our optimization technique. Our primary focus is to assess the extent to which our method enhances the analysis of semi-transparent, overlapping histograms. These multi-class histograms are commonly used for the analysis of distributions (e.g., comparing the distribution of a quantitative variable across multiple class stratification). We adopt three experimental tasks (see \autoref{fig:tasks}) and employ distributions with varying levels of smoothness and overlap. We compare our technique against three benchmarks: standard alpha blending ~\cite{porter1984compositing}, local color blending \cite{wang2008color}, and hue-preserving color blending ~\cite{chuang2009hue}. For these benchmarks, we employ color palettes selected from either Tableau~\cite{tableau} or generated through Colorgorical~\cite{Gramazio17}, employing uniform opacity and the same rendering order to simulate common usage conditions. 
By contrast, our optimization generates a color palette, opacity values, and rendering order tailored to the characteristics of the distributions. Additionally, because histograms are often shown accompanied by kernel density estimates~\cite{wilke2019fundamentals}, we test this setup in a second study \S\ref{sec:study2}. We conducted both studies on Prolific, recruiting a total of 160 participants\lk{, ranging in age from 18 to 65 years. The minimum education level among all participants was high school, with most participants having earned a bachelor's degree}. 

\begin{figure}[t]
	\centering
	\includegraphics[width=1\linewidth]{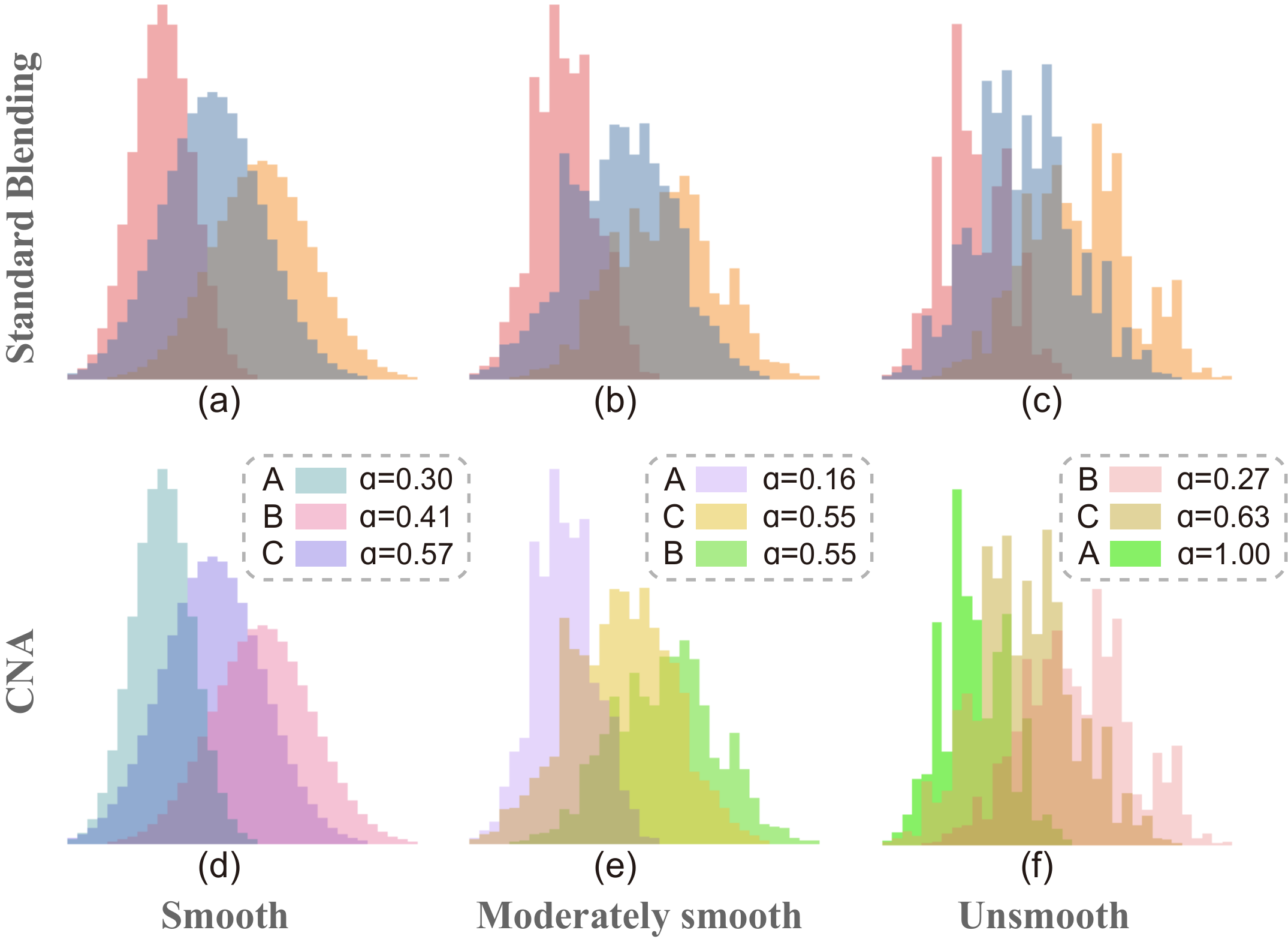}
	\caption{
 The impact of distribution smoothness. 
 (a, d) show smooth histograms representing ideal Gaussians. (b, e) are moderately smooth histograms produced by adding perturbation. (c, f) are unsmooth with an even higher level of perturbation, yielding more challenging stimuli.
	}
	\label{fig:shapeIllus}
	\vspace{-4mm}
\end{figure}

\label{smoothnessDefine}
\noindent{\textbf{Stimuli and Palette Generation.}}
We created a collection of 18 multi-class stimuli \revision{(see Figure 7 in the supplementary material)}, each comprising overlapping histograms with two, three, or four classes, with six stimuli for each class number. These histograms displayed a diverse range of characteristics, including varying degrees of distribution smoothness and overlap. To construct each stimulus, we sampled from a Gaussian distribution with a random standard deviation ($\sigma \in [3, 5]$) and a random scaling factor ($A \in [0.8, 1.2]$). After generating the initial Gaussian, we introduced random perturbations to induce discontinuities in the distribution. The level of perturbation dictates the difficulty of resolving the histogram; histograms that are closer to an idealized smooth Gaussian are easier to perceive, as they provide sufficient visual continuity and symmetry (see \autoref{fig:shapeIllus}). To quantity smoothness, we employed the Kullback-Leibler (KL) divergence to gauge the dissimilarity between the perturbed distribution and its ideal Gaussian. We categorized the smoothness into three levels: smooth, moderately smooth, and unsmooth. These levels correspond to KL values of $0$, [$0.02, 0.04$], and [$0.07, 0.1$], respectively. We factor each smoothness level with the number of classes (two repetitions each), giving us a total of 3 \emph{classes} $\times$ 3 \emph{smoothness levels} $\times$ 2 \emph{repetitions} $= 18$ overlapped histograms. \lk{Then for each histogram, we manually designed three response choices: One of the choices was the actual (i.e., correct) distribution. The other two options representing confounders correspond to the base histogram with overlapping segments either added or removed from the distribution.}

To color-code the overlapped histograms in the benchmark conditions, we utilized two sources: a designer-crafted palette and an auto-generated palette created using Colorgorical~\cite{Gramazio17}. For the designer-crafted palette, we employed the Tableau-10 categorical scheme~\cite{tableau}, known for its quality and discriminability. From this palette, we randomly selected subsets of two, three, or four colors to assign to the classes in the stimuli. For the Colorgorical-generated palette, we utilized the low-error setting, generating a palette with eight base colors. Similarly, we randomly sampled colors from this palette, depending on the number of classes required for the stimulus. \lk{We simplified the generation process of Colorgorical for a fair comparison with Tableau, but it is worth noting that Colorgorical is designed to return an optimal palette given the number of colors queried. It is possible that randomly selecting a subset of 4 colors from an 8-color palette will result in a worse palette than querying the tool for a 4-color palette. }

\subsection{Study I: Histograms}
\label{sec:study1}
In this experiment, we evaluate how well people perceive translucent overlapping histograms without additional visual cues. We compare our method against the three benchmarks outlined above. We utilize three tasks adapted from prior work on distribution analysis~\cite{blumenschein2020v, sahann2021histogram, Wang2018, Lu21, Lu23}:

\begin{itemize}[leftmargin=*]

    \item \emph{Distribution Estimation}:  Based on literature in histogram visualization~\cite{blumenschein2020v, sahann2021histogram}, we adopted a task of ``describing and identifying the shape and type of a distribution.'' In this task, participants are asked to report the shape of the distribution for a target class, given \lk{three options (including two confounders and one correct choice, as shown in} \autoref{fig:tasks}-a). We record participants' accuracy (a binary measure of whether the response is correct) and response time.
    
    \item \emph{Class Discrimination}: This task aims to evaluate participants' ability to accurately discern the number of distinct classes presented in the visualization, thus assessing color discriminability and the potential presence of false colors. As shown in \autoref{fig:tasks}-b, participants are asked to count and report the number of distinct categories they perceive in the visualizations, with options ranging from 1 to 10. We similarly analyze the response time and relative error (the difference between the true number and the participant's selection). To facilitate comparisons across stimuli, we normalize the error by the number of classes in the visualization. 

    \item \emph{User Preference}: This task measures the aesthetic appeal of our generated colors. Participants are presented with a pair of visualizations depicting the same histogram data but using two distinct color schemes: one derived from our optimization and the other from the benchmark sources. An example is shown in \autoref{fig:tasks}-c. Participants are prompted to select the visualization they find more appealing. 
\end{itemize}

\vspace{2mm}
\noindent{\textbf{Conditions.}} We conduct a comparison of our approach (referred to as \ourBlending) against three benchmarks: traditional alpha blending (\standardBlending), local color blending (\localBlending) and hue-preserving blending (\hueBlending). The latter two are visualization-specific blending approaches: the hue-preserving model is designed to minimize the appearance of false features by ensuring that the blended colors maintain a consistent hue with the original class color~\cite{chuang2009hue}. On the other hand, local blending seeks to reduce false colors by preserving the perceived order of color layers~\cite{wang2008color}. This is achieved by selectively desaturating `background' objects while retaining similar hue and lightness levels. For the three benchmarks, we sampled colors from the Tableau-10 scheme or Colorgorical (see `Palette Generation'). To assign opacity values, we followed the guidelines from Matplotlib~\cite{matplotlib}, adopting a uniform opacity of $0.5$. By contrast, our approach generates colors, opacity values, and rendering order algorithmically according to the optimization, with results composited using standard alpha blending. The four experimental conditions are illustrated in ~\autoref{fig:teaser}.

\vspace{2mm}
\noindent{\textbf{Hypotheses.}}
Our method is designed to increase both class cohesion and color discriminability. Hence, we expect our optimization to improve user performance in the two tasks:

    \vspace{2mm}\noindent\textbf{H1 --} In the \emph{Distribution Estimation} task, we anticipate our method to improve the perception of distribution features over the benchmarks. Therefore, we expect participants to perform better shape estimation with \ourBlending~than with \standardBlending, \localBlending, and \hueBlending.
    
    \vspace{2mm}\noindent \textbf{H2 --} In the \emph{Class Discrimination} task, we similarly predict our method will improve participants' performance in class discrimination, outperforming a fixed color assignment strategy (\standardBlending, \localBlending, and \hueBlending).

\vspace{2mm}\noindent Furthermore, we anticipate the optimized colorization from our approach to result in an aesthetically pleasing color scheme:

    \vspace{2mm}\noindent \textbf{H3 --} Participants will show comparable preference for \ourBlending~and the three benchmarks (\standardBlending, \localBlending, \hueBlending).

\vspace{2mm}
\noindent{\textbf{Experiment Design.}} 
\lk{The task was varied between subjects, with different participants recruited for each of the three tasks.} The experimental condition was a within-subject variable, with each participant experiencing all four colorization methods. To minimize potential learning effects, we displayed the stimuli in a random order. We recruited participants from Prolific, which is shown to generate higher quality responses as compared with other crowdsourcing platforms~\cite{douglas2023data}. \lk{We limited recruitment to residents of the UK and US, while balancing the proportion of male and female through Prolific recruitment options.}

To ensure the accuracy of the result, we asked participants to complete a color deficiency test at the beginning of the experiment and randomly inserted multiple engagement checks within the stimulus sequence. These checks featured two completely separated Gaussian distributions in easily distinguishable colors, making them straightforward to judge. Participants who did not correctly resolve these checks were excluded from the analysis.

After the participants passed the color deficiency test, they were presented with task instructions and completed three training trials. Following the training phase, they proceeded to complete the analyzed trials, which consisted of either 72 stimuli (18 histograms $\times$ 4 methods) for the distribution estimation task and class discrimination task, or 54 stimuli (18 histograms $\times$ 3 pairs) for the user preference task.  Each stimulus had a response time limit of 30 seconds, after which participants would be directed to the next trial.

\begin{figure}[t]
\centering
\includegraphics[width=1\linewidth]{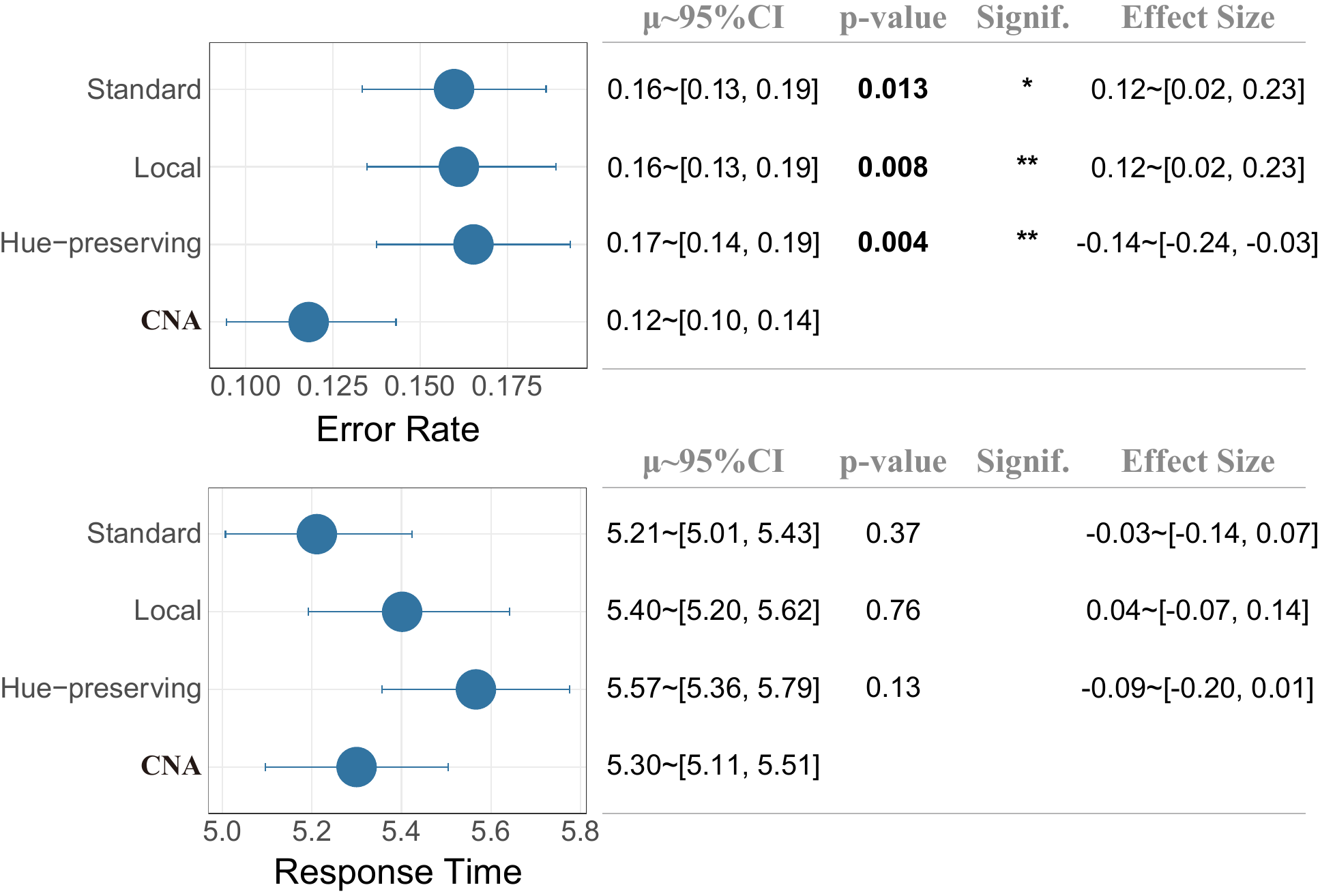}
\vspace{-2mm}
\caption{
Results for the {\emph{Distribution Estimation Task}} from {Study I}, including effect sizes and significance tests for our method against the benchmarks. Error bars ($\mu \sim$ 95\%CI), p-value, and significance level are calculated from the Mann-Whitney test. Effect size is calculated using Cohen's d.}
\label{fig:shapeResult}
\vspace{-4mm}
\end{figure}

\subsubsection{Results for Distribution Estimation} 
\label{sec:shape_task}

A total of 40 participants (18 males, 21 females, and 1 unspecified) were included in the analysis for this task. \lk{We initially excluded five participants who failed the attention checks, and replaced them with new recruits to reach our intended sample size of 40 responses}. On average, participants completed the experiment in 7.20 minutes ($\sigma=3.37$). \lk{We set the compensation at $\$1.5$ to exceed the US minimum hourly wage.} Our analysis includes 95\% confidence intervals, effect sizes (Cohen's $d$) for both error rate and response time, and p-values derived from Mann-Whitney tests (based on a positive Shapiro-Wilk test). Given the non-normality of the error data, we employed the GFD R-package ~\cite{gfd2017} to compute ANOVA-type interaction statistics.

\autoref{fig:shapeResult} illustrates the results for this task. We find that our proposed method (CNA) outperforms the benchmarks, which feature fixed color and opacity assignment (\standardBlending, \localBlending, \hueBlending), both in terms of error rate and response time. Notably, our method demonstrates a significantly lower error rate than all three benchmark conditions. We found an interaction between the colorization methods and the class number ($p=0.041$) \lk{for the error rate}. However, there was no significant interaction between the experimental conditions and smoothness ($p=0.244$). Hence, while the number of classes seems to affect the tested methods somewhat differently, the smoothness of the distribution appears to have a consistent effect, with unsmooth distributions uniformly more challenging across all tested techniques, including ours (see \autoref{fig:correlation}). 

\lk{We analyzed response times by first removing extreme outliers to reveal more systematic effects for the visualization technique. This was done by pruning participant trials that are $\pm2$ standard deviations away from their mean response time. There was no significant difference in response time between our approach and the three benchmarks, suggesting that our optimization does not significantly change response time.} These results are consistent with \textbf{H1}.

\begin{figure}[ht]
    \centering
    \includegraphics[width=1\linewidth]{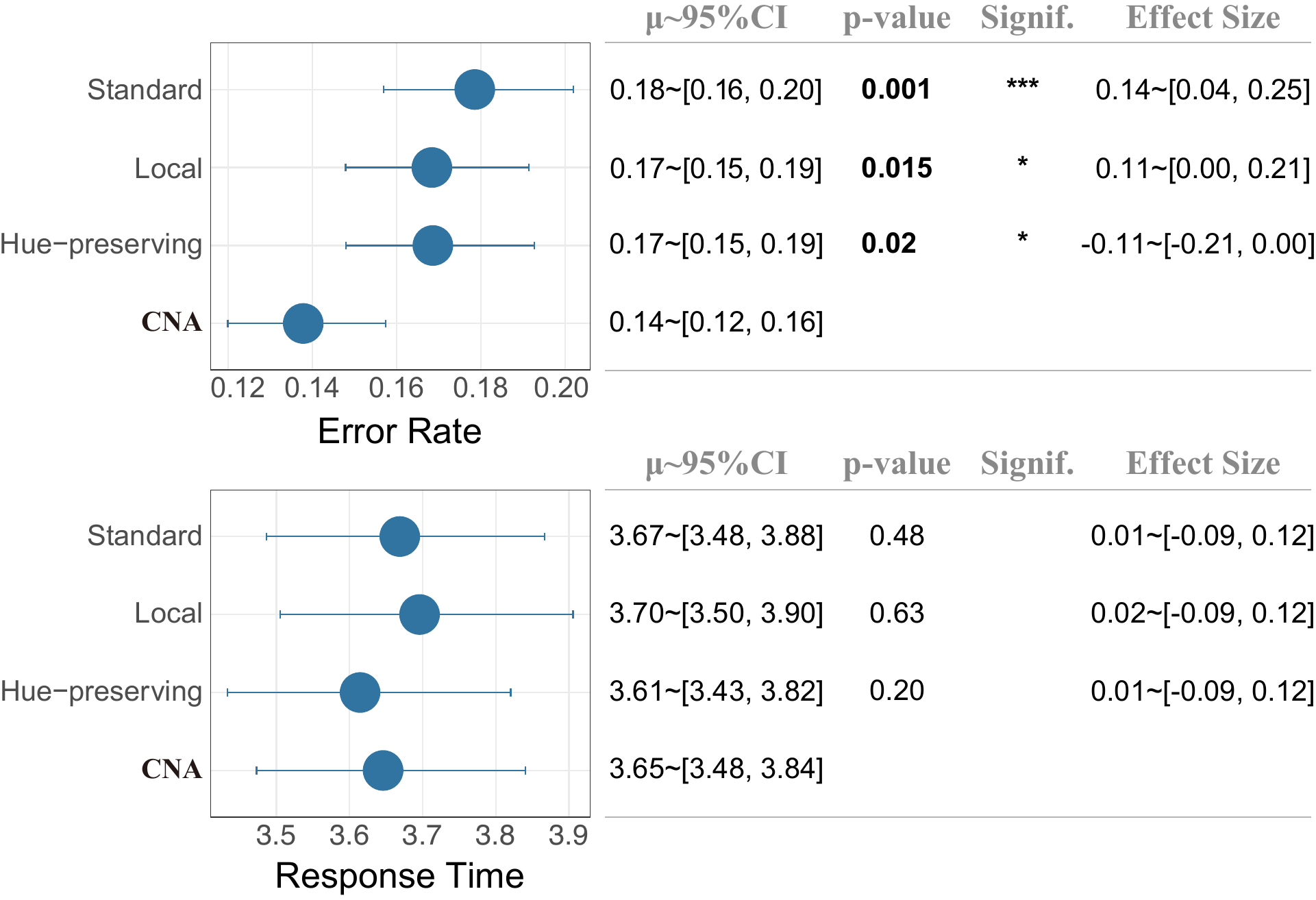}
    \vspace{-2mm}
    \caption{
    Results for the {\emph{Class Discrimination Task}} from {Study I}.
    }
    \label{fig:countResult}
    \vspace{-4mm}
\end{figure}

\subsubsection{Results for Class Discrimination}
We similarly recruited 40 participants (20 males, 20 females) for this task\lk{. Three participants were initially excluded for failing attention
checks, with additional participants recruited in replacement}. Each participant completed a total of 72 stimuli, with an average completion time of 5.26 minutes ($\sigma=3.62$) for the entire experiment. \lk{We compensated participants with a $\$1.2$ payment to exceed the US hourly minimum wage.} The results are depicted in \autoref{fig:countResult}. The error rate with our optimization method was consistently lower than that of the fixed assignment strategy. Specifically, our optimization achieved a significantly reduced error rate compared to the benchmarks (\standardBlending, \localBlending, \hueBlending). These results are consistent with \textbf{H2}.  We found no significant interaction effect between the experimental condition and the number of histogram classes ($p=0.267$), or with the distribution smoothness ($p=0.575$). This suggests that all tested techniques are similarly affected by the complexity of the visualization. In terms of response time, we found no significant difference between the four methods, even CNA led to a faster response, on average.

\subsubsection{Results for  User Preference}
\begin{figure}[h]
	\centering
	\includegraphics[width=1\linewidth]{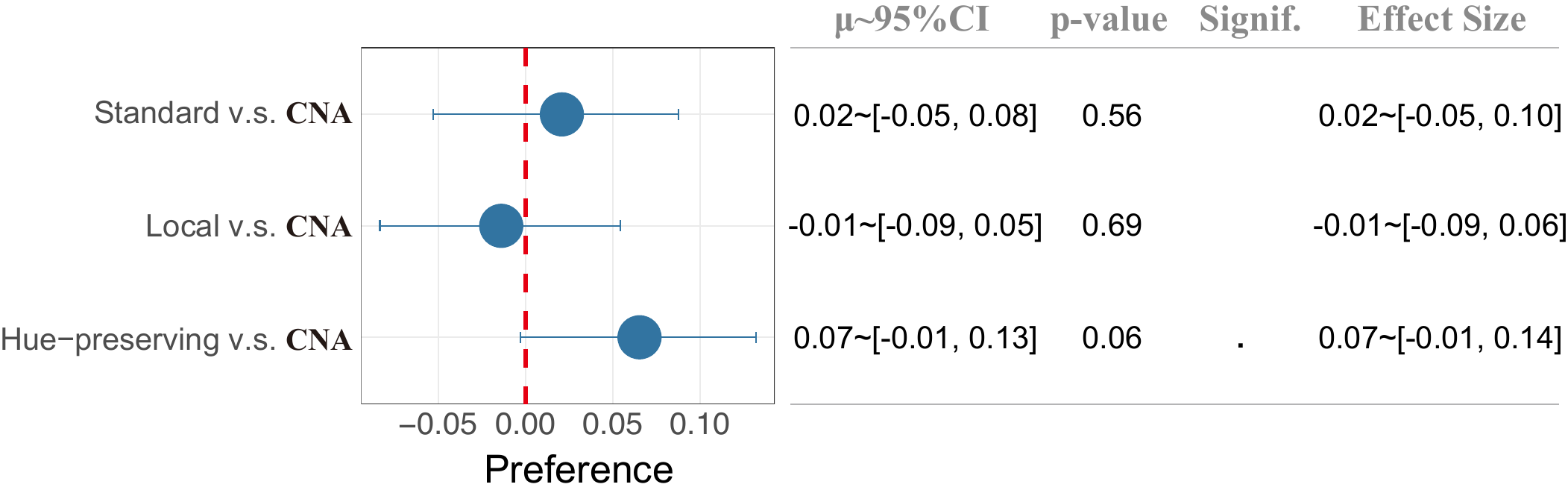}
	\vspace{-2mm}
	\caption{Results for the {\emph{User Preference Task}} from {Study I}. A preference score greater than 0 indicates that participants prefer our method over the respective benchmark.}
	\label{fig:preferResult}
	\vspace{-4mm}
\end{figure}

We enrolled 40 participants (20 males, and 20 females\lk{, with no participants excluded}). \lk{ Participants received a compensation of $\$0.75$}. For each comparison, we assigned a score of 1 if the participant preferred our approach and -1 for preferring one of the other benchmarks. Zero was assigned for a neutral choice when the participant indicated no clear preference. The results are depicted in \autoref{fig:preferResult}, with a positive score indicating a preference for our technique. \lk{Our method (\ourBlending) appears preferable to \hueBlending~($p=0.06$). Other comparisons did not reach statistical significance. We estimated the difference in preference between our technique relative to Standard ($0.02$, 95\% CI: $[-0.05,0.08]$) and Local blending ($-0.01$, 95\% CI: $[-0.09, 0.05]$). In both cases, confidence intervals are centered approximately around zero, suggesting that \ourBlending~is comparable to the benchmarks, if not more preferable.} These results are consistent with \textbf{H3}. 

\subsubsection{Relationship between Smoothness and Error}
\begin{figure}[ht]
\centering
\includegraphics[width=0.95\linewidth]{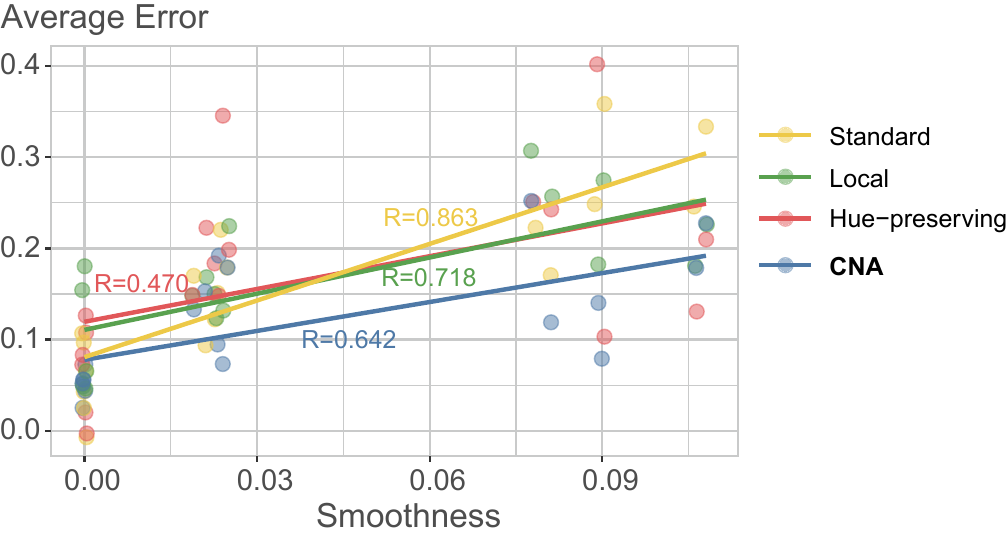}
	\vspace{-2mm}
\caption{
  Effect of smoothness on error in the distribution estimation task. Higher x-values indicate less smooth distributions. The R-value represents Pearson's correlation coefficient between the smoothness value and the average error rate. \lk{Each point represents one of the 18 histograms with certain condition (i.e., 18*4=72 points in total).}
}
\label{fig:correlation}
\vspace{-4mm}
\end{figure}

To further understand the role of distribution smoothness, we analyzed the relationship between the smoothness metric defined in \S\ref{smoothnessDefine} and the observed empirical error for the {distribution estimation task} (results for the {class discrimination} are similar). As expected, we found a significant positive correlation, with unsmooth distribution being significantly more difficult to interpret by participants ($F(1, 70)=49.34;p<0.001$), likely due to increased visual complexity. This correlation with error is slightly decreased with our optimization as compared to local and standard blending (see \autoref{fig:correlation}), although the interaction is not significant.

\subsection{Study II: Histograms + Kernel Density Curves}
\label{sec:study2}

Results from Study~I suggest that our optimization can enhance the perception of semi-transparent histograms, effectively compensating for the complex interaction between translucent colors. 
However, in practice, many visualization designers opt to augment these histograms with different visual representations, such as a curve representing kernel density estimate~\cite{wilke2019fundamentals}, or a hatching pattern~\cite{gutmann2021values}.
This additional visual cue can mitigate the challenges associated with translucent marks. We thus conducted a second study about the density curve, a most commonly used technique, to assess whether our optimization can still provide an advantage in this scenario.

\begin{figure}[t]
\centering
\includegraphics[width=1\linewidth]{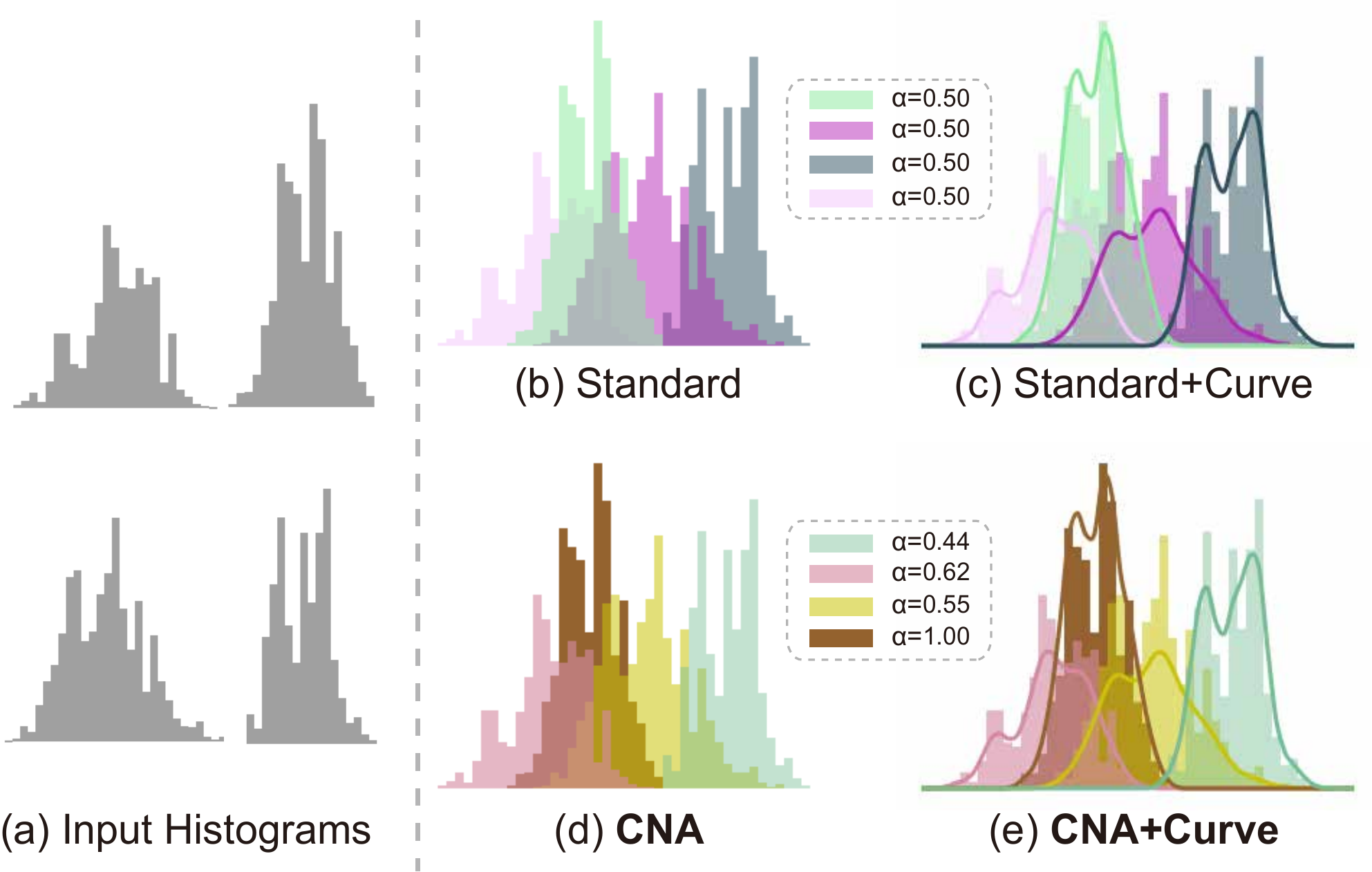}
\caption{Four example stimuli from {Study II} showing histograms with and without a kernel density estimate curve. \lk{(a) The input distributions with four classes.} The top row (b, c) represents standard alpha blending with a palette generated via Colorgorical. The bottom row (d, e) shows the results of our optimization.
}
\label{fig:outlineIllu}
\vspace{-4mm}
\end{figure}

\vspace{2mm}
\noindent{\textbf{Conditions, Stimuli, and Experiment Design.}}
Our primary focus in this study was the \emph{distribution estimation task}, which serves as a proxy for distribution analysis. \lk{Building on the findings from Study I}, we tested standard alpha blending against our method, adding two new additional variations: standard alpha blending with density curve (abbreviated as \standardOutline) and our method plus density curve (\ourOutline). These four conditions are illustrated in \autoref{fig:outlineIllu}. The experiment followed a within-subjects design, with all participants experiencing the four conditions above. Each participant completed a total of 72 trials, representing a factorization of 4 conditions $\times$ 18 histograms. The order of presentation for the conditions was randomized. 

\vspace{2mm}
\noindent{\textbf{Hypotheses.}}
We expected the addition of a kernel density curve to improve participant performance in all four conditions. However, we anticipate the benefit to be maximum when combined with our optimization. Thus, we pose the following hypotheses:

\vspace{2mm}\noindent \textbf{H4 --} We anticipate that the addition of a density curve will significantly improve user performance, no matter the condition. As a result, we expect \standardOutline~to exhibit better performance than \ourBlending~(i.e., a plain condition with no density cues).
    
\vspace{2mm}\noindent  \textbf{H5 --}  Our optimization with combined density curve will outperform all other conditions.

\vspace{2mm}
\noindent{\textbf{Participants \& Procedure.}}
For this study, we recruited 40 participants (20 males, and 20 females) from the Prolific. \lk{ To exceed the US minimum wage, we compensated each participant with $\$1.5$.} All participants passed the color vision deficiency test. The experimental procedure closely mirrored that of the previous study.

\begin{figure}[t]
	\centering
	\includegraphics[width=1\linewidth]{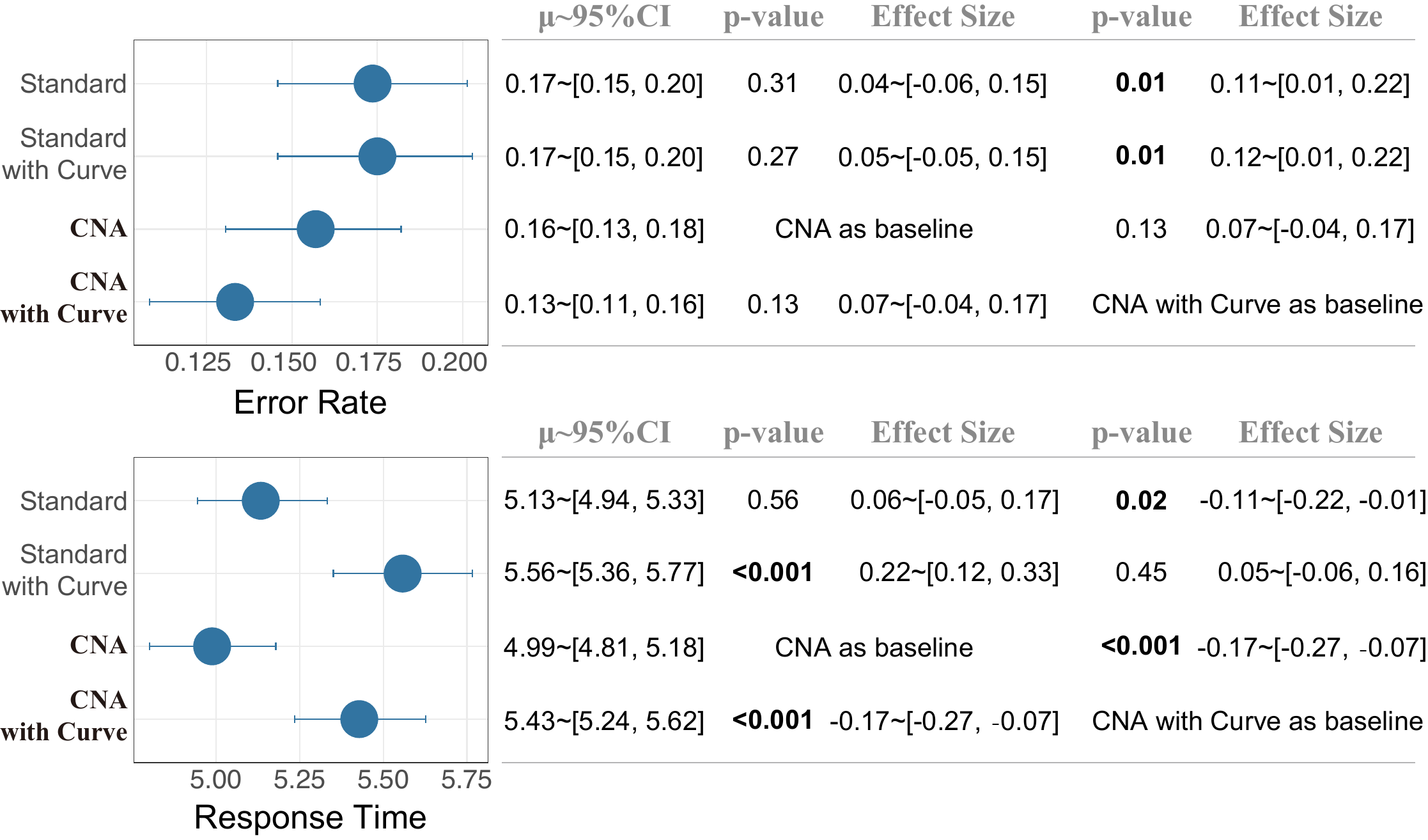}
	\caption{Results for the {\emph{Distribution Estimation Task}} in {Study II}. The top shows accuracy results whereas the bottom chart depicts response time. P-values and effects reflect comparisons with \ourBlending~ \lk{and \ourOutline, respectively}.
	}
	\label{fig:outlineResult}
	\vspace{-4mm}
\end{figure}

\vspace{2mm}
\noindent{\textbf{Results.}} 
The average completion time for participants was 6.94 minutes ($\sigma=2.56$). The experiment results are shown in \autoref{fig:outlineResult}. 
Our optimization, both with and without the inclusion of density curves (\ourBlending~and~\ourOutline), afforded higher \lk{average} accuracy compared to using a fixed color assignment strategy (\standardBlending~ and \standardOutline). This result suggests that even with the inclusion of enhancing visual cues, a conventional plot featuring overlapping histograms will not surpass the effectiveness of our technique. \lk{These results, however, are inconsistent with \textbf{H4}. That said, combining our method with a density curve (\ourOutline) appears to further reduce the error rate relative to the baseline technique (\ourBlending), yielding significantly lower error rates than both \standardBlending~ and \standardOutline~($p<0.01$).} These results support \textbf{H5}. 
This suggests that our method offers an additional advantage beyond what can be attained with histogram-enhancing kernel density estimates. There was no significant interaction between the representation method and the number of classes ($p=0.251$), or with distribution smoothness ($p=0.593$), suggesting consistent effects for the latter two factors.

\begin{figure}[ht]
\centering
\includegraphics[width=0.95\linewidth]{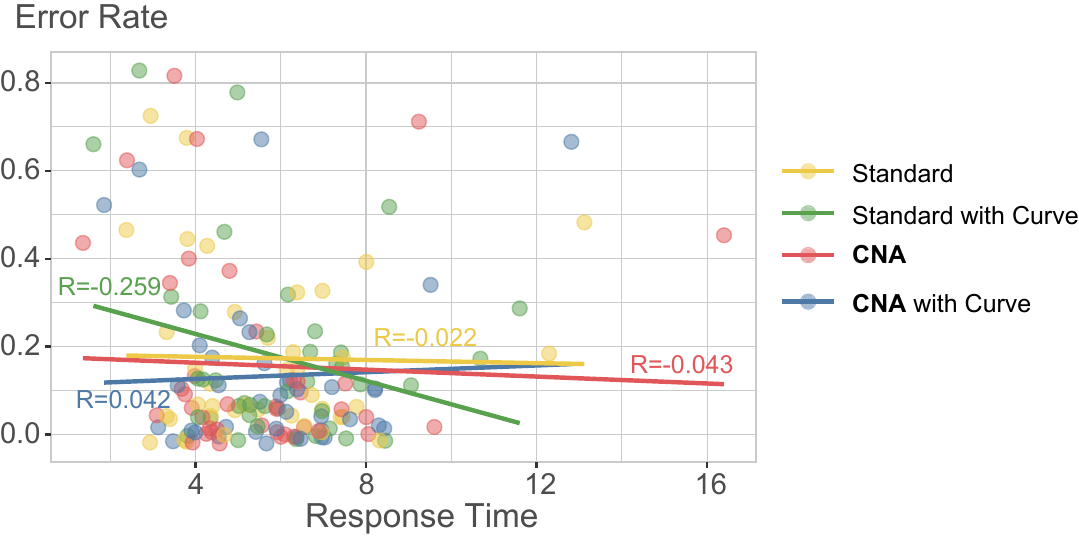}
	\vspace{-2mm}
\caption{
  \lk{Effect of response time on error in Study II. The R-value represents Pearson's correlation coefficient between the response time and the error rate for the \emph{Distribution Estimation Task}. Points represent mean participant performance in one of the four conditions.}
}
\label{fig:correlation-rt}
\vspace{-4mm}
\end{figure}

As for response time, we found that our optimization without a density curve (\ourBlending) led to a significantly faster response than \standardOutline~($p<0.001$) and \ourOutline~($p<0.001$). This suggests that adding a kernel density estimate increases the complexity of the visualization, potentially requiring more time for the viewer to parse it. \lk{We further looked for evidence of a correlation between error rates and response time to understand tradeoffs in speed vs. accuracy. we found that there is generally no significant correlation between the two (see \autoref{fig:correlation-rt}).} \revision{While \standardOutline~exhibits some negative correlation between error and response time as predicted ($R=-0.259$), this correlation is not statistically significant. A potential explanation for this result is that participants were instructed to respond as accurately as possible, leading them to allocate more time to each stimulus, thereby reducing variability in response times.}

\subsection{Discussion}

Our findings suggest that the proposed method offers significant benefits for interpreting complex, translucent histogram plots, even when compared to blending methods that are designed for transparency conditions. In the first study, we specifically compared our method to three other benchmarks representing a standard alpha blending approach and two more specialized methods for preserving hues and local color properties. Findings from the two analytical tasks show that our optimization led participants to outperform the alternatives, \lk{providing support for H1 and H2}. Notably, the hue-preserving color blending model (\hueBlending) exhibited the poorest performance in both tasks. This subpar performance is likely attributable to the generation of neutral gray colors in the overlapping areas, which could potentially lead to a breakdown in color nameability. By contrast, our optimization reinforces color-name associations between the overlapping segments and their original class color. 

Our approach similarly outperformed a local blending model, intended to enhance depth perception by reducing the saturation of `deeper' objects. However, this approach may impede the perception of wholes from parts in overlapping distributions. In contrast, enhancing color name similarity across all parts of a histogram appears to be a more appropriate strategy. Our technique leverages color nameability while also guaranteeing perceptual discriminability against both the background and other overlapping distributional features within the visualizations. Such optimization does not seem to diminish the aesthetic appeal of the visualization \lk{(consistent with H3)}. Indeed, participants expressed a comparable preference for our technique compared to the benchmark.

\begin{figure*}[t]
\centering
\includegraphics[width=0.98\linewidth]{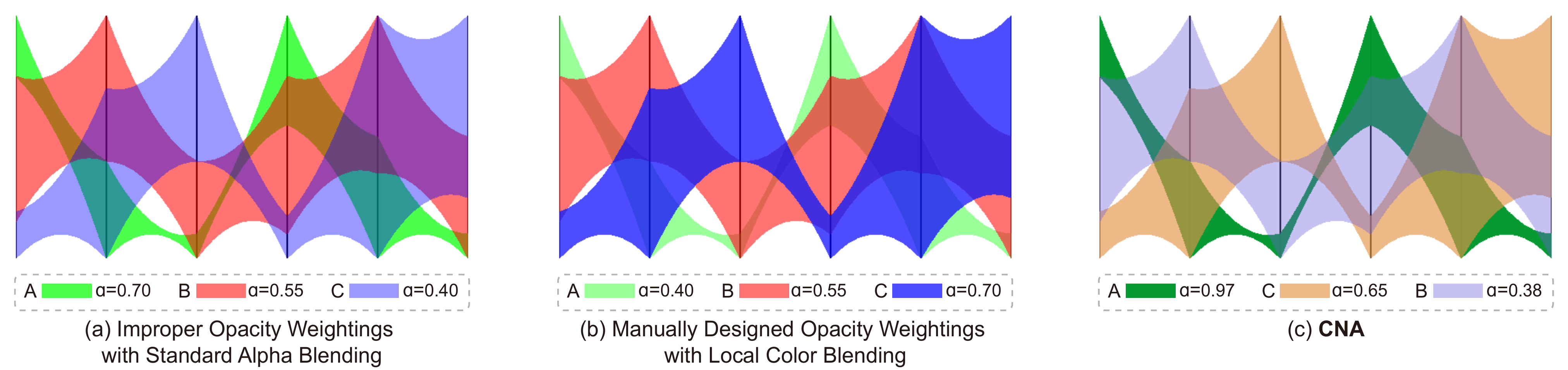}
\caption{
Visualizing the Scan Bio dataset with illustrative parallel coordinates. results generated by using (a) default opacity weightings and standard alpha blending,  (b) local color blending, and (c) our optimization.
}
\vspace*{-4mm}
\label{fig:caseStudy-parallel}
\end{figure*}

In the second study, we evaluated the perception of histograms augmented with curves representing kernel density estimates. Here, our optimization outperformed a straightforward approach of using fixed color and opacity assignments, even when the effects were not always significant. This finding extends the results of the first study, indicating that our technique can effectively substitute kernel density curves to enhance perception. Thus, our approach could be preferable when designers want to maintain the simplicity of the visualization without introducing additional cues \lk{(H4 not confirmed)}. Moreover, for an even greater performance improvement, combining a density estimate with our optimization results in the highest accuracy \lk{(H5 confirmed)}. To illustrate this effect, consider the pink distribution in \autoref{fig:outlineIllu}-a. This visualization remains challenging to interpret even with the addition of a density curve (b). However, when colors and transparency are optimized using our method, the same histogram becomes easier to discern. Adding a density outline (d) yields to the best empirically observed performance.

\subsection{Study Limitations}

There are some limitations to our experiments that should be contextualized. First, the diversity of our histogram datasets may be limited. Although we attempted to vary stimuli characteristics, such as the number of classes, the degree of overlap, and the smoothness of the distributions, other factors could also influence perception. These include the distribution family (e.g., Gaussian vs. exponential). Future work could study the impact of these factors. Moreover, there are additional techniques for improving the readability of the translucent histograms, including stepped histograms~\cite{claire2019millisecond} and hatching pattern~\cite{gutmann2021values}. While our study employs commonly used techniques, future work could compare or combine our optimization with other representations. Lastly, our evaluation was limited to a shape estimation task for a single distribution. Future work should investigate the potential of our method in more complex, comparative tasks, such as comparing shapes or features across multiple distributions~\cite{blumenschein2020v}.

\section{Extensions}
\label{sec:casestudy}

Since our optimization is computed in the screen space, it can be readily applied to other chart types, such as illustrative parallel coordinates and Venn diagrams. This is done by first rendering each layer in a separate canvas to obtain the corresponding pixel layout. We then optimize color assignment and opacity with our algorithm, considering the overlap between the layers, their neighborhood graph, and the size of each color segment. To demonstrate the effectiveness of our technique in these scenarios, we present case studies with real-world datasets. 

\subsection{Illustrative Parallel Coordinates}
We visualize six dimensions of the ScanBio dataset\footnote{Obtained from \url{http://davis.wpi.edu/~xmdv}.} with illustrative parallel coordinates~\cite{mcdonnell2008illustrative}. There are three layers to this chart colorized as green, red, and blue ribbons (see \autoref{fig:caseStudy-parallel}). We show this visualization with default colors and transparency as appearing in the original work~\cite{mcdonnell2008illustrative}. We compare with local color blending and our optimization. Viewed with default opacity settings using standard alpha blending the result introduces new false colors, including what might be interpreted as a purple class. Using the suggested opacity weightings derived with local color blending (b), the `depth' of the chart is perceived more clearly, but the overlapping segments seem difficult to discriminate individually. Our auto-generated color and transparency levels (c) yield a qualitatively improved visualization, enhancing the ribbon's visual continuity while still allowing the identification of overlapping segments.

\begin{figure}[ht]
\centering
\includegraphics[width=0.98\linewidth]{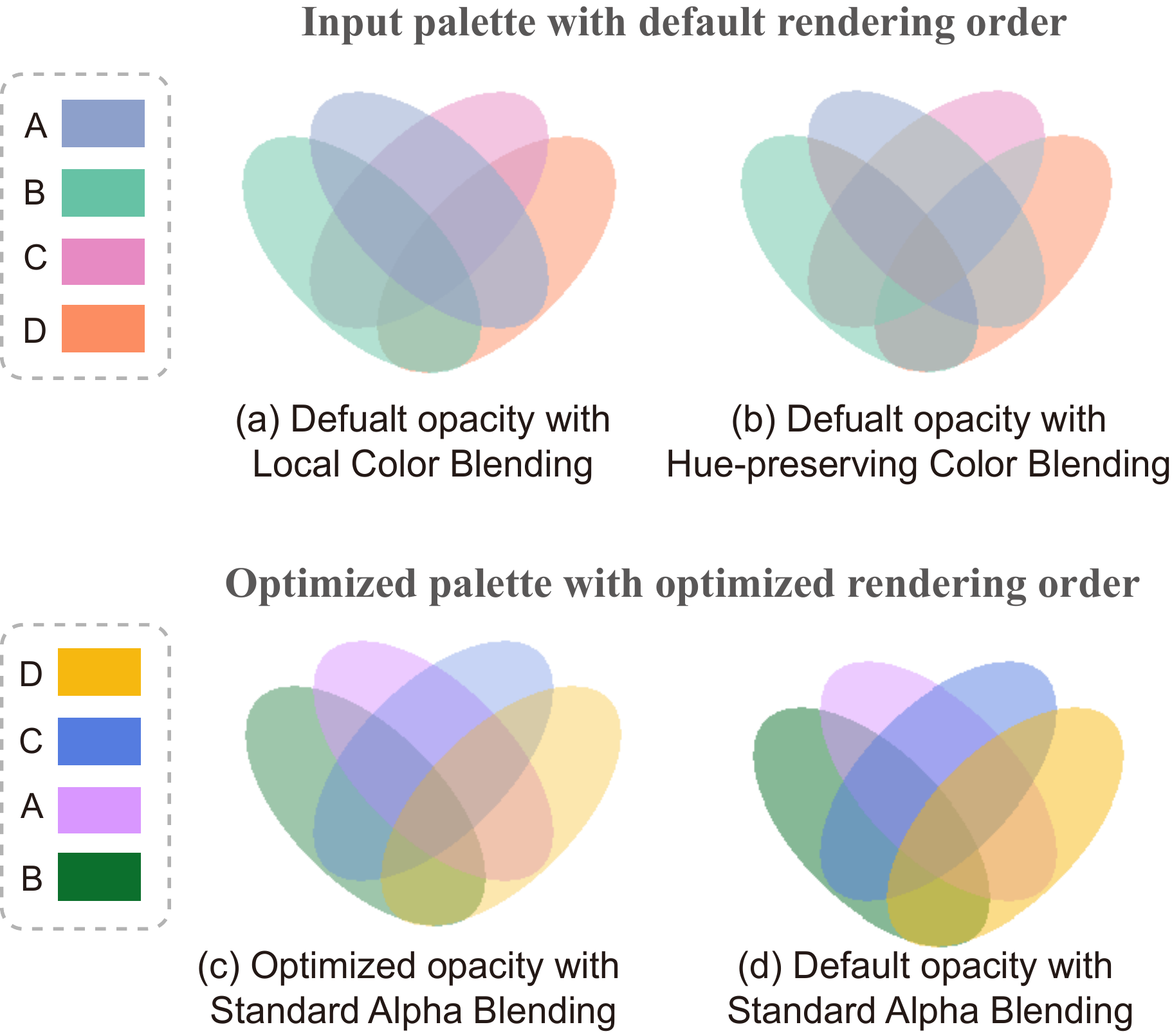}
\caption{
An elliptical Venn diagram shown with (a) ColorBrewer palette and uniform opacity (0.5) using local blending, and (b) hue-preserving blending. Both visualizations lead to suboptimal compositing. 
By contrast, our optimization (c) produces a more perceptible set visualization which guarantees minimum color separability for all parts. \lk{(d) Shows the same colors as (c) but with uniform, non-optimized opacities.} 
}

\vspace*{-4mm}
\label{fig:caseStudy-venn}
\end{figure}
\subsection{Venn Diagrams}
Venn diagrams are extensively utilized for visualizing sets~\cite{jia2021venn}, providing an intuitive representation to identify set intersections, unions, and differences. Incorporating color enhances the perception of set boundaries, but color interactions could also compromise the diagram's effectiveness~\cite {mann2015calcified}. To illustrate the applicability of our method in this domain, we generated a Venn diagram with four intersecting elliptical sets using Venny~\cite{oliveros2007venny}. We colorized the ellipses using a 4-class palette generated from ColorBrewer (uniform opacity of $\alpha=0.5$), and then applied local blending model and hue-preserving model. Results are shown in \autoref{fig:caseStudy-venn}. Observe how the intersection between the blue and red sets is challenging to associate with the red component. Similarly, employing a hue-preserving color blending model generates an abundance of grey tones. With our approach (c), the resulting color and transparency appear to qualitatively improve the visualization. For instance, the overlapping parts of the green set are more readily identifiable as green. This visualization can overall be read with minimal ambiguity. 

\section {Conclusion and Future Work}
We proposed a color-name aware optimization approach that automatically generates color assignment and determines proper transparency settings for overlapping histograms. Compared to existing blending models, our approach allows for a more accurate estimation of distributions and color classes. We achieved these results by optimizing for both within-class name association and between-class disassociation, and by also considering region-based color contrast. These factors are scored using an objective function and optimized using a custom simulated annealing algorithm. Our method allows for rapidly generating new color palettes and transparency assignments to reduce misinterpretation errors. We demonstrated the effectiveness of our method in two experiments. We also show extensions to other multi-class visualizations with overlaps, such as illustrative parallel coordinates and Venn diagrams. To help disseminate this method, we implemented our technique in a web-based tool that integrates our optimization.

Although effective, the quality of our optimization may be influenced by the color naming model used. In this study, we utilized the Heer and Stone name model~\cite{heer2012color}, which at times showed inconsistent accuracy. For example, the name distance between bright hues (\eg light blue and light green) displayed higher variation than expected. Future research could explore alternative color naming models. \revision{Another limitation lies in the variability of the algorithm's results across different runs of simulated annealing, which stems from its inherent stochastic nature. To address this issue in future work, alternative optimization algorithms, such as genetic algorithms or particle swarm optimization, could be explored to yield deterministic solutions. Another limiting factor is the need to account for the effect of the algorithm on emotional valence. Studies suggest that individual colors are often associated with specific emotional responses~\cite{anderson2021affective,bartram2017affective}, whereas blended colors can alter, amplify, or even give rise to entirely different emotional associations. To address this challenge, future work should incorporate the evaluation of emotional responses to color blending, ensuring that visualization designs are both semantically resonant and emotionally appropriate.}
Additionally, we aim to expand our optimization framework to generate color palettes friendly to color vision deficiency (CVD). \revision{Our method can be further extended to 3D scenarios, such as applications involving Gaussian Mixture Models~\cite{gmm9903677}, by incorporating relevant visual cues and spatial properties.} Finally, we would like to more thoroughly investigate how data characteristics (e.g., distribution family and shape) might impact task accuracy.

\section*{Acknowledgments}
\revisionB{This work is supported by the grants of the National Key R\&D Program of China under Grant 2022ZD0160805, NSFC (No.62132017), the Shandong Provincial Natural Science Foundation (No.ZQ2022JQ32), the Fundamental Research Funds for the Central Universities, the Beijing Natural Science Foundation (L247027), and the Research Funds of Renmin University of China.
The authors would like to thank Daniel Weiskopf (University of Stuttgart) for his valuable insights and guidance.}

\bibliographystyle{abbrv-doi-hyperref}
\bibliography{opacity}

\begin{thebibliography}{10}

\bibitem{aarts1989stochastic}
E.~Aarts.
\newblock A stochastic approach to combinatorial optimization and neural computing.
\newblock {\em Simulated Annealing and Boltzmann Machines}, 1989.

\bibitem{adelson1990ordinal}
E.~H. Adelson and P.~Anandan.
\newblock {\em Ordinal characteristics of transparency}.
\newblock Citeseer, 1990.

\bibitem{albers2014task}
D.~Albers, M.~Correll, and M.~Gleicher.
\newblock Task-driven evaluation of aggregation in time series visualization.
\newblock In {\em Proceedings of the SIGCHI Conference on Human Factors in Computing Systems}, pp. 551--560, 2014. \href{https://doi.org/10.1145/2556288.2557200}
{doi: {{%
10\hspace{.1pt}\discretionary{.}{%
}{.}\hspace{.4pt}1145\discretionary{/}{%
}{/}2556288\hspace{.1pt}\discretionary{.}{%
}{.}\hspace{.4pt}2557200}}}


\bibitem{anderson2021affective}
C.~L. Anderson and A.~C. Robinson.
\newblock Affective congruence in visualization design: Influences on reading categorical maps.
\newblock {\em IEEE Transactions on Visualization and Computer Graphics}, 28(8):2867--2878, 2021. \href{https://doi.org/10.1109/TVCG.2021.3050118}
{doi: {{%
10\hspace{.1pt}\discretionary{.}{%
}{.}\hspace{.4pt}1109\discretionary{/}{%
}{/}TVCG\hspace{.1pt}\discretionary{.}{%
}{.}\hspace{.4pt}2021\hspace{.1pt}\discretionary{.}{%
}{.}\hspace{.4pt}3050118}}}


\bibitem{bartram2017affective}
L.~Bartram, A.~Patra, and M.~Stone.
\newblock Affective color in visualization.
\newblock In {\em Proceedings of the 2017 CHI conference on human factors in computing systems}, pp. 1364--1374, 2017. \href{https://doi.org/10.1145/3025453.302604}
{doi: {{%
10\hspace{.1pt}\discretionary{.}{%
}{.}\hspace{.4pt}1145\discretionary{/}{%
}{/}3025453\hspace{.1pt}\discretionary{.}{%
}{.}\hspace{.4pt}302604}}}


\bibitem{beck1984perception}
J.~Beck, K.~Prazdny, and R.~Ivry.
\newblock The perception of transparency with achromatic colors.
\newblock {\em Perception \& psychophysics}, 35(5):407--422, 1984.

\bibitem{blumenschein2020v}
M.~Blumenschein, L.~J. Debbeler, N.~C. Lages, B.~Renner, D.~A. Keim, and M.~El-Assady.
\newblock v-plots: Designing hybrid charts for the comparative analysis of data distributions.
\newblock In {\em Computer Graphics Forum}, vol.~39, pp. 565--577. Wiley Online Library, 2020. \href{https://doi.org/10.1111/cgf.14002}
{doi: {{%
10\hspace{.1pt}\discretionary{.}{%
}{.}\hspace{.4pt}1111\discretionary{/}{%
}{/}cgf\hspace{.1pt}\discretionary{.}{%
}{.}\hspace{.4pt}14002}}}


\bibitem{chuang2009hue}
J.~Chuang, D.~Weiskopf, and T.~Moller.
\newblock Hue-preserving color blending.
\newblock {\em IEEE Transactions on Visualization and Computer Graphics}, 15(6):1275--1282, 2009. \href{https://doi.org/10.1109/TVCG.2009.150}
{doi: {{%
10\hspace{.1pt}\discretionary{.}{%
}{.}\hspace{.4pt}1109\discretionary{/}{%
}{/}TVCG\hspace{.1pt}\discretionary{.}{%
}{.}\hspace{.4pt}2009\hspace{.1pt}\discretionary{.}{%
}{.}\hspace{.4pt}150}}}


\bibitem{claire2019millisecond}
S.~Y. Claire, K.~Kremer, S.~Chatterjee, C.~L. Rodriguez, and F.~A. Rasio.
\newblock Millisecond pulsars and black holes in globular clusters.
\newblock {\em The Astrophysical Journal}, 877(2):122, 2019. \href{https://doi.org/10.3847/1538-4357/ab1b21}
{doi: {{%
10\hspace{.1pt}\discretionary{.}{%
}{.}\hspace{.4pt}3847\discretionary{/}{%
}{/}1538\discretionary{%
}{-}{-}4357\discretionary{/}{%
}{/}ab1b21}}}


\bibitem{correll2018looks}
M.~Correll, M.~Li, G.~Kindlmann, and C.~Scheidegger.
\newblock Looks good to me: Visualizations as sanity checks.
\newblock {\em IEEE Transactions on Visualization and Computer Graphics}, 25(1):830--839, 2018. \href{https://doi.org/10.1109/TVCG.2018.2864907}
{doi: {{%
10\hspace{.1pt}\discretionary{.}{%
}{.}\hspace{.4pt}1109\discretionary{/}{%
}{/}TVCG\hspace{.1pt}\discretionary{.}{%
}{.}\hspace{.4pt}2018\hspace{.1pt}\discretionary{.}{%
}{.}\hspace{.4pt}2864907}}}


\bibitem{douglas2023data}
B.~D. Douglas, P.~J. Ewell, and M.~Brauer.
\newblock Data quality in online human-subjects research: Comparisons between mturk, prolific, cloudresearch, qualtrics, and sona.
\newblock {\em Plos one}, 18(3):e0279720, 2023. \href{https://doi.org/10.1371/journal.pone.0279720}
{doi: {{%
10\hspace{.1pt}\discretionary{.}{%
}{.}\hspace{.4pt}1371\discretionary{/}{%
}{/}journal\hspace{.1pt}\discretionary{.}{%
}{.}\hspace{.4pt}pone\hspace{.1pt}\discretionary{.}{%
}{.}\hspace{.4pt}0279720}}}


\bibitem{el2022semantic}
M.~El-Assady, R.~Kehlbeck, Y.~Metz, U.~Schlegel, R.~Sevastjanova, F.~Sperrle, and T.~Spinner.
\newblock Semantic color mapping: A pipeline for assigning meaningful colors to text.
\newblock In {\em 2022 IEEE 4th Workshop on Visualization Guidelines in Research, Design, and Education (VisGuides)}, pp. 16--22. IEEE, 2022.

\bibitem{foley1996computer}
J.~D. Foley.
\newblock {\em Computer graphics: principles and practice}, vol. 12110.
\newblock Addison-Wesley Professional, 1996.

\bibitem{gfd2017}
S.~Friedrich, F.~Konietschke, and M.~Pauly.
\newblock {GFD}: An {R} package for the analysis of general factorial designs.
\newblock {\em Journal of Statistical Software, Code Snippets}, 79(1):1--18, 2017. \href{https://doi.org/10.18637/jss.v079.c01}
{doi: {{%
10\hspace{.1pt}\discretionary{.}{%
}{.}\hspace{.4pt}18637\discretionary{/}{%
}{/}jss\hspace{.1pt}\discretionary{.}{%
}{.}\hspace{.4pt}v079\hspace{.1pt}\discretionary{.}{%
}{.}\hspace{.4pt}c01}}}


\bibitem{gama2014guidelines}
S.~Gama and D.~Gon{\c{c}}alves.
\newblock Guidelines for using color blending in data visualization.
\newblock In {\em Proceedings of the 2014 International Working Conference on Advanced Visual Interfaces}, pp. 363--364, 2014. \href{https://doi.org/10.1145/2598153.2600039}
{doi: {{%
10\hspace{.1pt}\discretionary{.}{%
}{.}\hspace{.4pt}1145\discretionary{/}{%
}{/}2598153\hspace{.1pt}\discretionary{.}{%
}{.}\hspace{.4pt}2600039}}}


\bibitem{gama2014studying}
S.~Gama and D.~Goncalves.
\newblock Studying color blending perception for data visualization.
\newblock 2014. \href{https://doi.org/10.2312/eurovisshort.20141168}
{doi: {{%
10\hspace{.1pt}\discretionary{.}{%
}{.}\hspace{.4pt}2312\discretionary{/}{%
}{/}eurovisshort\hspace{.1pt}\discretionary{.}{%
}{.}\hspace{.4pt}20141168}}}


\bibitem{gigilashvili2021translucency}
D.~Gigilashvili, J.-B. Thomas, J.~Y. Hardeberg, and M.~Pedersen.
\newblock Translucency perception: A review.
\newblock {\em Journal of Vision}, 21(8):4--4, 2021.

\bibitem{Gleicher11}
M.~Gleicher, D.~Albers, R.~Walker, I.~Jusufi, C.~D. Hansen, and J.~C. Roberts.
\newblock Visual comparison for information visualization.
\newblock {\em Information Visualization}, 10(4):289--309,  21 pages, 2011. \href{https://doi.org/10.1177/1473871611416549}
{doi: {{%
10\hspace{.1pt}\discretionary{.}{%
}{.}\hspace{.4pt}1177\discretionary{/}{%
}{/}1473871611416549}}}


\bibitem{Gramazio17}
C.~C. {Gramazio}, D.~H. {Laidlaw}, and K.~B. {Schloss}.
\newblock Colorgorical: creating discriminable and preferable color palettes for information visualization.
\newblock {\em IEEE Transactions on Visualization and Computer Graphics}, 23(1):521--530, 2017. \href{https://doi.org/10.1109/TVCG.2016.2598918}
{doi: {{%
10\hspace{.1pt}\discretionary{.}{%
}{.}\hspace{.4pt}1109\discretionary{/}{%
}{/}TVCG\hspace{.1pt}\discretionary{.}{%
}{.}\hspace{.4pt}2016\hspace{.1pt}\discretionary{.}{%
}{.}\hspace{.4pt}2598918}}}


\bibitem{gunther2013opacity}
T.~G{\"u}nther, C.~R{\"o}ssl, and H.~Theisel.
\newblock Opacity optimization for 3d line fields.
\newblock {\em ACM Transactions on Graphics (TOG)}, 32(4):1--8, 2013. \href{https://doi.org/10.1145/2461912.2461930}
{doi: {{%
10\hspace{.1pt}\discretionary{.}{%
}{.}\hspace{.4pt}1145\discretionary{/}{%
}{/}2461912\hspace{.1pt}\discretionary{.}{%
}{.}\hspace{.4pt}2461930}}}


\bibitem{gunther2014opacity}
T.~G{\"u}nther, M.~Schulze, J.~M. Esturo, C.~R{\"o}ssl, and H.~Theisel.
\newblock Opacity optimization for surfaces.
\newblock In {\em Computer Graphics Forum}, vol.~33, pp. 11--20. Wiley Online Library, 2014. \href{https://doi.org/10.1111/cgf.12357}
{doi: {{%
10\hspace{.1pt}\discretionary{.}{%
}{.}\hspace{.4pt}1111\discretionary{/}{%
}{/}cgf\hspace{.1pt}\discretionary{.}{%
}{.}\hspace{.4pt}12357}}}


\bibitem{gutmann2021values}
B.~Gutmann and T.~Stelzer.
\newblock Values affirmation replication at the university of illinois.
\newblock {\em Physical Review Physics Education Research}, 17(2):020121, 2021. \href{https://doi.org/10.1103/PhysRevPhysEducRes.17.020121}
{doi: {{%
10\hspace{.1pt}\discretionary{.}{%
}{.}\hspace{.4pt}1103\discretionary{/}{%
}{/}PhysRevPhysEducRes\hspace{.1pt}\discretionary{.}{%
}{.}\hspace{.4pt}17\hspace{.1pt}\discretionary{.}{%
}{.}\hspace{.4pt}020121}}}


\bibitem{hagh2007weaving}
H.~Hagh-Shenas, S.~Kim, V.~Interrante, and C.~Healey.
\newblock Weaving versus blending: a quantitative assessment of the information carrying capacities of two alternative methods for conveying multivariate data with color.
\newblock {\em IEEE Transactions on Visualization and Computer Graphics}, 13(6):1270--1277, 2007. \href{https://doi.org/10.1109/TVCG.2007.70623}
{doi: {{%
10\hspace{.1pt}\discretionary{.}{%
}{.}\hspace{.4pt}1109\discretionary{/}{%
}{/}TVCG\hspace{.1pt}\discretionary{.}{%
}{.}\hspace{.4pt}2007\hspace{.1pt}\discretionary{.}{%
}{.}\hspace{.4pt}70623}}}


\bibitem{harrower2003colorbrewer}
M.~Harrower and C.~A. Brewer.
\newblock {ColorBrewer}.org: an online tool for selecting colour schemes for maps.
\newblock {\em The Cartographic Journal}, 40(1):27--37, 2003. \href{https://doi.org/10.1179/000870403235002042}
{doi: {{%
10\hspace{.1pt}\discretionary{.}{%
}{.}\hspace{.4pt}1179\discretionary{/}{%
}{/}000870403235002042}}}


\bibitem{heer2012color}
J.~Heer and M.~Stone.
\newblock Color naming models for color selection, image editing and palette design.
\newblock In {\em Proceedings of the CHI Conference on Human Factors in Computing Systems}, pp. 1007--1016, 2012. \href{http://dx.doi.org/10.1145/2207676.2208547}
{doi: {{%
10\hspace{.1pt}\discretionary{.}{%
}{.}\hspace{.4pt}1145\discretionary{/}{%
}{/}2207676\hspace{.1pt}\discretionary{.}{%
}{.}\hspace{.4pt}2208547}}}


\bibitem{jia2021venn}
A.~Jia, L.~Xu, and Y.~Wang.
\newblock Venn diagrams in bioinformatics.
\newblock {\em Briefings in bioinformatics}, 22(5):bbab108, 2021. \href{https://doi.org/10.1093/bib/bbab108}
{doi: {{%
10\hspace{.1pt}\discretionary{.}{%
}{.}\hspace{.4pt}1093\discretionary{/}{%
}{/}bib\discretionary{/}{%
}{/}bbab108}}}


\bibitem{johansson2005revealing}
J.~Johansson, P.~Ljung, M.~Jern, and M.~Cooper.
\newblock Revealing structure within clustered parallel coordinates displays.
\newblock In {\em IEEE Symposium on Information Visualization, 2005. INFOVIS 2005.}, pp. 125--132. IEEE, 2005. \href{https://doi.org/10.1109/INFVIS.2005.1532138}
{doi: {{%
10\hspace{.1pt}\discretionary{.}{%
}{.}\hspace{.4pt}1109\discretionary{/}{%
}{/}INFVIS\hspace{.1pt}\discretionary{.}{%
}{.}\hspace{.4pt}2005\hspace{.1pt}\discretionary{.}{%
}{.}\hspace{.4pt}1532138}}}


\bibitem{kaplan2014investigating}
J.~J. Kaplan, J.~G. Gabrosek, P.~Curtiss, and C.~Malone.
\newblock Investigating student understanding of histograms.
\newblock {\em Journal of Statistics Education}, 22(2), 2014. \href{https://doi.org/10.1080/10691898.2014.11889701}
{doi: {{%
10\hspace{.1pt}\discretionary{.}{%
}{.}\hspace{.4pt}1080\discretionary{/}{%
}{/}10691898\hspace{.1pt}\discretionary{.}{%
}{.}\hspace{.4pt}2014\hspace{.1pt}\discretionary{.}{%
}{.}\hspace{.4pt}11889701}}}


\bibitem{kuhne2012data}
L.~K{\"u}hne, J.~Giesen, Z.~Zhang, S.~Ha, and K.~Mueller.
\newblock A data-driven approach to hue-preserving color-blending.
\newblock {\em IEEE Transactions on Visualization and Computer Graphics}, 18(12):2122--2129, 2012. \href{https://doi.org/10.1109/TVCG.2012.186}
{doi: {{%
10\hspace{.1pt}\discretionary{.}{%
}{.}\hspace{.4pt}1109\discretionary{/}{%
}{/}TVCG\hspace{.1pt}\discretionary{.}{%
}{.}\hspace{.4pt}2012\hspace{.1pt}\discretionary{.}{%
}{.}\hspace{.4pt}186}}}


\bibitem{languenou2023contrast}
E.~Languenou.
\newblock Contrast driven color-group assignment in categorical data visualization.
\newblock 2023. \href{https://doi.org/10.5220/0011615000003417}
{doi: {{%
10\hspace{.1pt}\discretionary{.}{%
}{.}\hspace{.4pt}5220\discretionary{/}{%
}{/}0011615000003417}}}


\bibitem{gmm9903677}
K.~Lawonn, M.~Meuschke, P.~Eulzer, M.~Mitterreiter, J.~Giesen, and T.~Günther.
\newblock Gray: Ray casting for visualization and interactive data exploration of gaussian mixture models.
\newblock {\em IEEE Transactions on Visualization and Computer Graphics}, 29(1):526--536, 2023. \href{https://doi.org/10.1109/TVCG.2022.3209374}
{doi: {{%
10\hspace{.1pt}\discretionary{.}{%
}{.}\hspace{.4pt}1109\discretionary{/}{%
}{/}TVCG\hspace{.1pt}\discretionary{.}{%
}{.}\hspace{.4pt}2022\hspace{.1pt}\discretionary{.}{%
}{.}\hspace{.4pt}3209374}}}


\bibitem{lem2013misinterpretation}
S.~Lem, P.~Onghena, L.~Verschaffel, and W.~Van~Dooren.
\newblock On the misinterpretation of histograms and box plots.
\newblock {\em Educational Psychology}, 33(2):155--174, 2013. \href{https://doi.org/10.1080/01443410.2012.674006}
{doi: {{%
10\hspace{.1pt}\discretionary{.}{%
}{.}\hspace{.4pt}1080\discretionary{/}{%
}{/}01443410\hspace{.1pt}\discretionary{.}{%
}{.}\hspace{.4pt}2012\hspace{.1pt}\discretionary{.}{%
}{.}\hspace{.4pt}674006}}}


\bibitem{lin2013selecting}
S.~Lin, J.~Fortuna, C.~Kulkarni, M.~Stone, and J.~Heer.
\newblock Selecting semantically-resonant colors for data visualization.
\newblock {\em Computer Graphics Forum}, 32(3):401--410, 2013. \href{https://doi.org/10.1111/cgf.12127}
{doi: {{%
10\hspace{.1pt}\discretionary{.}{%
}{.}\hspace{.4pt}1111\discretionary{/}{%
}{/}cgf\hspace{.1pt}\discretionary{.}{%
}{.}\hspace{.4pt}12127}}}


\bibitem{Lu21}
K.~{Lu}, M.~{Feng}, X.~{Chen}, M.~{Sedlmair}, O.~{Deussen}, D.~{Lischinski}, Z.~{Cheng}, and Y.~{Wang}.
\newblock Palettailor: discriminable colorization for categorical data.
\newblock {\em IEEE Transactions on Visualization and Computer Graphics}, 27(2):475--484, 2021. \href{https://doi.org/10.1109/TVCG.2020.3030406}
{doi: {{%
10\hspace{.1pt}\discretionary{.}{%
}{.}\hspace{.4pt}1109\discretionary{/}{%
}{/}TVCG\hspace{.1pt}\discretionary{.}{%
}{.}\hspace{.4pt}2020\hspace{.1pt}\discretionary{.}{%
}{.}\hspace{.4pt}3030406}}}


\bibitem{Lu23}
K.~Lu, K.~Reda, O.~Deussen, and Y.~Wang.
\newblock Interactive context-preserving color highlighting for multiclass scatterplots.
\newblock In {\em Proceedings of the 2023 CHI Conference on Human Factors in Computing Systems}, pp. 1--15, 2023. \href{https://doi.org/10.1145/3544548.3580734}
{doi: {{%
10\hspace{.1pt}\discretionary{.}{%
}{.}\hspace{.4pt}1145\discretionary{/}{%
}{/}3544548\hspace{.1pt}\discretionary{.}{%
}{.}\hspace{.4pt}3580734}}}


\bibitem{lu2021curve}
Y.~Lu, L.~Cheng, T.~Isenberg, C.-W. Fu, G.~Chen, H.~Liu, O.~Deussen, and Y.~Wang.
\newblock Curve complexity heuristic kd-trees for neighborhood-based exploration of 3d curves.
\newblock In {\em Computer Graphics Forum}, vol.~40, pp. 461--474. Wiley Online Library, 2021. \href{https://doi.org/10.1111/cgf.142647}
{doi: {{%
10\hspace{.1pt}\discretionary{.}{%
}{.}\hspace{.4pt}1111\discretionary{/}{%
}{/}cgf\hspace{.1pt}\discretionary{.}{%
}{.}\hspace{.4pt}142647}}}


\bibitem{luboschik2010new}
M.~Luboschik, A.~Radloff, and H.~Schumann.
\newblock A new weaving technique for handling overlapping regions.
\newblock In {\em Proceedings of the International Conference on Advanced Visual Interfaces}, pp. 25--32, 2010. \href{https://doi.org/10.1145/1842993.1842999}
{doi: {{%
10\hspace{.1pt}\discretionary{.}{%
}{.}\hspace{.4pt}1145\discretionary{/}{%
}{/}1842993\hspace{.1pt}\discretionary{.}{%
}{.}\hspace{.4pt}1842999}}}


\bibitem{jnd1994}
M.~Mahy, L.~Van~Eycken, and A.~Oosterlinck.
\newblock Evaluation of uniform color spaces developed after the adoption of cielab and cieluv.
\newblock {\em Color Research \& Application}, 19(2):105--121, 1994. \href{https://doi.org/10.1111/j.1520-6378.1994.tb00070.x}
{doi: {{%
10\hspace{.1pt}\discretionary{.}{%
}{.}\hspace{.4pt}1111\discretionary{/}{%
}{/}j\hspace{.1pt}\discretionary{.}{%
}{.}\hspace{.4pt}1520\discretionary{%
}{-}{-}6378\hspace{.1pt}\discretionary{.}{%
}{.}\hspace{.4pt}1994\hspace{.1pt}\discretionary{.}{%
}{.}\hspace{.4pt}tb00070\hspace{.1pt}\discretionary{.}{%
}{.}\hspace{.4pt}x}}}


\bibitem{mann2015calcified}
K.~Mann.
\newblock The calcified eggshell matrix proteome of a songbird, the zebra finch (taeniopygia guttata).
\newblock {\em Proteome science}, 13:1--20, 2015. \href{https://doi.org/10.1186/s12953-015-0086-1}
{doi: {{%
10\hspace{.1pt}\discretionary{.}{%
}{.}\hspace{.4pt}1186\discretionary{/}{%
}{/}s12953\discretionary{%
}{-}{-}015\discretionary{%
}{-}{-}0086\discretionary{%
}{-}{-}1}}}


\bibitem{matejka2015dynamic}
J.~Matejka, F.~Anderson, and G.~Fitzmaurice.
\newblock Dynamic opacity optimization for scatter plots.
\newblock In {\em Proceedings of the 33rd Annual ACM Conference on Human Factors in Computing Systems}, pp. 2707--2710, 2015. \href{https://doi.org/10.1145/2702123.2702585}
{doi: {{%
10\hspace{.1pt}\discretionary{.}{%
}{.}\hspace{.4pt}1145\discretionary{/}{%
}{/}2702123\hspace{.1pt}\discretionary{.}{%
}{.}\hspace{.4pt}2702585}}}


\bibitem{mcdonnell2008illustrative}
K.~T. McDonnell and K.~Mueller.
\newblock Illustrative parallel coordinates.
\newblock In {\em Computer Graphics Forum}, vol.~27, pp. 1031--1038. Wiley Online Library, 2008. \href{https://doi.org/10.1111/j.1467-8659.2008.01239.x}
{doi: {{%
10\hspace{.1pt}\discretionary{.}{%
}{.}\hspace{.4pt}1111\discretionary{/}{%
}{/}j\hspace{.1pt}\discretionary{.}{%
}{.}\hspace{.4pt}1467\discretionary{%
}{-}{-}8659\hspace{.1pt}\discretionary{.}{%
}{.}\hspace{.4pt}2008\hspace{.1pt}\discretionary{.}{%
}{.}\hspace{.4pt}01239\hspace{.1pt}\discretionary{.}{%
}{.}\hspace{.4pt}x}}}


\bibitem{metelli1974perception}
F.~Metelli.
\newblock The perception of transparency.
\newblock {\em Scientific American}, 230(4):90--99, 1974. \href{https://doi.org/10.1038/scientificamerican0474-90}
{doi: {{%
10\hspace{.1pt}\discretionary{.}{%
}{.}\hspace{.4pt}1038\discretionary{/}{%
}{/}scientificamerican0474\discretionary{%
}{-}{-}90}}}


\bibitem{micallef2017towards}
L.~Micallef, G.~Palmas, A.~Oulasvirta, and T.~Weinkauf.
\newblock Towards perceptual optimization of the visual design of scatterplots.
\newblock {\em IEEE Transactions on Visualization and Computer Graphics}, 23(6):1588--1599, 2017. \href{https://doi.org/10.1109/TVCG.2017.2674978}
{doi: {{%
10\hspace{.1pt}\discretionary{.}{%
}{.}\hspace{.4pt}1109\discretionary{/}{%
}{/}TVCG\hspace{.1pt}\discretionary{.}{%
}{.}\hspace{.4pt}2017\hspace{.1pt}\discretionary{.}{%
}{.}\hspace{.4pt}2674978}}}


\bibitem{mukherjee2021context}
K.~Mukherjee, B.~Yin, B.~E. Sherman, L.~Lessard, and K.~B. Schloss.
\newblock Context matters: A theory of semantic discriminability for perceptual encoding systems.
\newblock {\em IEEE Transactions on Visualization and Computer Graphics}, 28(1):697--706, 2021.

\bibitem{oliveros2007venny}
J.~C. Oliveros.
\newblock Venny. an interactive tool for comparing lists with venn diagrams.
\newblock {\em http://bioinfogp. cnb. csic. es/tools/venny/index. html}, 2007.

\bibitem{porter1984compositing}
T.~Porter and T.~Duff.
\newblock Compositing digital images.
\newblock In {\em Proceedings of the 11th annual conference on Computer graphics and interactive techniques}, pp. 253--259, 1984. \href{https://doi.org/10.1145/800031.808606}
{doi: {{%
10\hspace{.1pt}\discretionary{.}{%
}{.}\hspace{.4pt}1145\discretionary{/}{%
}{/}800031\hspace{.1pt}\discretionary{.}{%
}{.}\hspace{.4pt}808606}}}


\bibitem{quadri2022automatic}
G.~J. Quadri, J.~A. Nieves, B.~M. Wiernik, and P.~Rosen.
\newblock Automatic scatterplot design optimization for clustering identification.
\newblock {\em IEEE Transactions on Visualization and Computer Graphics}, 2022. \href{https://doi.org/10.1109/TVCG.2022.3189883}
{doi: {{%
10\hspace{.1pt}\discretionary{.}{%
}{.}\hspace{.4pt}1109\discretionary{/}{%
}{/}TVCG\hspace{.1pt}\discretionary{.}{%
}{.}\hspace{.4pt}2022\hspace{.1pt}\discretionary{.}{%
}{.}\hspace{.4pt}3189883}}}


\bibitem{reda2022rainbow}
K.~Reda.
\newblock Rainbow colormaps: What are they good and bad for?
\newblock {\em IEEE Transactions on Visualization and Computer Graphics}, 29(12):5496--5510, 2022.

\bibitem{reda2021color}
K.~Reda, A.~A. Salvi, J.~Gray, and M.~E. Papka.
\newblock Color nameability predicts inference accuracy in spatial visualizations.
\newblock In {\em Computer Graphics Forum}, vol.~40, pp. 49--60. Wiley Online Library, 2021. \href{https://doi.org/10.1111/cgf.14288}
{doi: {{%
10\hspace{.1pt}\discretionary{.}{%
}{.}\hspace{.4pt}1111\discretionary{/}{%
}{/}cgf\hspace{.1pt}\discretionary{.}{%
}{.}\hspace{.4pt}14288}}}


\bibitem{reda2020rainbows}
K.~Reda and D.~A. Szafir.
\newblock Rainbows revisited: Modeling effective colormap design for graphical inference.
\newblock {\em IEEE Transactions on Visualization and Computer Graphics}, 27(2):1032--1042, 2020.

\bibitem{matplotlib}
riyaaggarwal.
\newblock Overlapping histograms with matplotlib in python.
\newblock \url{https://www.geeksforgeeks.org/overlapping-histograms-with-matplotlib-in-python/}, 2020.
\newblock Accessed: 2020-11-26.

\bibitem{sahann2021histogram}
R.~Sahann, T.~M{\"u}ller, and J.~Schmidt.
\newblock Histogram binning revisited with a focus on human perception.
\newblock In {\em 2021 IEEE Visualization Conference (VIS)}, pp. 66--70. IEEE, 2021. \href{https://doi.org/10.1109/VIS49827.2021.9623301}
{doi: {{%
10\hspace{.1pt}\discretionary{.}{%
}{.}\hspace{.4pt}1109\discretionary{/}{%
}{/}VIS49827\hspace{.1pt}\discretionary{.}{%
}{.}\hspace{.4pt}2021\hspace{.1pt}\discretionary{.}{%
}{.}\hspace{.4pt}9623301}}}


\bibitem{schloss2020semantic}
K.~B. Schloss, Z.~Leggon, and L.~Lessard.
\newblock Semantic discriminability for visual communication.
\newblock {\em Ieee transactions on visualization and computer graphics}, 27(2):1022--1031, 2020.

\bibitem{schloss2018color}
K.~B. Schloss, L.~Lessard, C.~S. Walmsley, and K.~Foley.
\newblock Color inference in visual communication: the meaning of colors in recycling.
\newblock {\em Cognitive research: principles and implications}, 3:1--17, 2018.

\bibitem{setlur2016linguistic}
V.~{Setlur} and M.~C. {Stone}.
\newblock A linguistic approach to categorical color assignment for data visualization.
\newblock {\em IEEE Transactions on Visualization and Computer Graphics}, 22(1):698--707, 2016. \href{https://doi.org/10.1109/TVCG.2015.2467471}
{doi: {{%
10\hspace{.1pt}\discretionary{.}{%
}{.}\hspace{.4pt}1109\discretionary{/}{%
}{/}TVCG\hspace{.1pt}\discretionary{.}{%
}{.}\hspace{.4pt}2015\hspace{.1pt}\discretionary{.}{%
}{.}\hspace{.4pt}2467471}}}


\bibitem{sharma2005ciede2000}
G.~Sharma, W.~Wu, and E.~N. Dalal.
\newblock The {CIEDE}2000 color-difference formula: implementation notes, supplementary test data, and mathematical observations.
\newblock {\em Color Research \& Application}, 30(1):21--30, 2005. \href{https://doi.org/10.1002/col.20070}
{doi: {{%
10\hspace{.1pt}\discretionary{.}{%
}{.}\hspace{.4pt}1002\discretionary{/}{%
}{/}col\hspace{.1pt}\discretionary{.}{%
}{.}\hspace{.4pt}20070}}}


\bibitem{singh2002toward}
M.~Singh and B.~L. Anderson.
\newblock Toward a perceptual theory of transparency.
\newblock {\em Psychological review}, 109(3):492, 2002.

\bibitem{tableau}
T.~Software.
\newblock The tableau visualization system.
\newblock \url{http://www. tableausoftware.com/.}

\bibitem{szafir2017modeling}
D.~A. Szafir.
\newblock Modeling color difference for visualization design.
\newblock {\em IEEE Transactions on Visualization and Computer Graphics}, 24(1):392--401, 2018. \href{https://doi.org/10.1109/TVCG.2017.2744359}
{doi: {{%
10\hspace{.1pt}\discretionary{.}{%
}{.}\hspace{.4pt}1109\discretionary{/}{%
}{/}TVCG\hspace{.1pt}\discretionary{.}{%
}{.}\hspace{.4pt}2017\hspace{.1pt}\discretionary{.}{%
}{.}\hspace{.4pt}2744359}}}


\bibitem{urness2003effectively}
T.~Urness, V.~Interrante, I.~Marusic, E.~Longmire, and B.~Ganapathisubramani.
\newblock Effectively visualizing multi-valued flow data using color and texture.
\newblock In {\em IEEE Visualization, 2003. VIS 2003.}, pp. 115--121. IEEE, 2003. \href{https://doi.org/10.1109/VISUAL.2003.1250362}
{doi: {{%
10\hspace{.1pt}\discretionary{.}{%
}{.}\hspace{.4pt}1109\discretionary{/}{%
}{/}VISUAL\hspace{.1pt}\discretionary{.}{%
}{.}\hspace{.4pt}2003\hspace{.1pt}\discretionary{.}{%
}{.}\hspace{.4pt}1250362}}}


\bibitem{wang2008color}
L.~Wang, J.~Giesen, K.~T. McDonnell, P.~Zolliker, and K.~Mueller.
\newblock Color design for illustrative visualization.
\newblock {\em IEEE Transactions on Visualization and Computer Graphics}, 14(6):1739--1754, 2008. \href{https://doi.org/10.1109/TVCG.2008.118}
{doi: {{%
10\hspace{.1pt}\discretionary{.}{%
}{.}\hspace{.4pt}1109\discretionary{/}{%
}{/}TVCG\hspace{.1pt}\discretionary{.}{%
}{.}\hspace{.4pt}2008\hspace{.1pt}\discretionary{.}{%
}{.}\hspace{.4pt}118}}}


\bibitem{Wang2018}
Y.~Wang, X.~Chen, T.~Ge, C.~Bao, M.~Sedlmair, C.-W. Fu, O.~Deussen, and B.~Chen.
\newblock Optimizing color assignment for perception of class separability in multiclass scatterplots.
\newblock {\em IEEE Transactions on Visualization and Computer Graphics}, 25(1):820--829, 2019. \href{https://doi.org/10.1109/TVCG.2018.2864912}
{doi: {{%
10\hspace{.1pt}\discretionary{.}{%
}{.}\hspace{.4pt}1109\discretionary{/}{%
}{/}TVCG\hspace{.1pt}\discretionary{.}{%
}{.}\hspace{.4pt}2018\hspace{.1pt}\discretionary{.}{%
}{.}\hspace{.4pt}2864912}}}


\bibitem{wang2011efficient}
Y.~Wang, J.~Zhang, W.~Chen, H.~Zhang, and X.~Chi.
\newblock Efficient opacity specification based on feature visibilities in direct volume rendering.
\newblock In {\em Computer Graphics Forum}, vol.~30, pp. 2117--2126. Wiley Online Library, 2011. \href{https://doi.org/10.1111/j.1467-8659.2011.02045.x}
{doi: {{%
10\hspace{.1pt}\discretionary{.}{%
}{.}\hspace{.4pt}1111\discretionary{/}{%
}{/}j\hspace{.1pt}\discretionary{.}{%
}{.}\hspace{.4pt}1467\discretionary{%
}{-}{-}8659\hspace{.1pt}\discretionary{.}{%
}{.}\hspace{.4pt}2011\hspace{.1pt}\discretionary{.}{%
}{.}\hspace{.4pt}02045\hspace{.1pt}\discretionary{.}{%
}{.}\hspace{.4pt}x}}}


\bibitem{weiskopf2007gpu}
D.~Weiskopf et~al.
\newblock {\em GPU-based interactive visualization techniques}.
\newblock Springer, 2007.

\bibitem{wertheimer1938gestalt}
M.~Wertheimer.
\newblock Gestalt theory.
\newblock 1938.

\bibitem{wilke2019fundamentals}
C.~O. Wilke.
\newblock {\em Fundamentals of data visualization: a primer on making informative and compelling figures}.
\newblock O'Reilly Media, 2019.

\bibitem{yang2012color}
Y.~Yang, J.~Ming, and N.~Yu.
\newblock Color image quality assessment based on ciede2000.
\newblock {\em Advances in Multimedia}, 2012:11--11, 2012. \href{https://doi.org/10.1155/2012/273723}
{doi: {{%
10\hspace{.1pt}\discretionary{.}{%
}{.}\hspace{.4pt}1155\discretionary{/}{%
}{/}2012\discretionary{/}{%
}{/}273723}}}


\bibitem{zheng2012perceptually}
L.~Zheng, Y.~Wu, and K.-L. Ma.
\newblock Perceptually-based depth-ordering enhancement for direct volume rendering.
\newblock {\em IEEE Transactions on Visualization and Computer Graphics}, 19(3):446--459, 2012. \href{https://doi.org/10.1109/TVCG.2012.144}
{doi: {{%
10\hspace{.1pt}\discretionary{.}{%
}{.}\hspace{.4pt}1109\discretionary{/}{%
}{/}TVCG\hspace{.1pt}\discretionary{.}{%
}{.}\hspace{.4pt}2012\hspace{.1pt}\discretionary{.}{%
}{.}\hspace{.4pt}144}}}


\bibitem{zhou2008visual}
H.~Zhou, X.~Yuan, H.~Qu, W.~Cui, and B.~Chen.
\newblock Visual clustering in parallel coordinates.
\newblock In {\em Computer Graphics Forum}, vol.~27, pp. 1047--1054. Wiley Online Library, 2008. \href{https://doi.org/10.1111/j.1467-8659.2008.01241.x}
{doi: {{%
10\hspace{.1pt}\discretionary{.}{%
}{.}\hspace{.4pt}1111\discretionary{/}{%
}{/}j\hspace{.1pt}\discretionary{.}{%
}{.}\hspace{.4pt}1467\discretionary{%
}{-}{-}8659\hspace{.1pt}\discretionary{.}{%
}{.}\hspace{.4pt}2008\hspace{.1pt}\discretionary{.}{%
}{.}\hspace{.4pt}01241\hspace{.1pt}\discretionary{.}{%
}{.}\hspace{.4pt}x}}}


\end{thebibliography}

\vspace{-20pt}
\begin{IEEEbiography}[{\includegraphics[width=1in,height=1.25in,clip,keepaspectratio]{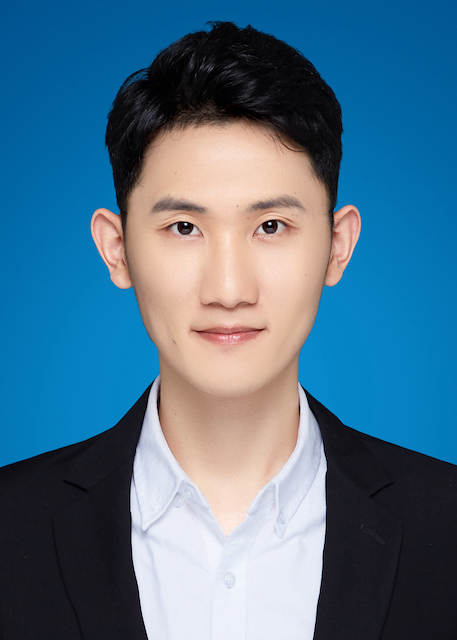}}]{Kecheng Lu}
	is a postdoctoral researcher at the School of Information, Renmin University of China. He completed his Ph.D. in Computer Science at Shandong University in 2023. His research focuses on information visualization and human-computer interaction.
\end{IEEEbiography}

\vspace{-33pt}
\begin{IEEEbiography}[{\includegraphics[width=1in,height=1.25in,clip,keepaspectratio]{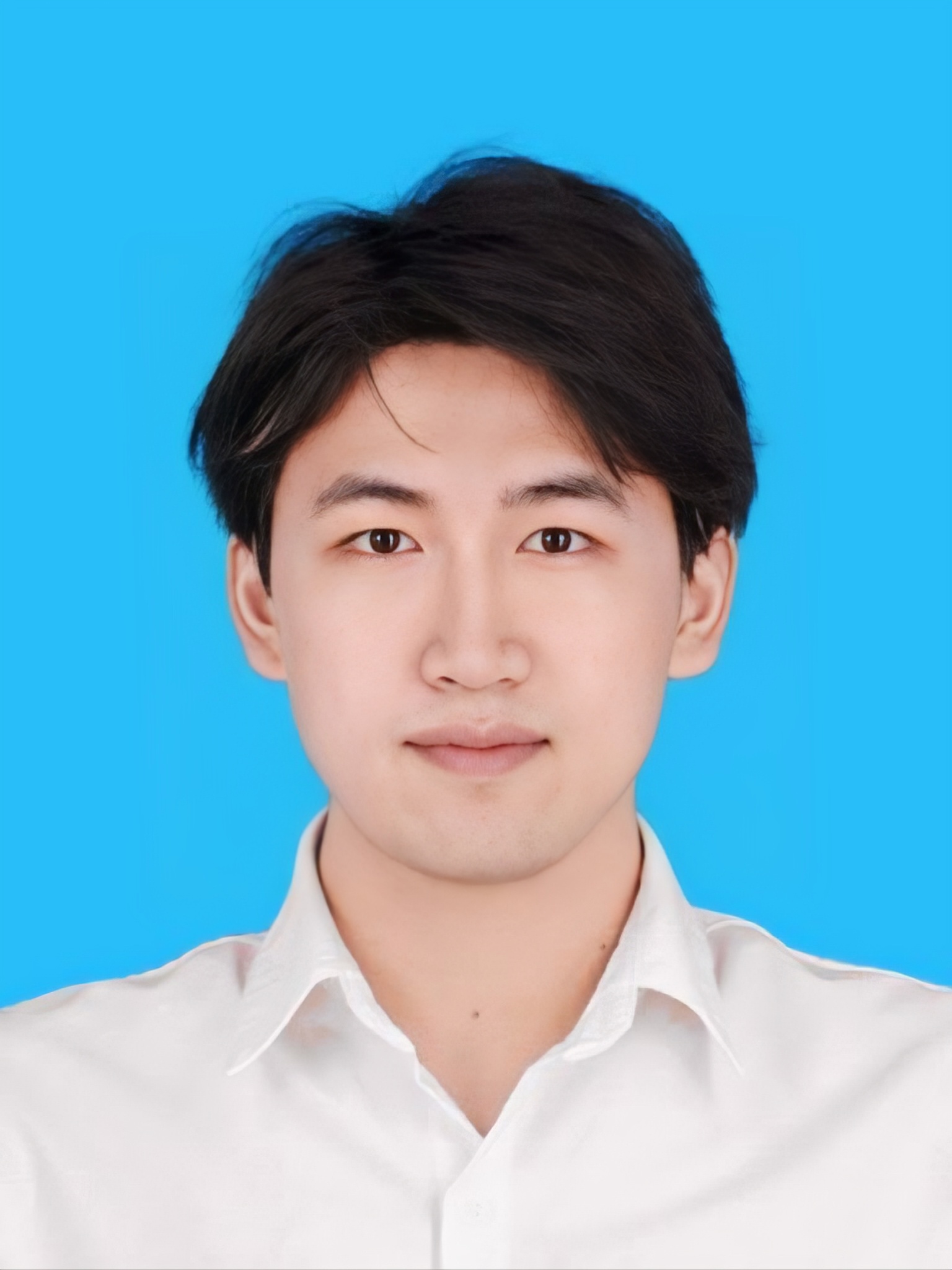}}]{Lihang Zhu} is currently a second-year Master student in School Of Computer Science and Technology, Shandong University. His research interests include information visualization and human computer interaction.
\end{IEEEbiography}

\vspace{-33pt}
\begin{IEEEbiography}[{\includegraphics[width=1in,height=1.25in,clip,keepaspectratio]{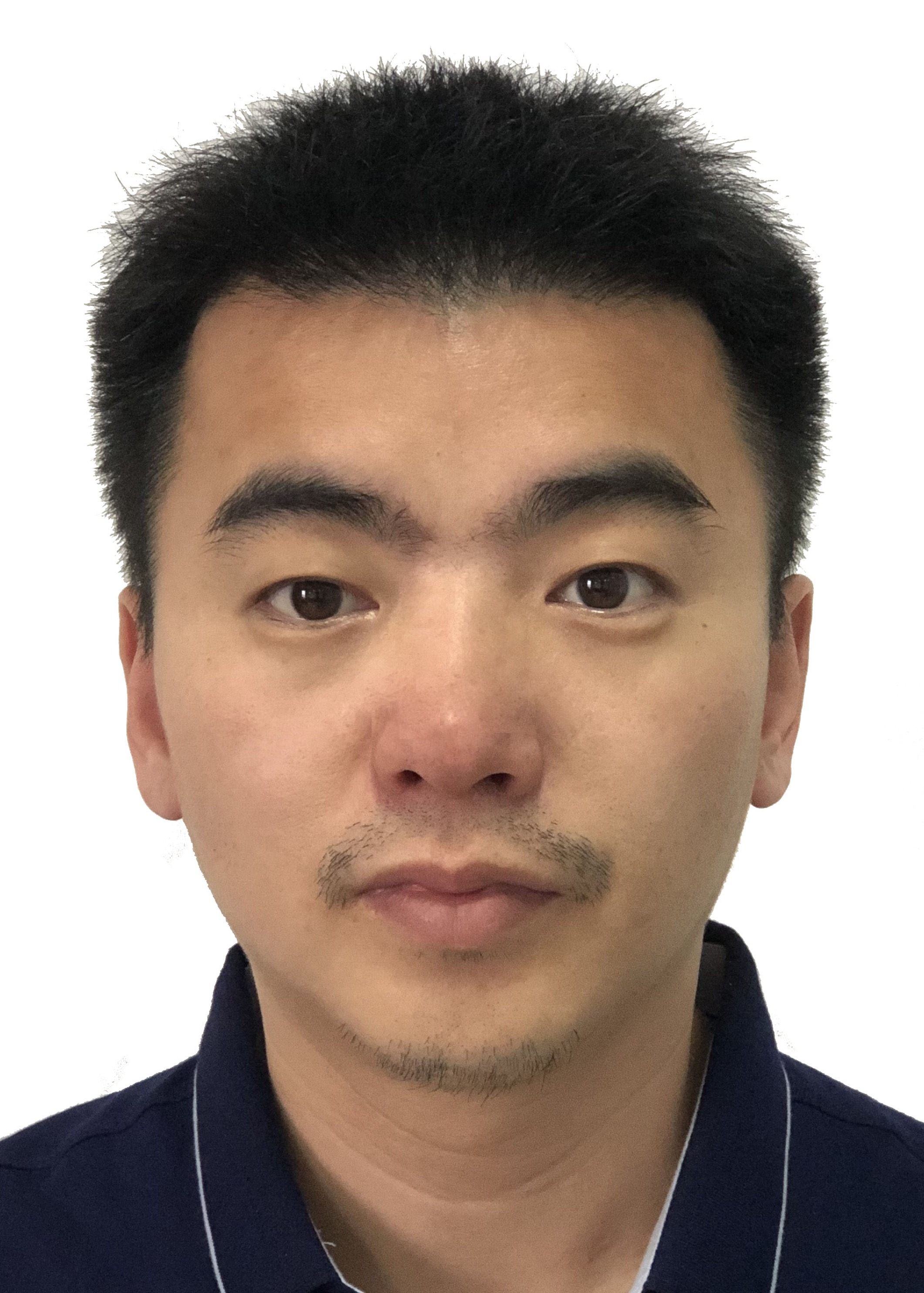}}]{Yunhai Wang}
	is a professor in School of Information, Renmin University of China. He serves as the associate editor of IEEE Transactions on Visualization and Computer Graphics, IEEE Computer Graphics and Applications, and Computer Graphics Forum. His interests include scientific visualization, information visualization, and computer graphics.
\end{IEEEbiography}

\vspace{-33pt}
\begin{IEEEbiography}[{\includegraphics[width=1in,height=1.25in,clip,keepaspectratio]{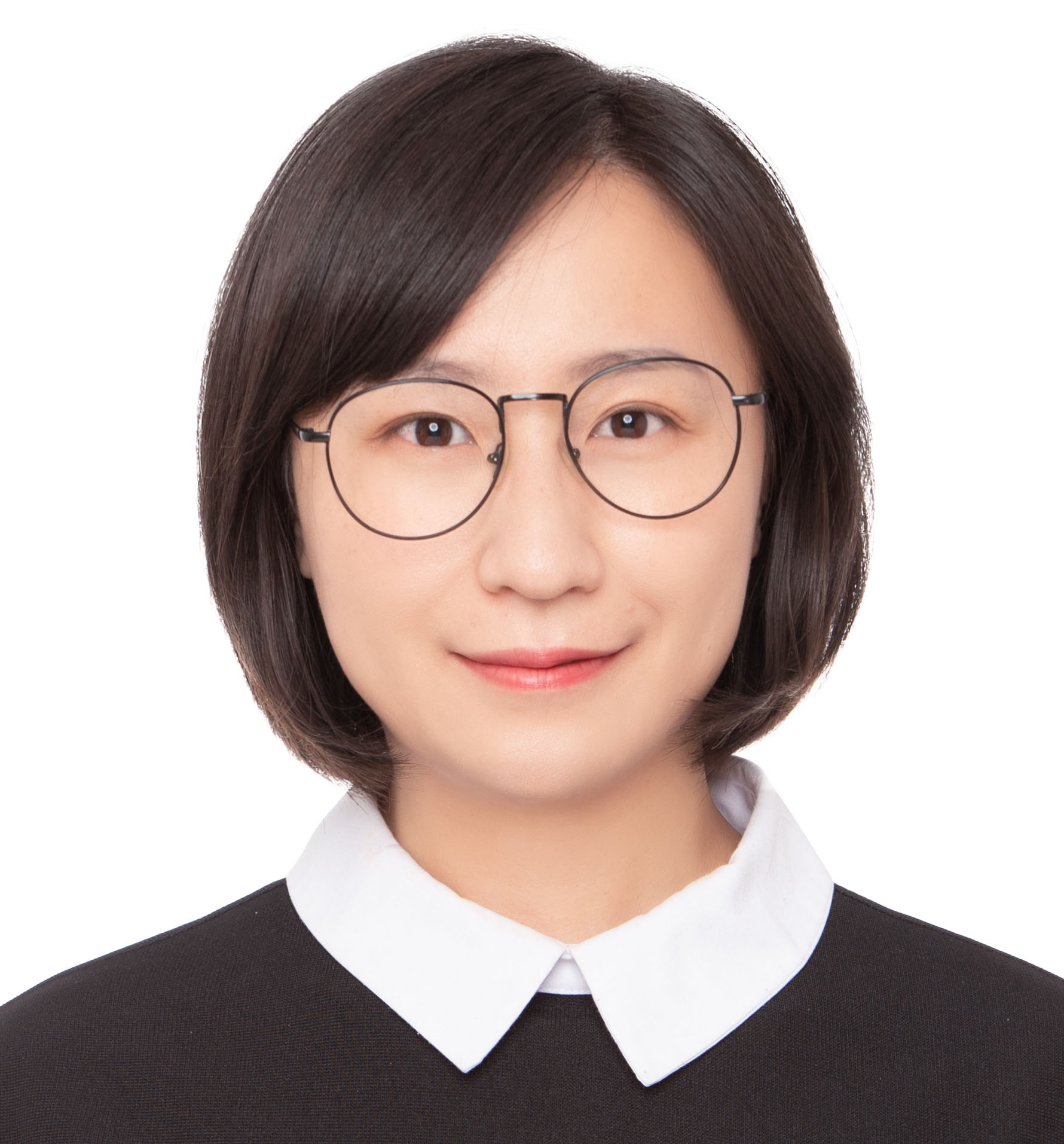}}]{Qiong Zeng} is an associate professor in the School of Computer Science and Technology at Shandong University, China. Her research interests include scientific visualization and computer graphics. Qiong obtained her Ph.D. in Software Engineering from Shandong University, China, in 2015. She is a member of the IEEE Computer Society. Contact her at qiong.zn@sdu.edu.cn.
\end{IEEEbiography}

\vspace{-33pt}
\begin{IEEEbiography}[{\includegraphics[width=1in,height=1.25in,clip,keepaspectratio]{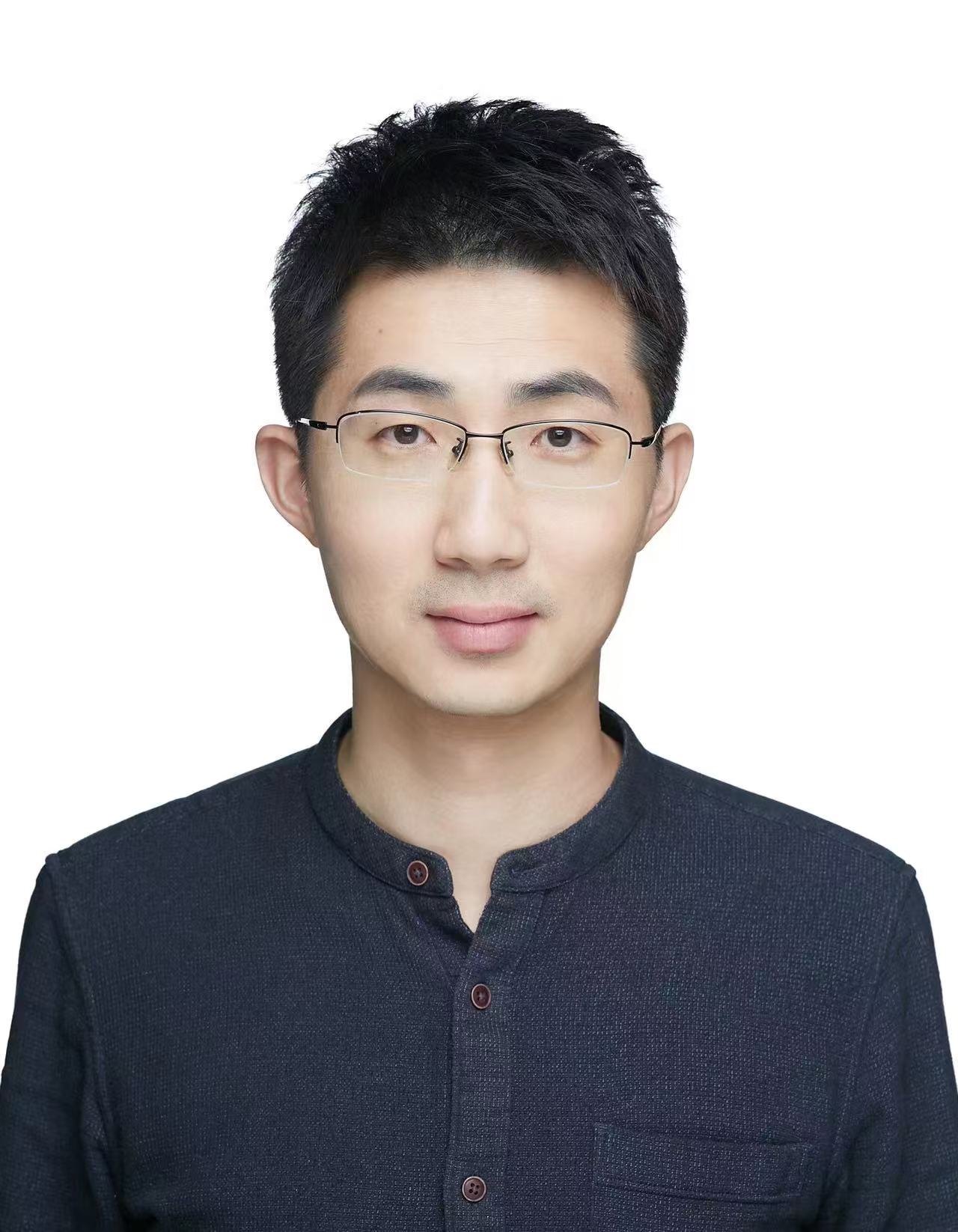}}]{Weitao Song} received the BSc and PhD degrees in optics from the Beijing Institute of Technology, China, in 2010 and 2016, respectively. He is currently a professor with the School of Optics and Photonics, Beijing Institute of Technology, China. His research interests include color science, optical instrumentation, virtual reality, and augmented reality.
\end{IEEEbiography}


\vspace{-28pt}
\begin{IEEEbiography}[{\includegraphics[width=1in,height=1.25in,clip,keepaspectratio]{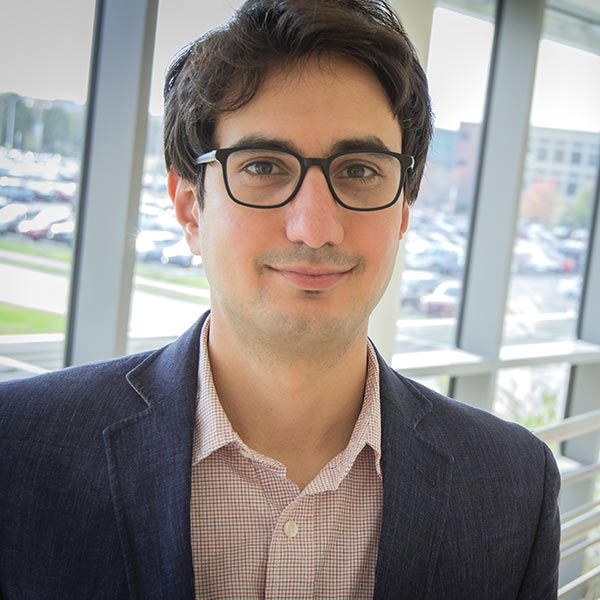}}]{Khairi Reda} is an Associate Professor in the School of Informatics and Computing at Indiana University Indianapolis. His
research lies at the intersection of Human-Computer Interaction and Data Science, with interests in visual analytics and data cognition.
	
\end{IEEEbiography}

\vfill
\end{document}



\title{Color-Name Aware Optimization to Enhance the Perception of Transparent Overlapped Charts}

\vspace{5mm}


\maketitle

This \textbf{supplementary material} provides additional experimental results for our submitted paper. Experimental data and analysis code can be accessed at \url{https://osf.io/xevk9/?view_only=1b79aeefec774209ad60f1a5b0cceda2}.

\paragraph{Navigation:}
\begin{enumerate}[start=1]
\item \LinkTo{similarity}
\item \LinkTo{evaluation}
\item \LinkTo{caseHist}
\item \LinkTo{caseHull}
\item \LinkTo{stimuli}

\end{enumerate}


\CreateLink{similarity}{Comparing different similarity measures}
\section{Comparing different similarity measures}

\definecolor{mpink}{RGB}{238,204,211}
\definecolor{mgrey}{RGB}{197,195,186}
\definecolor{morange}{RGB}{255,192,131}
\definecolor{mgreen}{RGB}{180,190,100}
\definecolor{mred}{RGB}{181,104,96}
\definecolor{mgreen1}{RGB}{173,185,81}
\definecolor{mbrown}{RGB}{122,93,44}

\begin{figure}[!ht]
	\centering
	\includegraphics[width=0.98\linewidth]{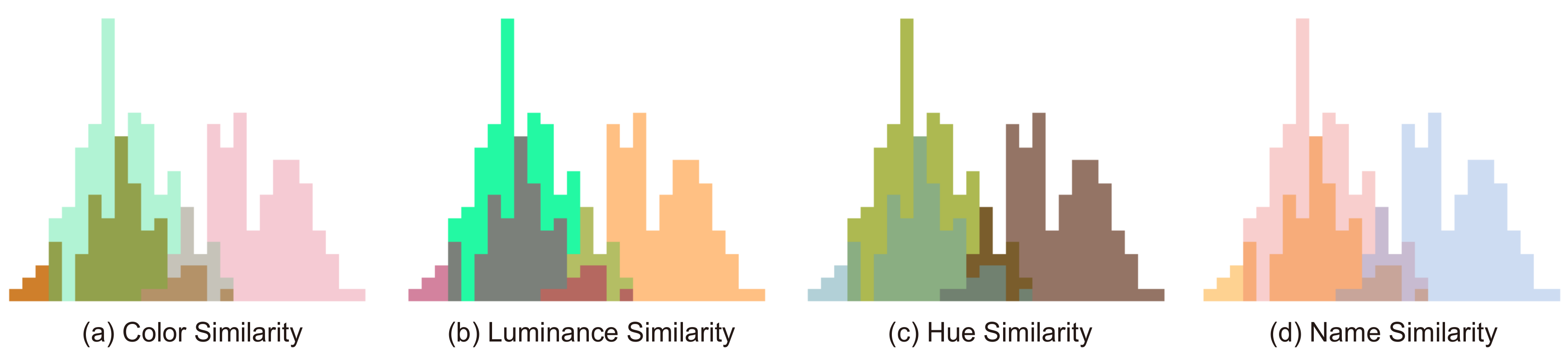}
	\caption{The optimization results by using different similarity measures for $S(c_i, c_j)$. 
	}
	\label{fig:simialrity}
\end{figure}

In this paper, we tried multiple measures for quantifying the association between different blended colors of one class, including color similarity, luminance similarity, hue similarity, and name similarity. Color Similarity(CS) is calculated by $1-CD(c_i, c_j)$, where $CD(c_i, c_j)$ is a normalized CIEDE2000 color difference~\cite{sharma2005ciede2000}.This metric often resulted in colors with a small color distance but significant differences in their names. For instance, the pink and grey colors in \autoref{fig:simialrity}-a exhibit high color similarity ($CS(\tikz\draw[mpink,fill=mpink] (0,0) circle (0.9ex);, \tikz\draw[mgrey,fill=mgrey] (0,0) circle (.9ex);)=0.61$) despite their distinct appearances. The Luminance Similarity(LS) is calculated by $1-0.01*LD(c_i, c_j)$, where LD is the absolute difference between two colors' luminance values in CIE Lab space. However, large luminance similarity does not necessarily indicate a strong association between colors, as shown in \autoref{fig:simialrity}-b, the colors of components exhibit highly similar luminance($LS(\tikz\draw[morange,fill=morange] (0,0) circle (0.9ex);, \tikz\draw[mred,fill=mred] (0,0) circle (.9ex);)=0.70$, $LS(\tikz\draw[morange,fill=morange] (0,0) circle (0.9ex);, \tikz\draw[mgreen,fill=mgreen] (0,0) circle (.9ex);)=0.92$, $LS(\tikz\draw[mgreen,fill=mgreen] (0,0) circle (0.9ex);, \tikz\draw[mred,fill=mred] (0,0) circle (.9ex);)=0.78$), yet grouping them into a single class poses a challenge. Hue Similarity(HS) is calculated by $1-HD(c_i, c_j)/180$, where HD is the absolute difference between two colors' hue values in CIE Lab space. However, a high hue similarity does not guarantee similar colors. For example, as shown in \autoref{fig:simialrity}-c, the green color has a similar hue to the brown color($HS(\tikz\draw[mgreen1,fill=mgreen1] (0,0) circle (0.9ex);, \tikz\draw[mbrown,fill=mbrown] (0,0) circle (.9ex);)=0.83$), but people often find it challenging to associate the two colors together. 

Based on these attempts, we finally decided to use the name similarity as the final similarity measure. As shown in \autoref{fig:simialrity}-d, the blue class is more easily distinguishable using name similarity compared to the other metrics, due to the stronger association between the names of the blended colors.

 \CreateLink{evaluation}{Detailed discussion of the results}
 \section{Detailed Discussion of the Results}
We assessed the effectiveness of our approach against benchmark conditions (traditional alpha blending, local color blending, and hue-preserving blending) using colors from the Tableau-10 scheme or Colorgorical through three crowdsourced experiments. To color-code the stimuli, we randomly selected subsets of two, three, or four colors from the corresponding palette to assign to the classes in the stimuli. This random selection process ensured a diverse range of color combinations, providing a comprehensive evaluation of our approach's performance across different color schemes.

Each participant completed a total of 72 stimuli, consisting of 3 \emph{classes} $\times$ 3 \emph{smoothness levels} $\times$ 2 \emph{benchmarks} $\times$ 2 \emph{repetitions}, with 2 \emph{repetitions} corresponding to Colorgorical and Tableau palettes. This design ensured that each palette was equally compared with our method, allowing for a balanced assessment of our approach's effectiveness in different color environments.

In this study, we use Tableau to represent designer-crafted palettes and Colorgorical to represent auto-generated palettes. Our comparative analysis demonstrates that the results generated by our algorithm are superior in several aspects to the existing designer-crafted and auto-generated palettes. Specifically, our approach consistently achieved lower error rates and faster response times, highlighting its effectiveness in providing clear and efficient visualizations.

\subsection{Results for Distribution Estimation} 
\begin{figure}[!ht]
	\centering
	\includegraphics[width=0.98\linewidth]{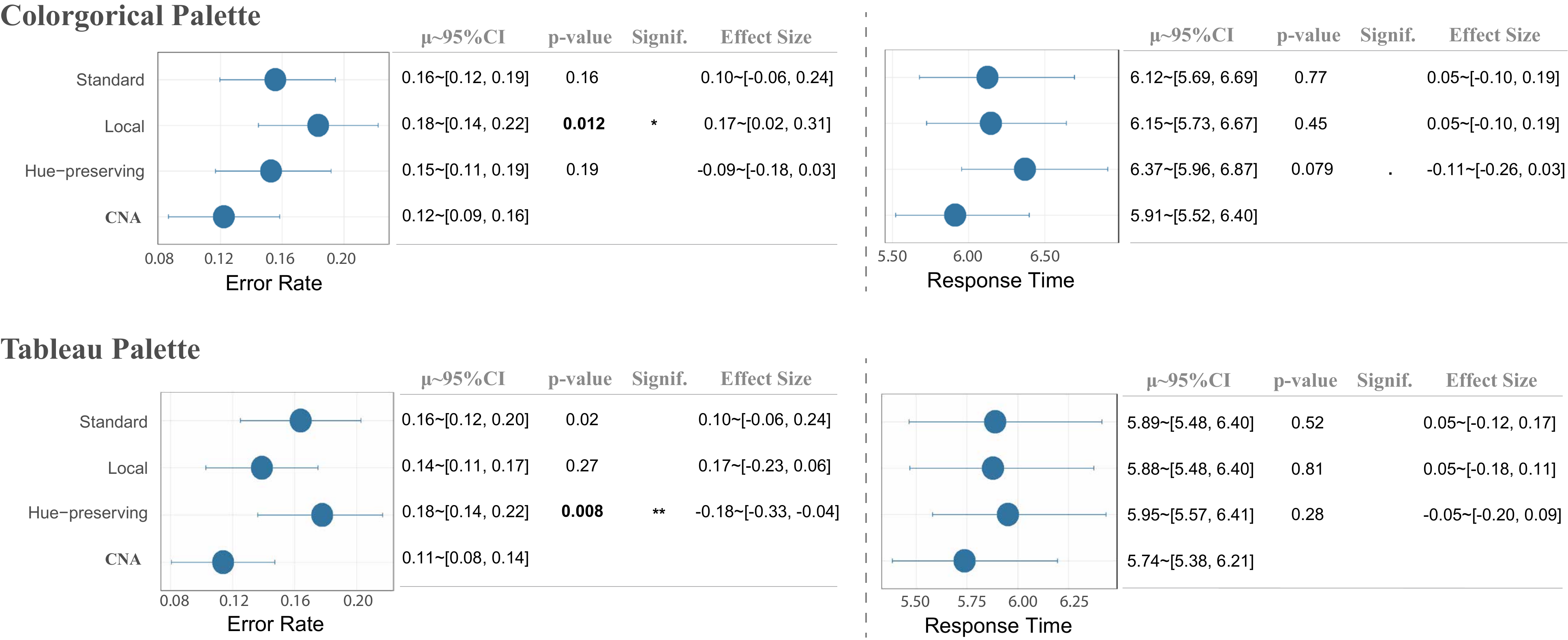}
	\caption{
  Results for the {\emph{Distribution Estimation Task}}, including effect sizes and significance tests for our method against the benchmarks. Error bars ($\mu \sim$ 95\%CI), p-value, and significance level are calculated from the Mann-Whitney test. Effect size is calculated using Cohen's d
	}
	\label{fig:palette_shape}
\end{figure}

\autoref{fig:palette_shape} presents the results for this task in comparison with benchmarks using the Colorgorical palette. Our proposed method surpasses the benchmarks characterized by fixed color and opacity assignment, in both error rate and response time. Notably, our algorithm significantly outperforms certain benchmark conditions(\emph{Local}) in reducing error rate. Although the difference is not pronounced, our algorithm still maintains an advantage in terms of time efficiency compared to other methods.

It also showcases the results for this task compared with benchmarks using the Tableau palette. Specifically, our optimization achieved a significantly reduced error rate compared to certain benchmarks (\emph{Hue-preserving}). While our optimization does not significantly outperform all the other benchmarks in terms of error rate and response time, it still holds a slight advantage in terms of overall results.

\subsection{Results for Class Discrimination} 

\begin{figure}[!ht]
	\centering
	\includegraphics[width=0.98\linewidth]{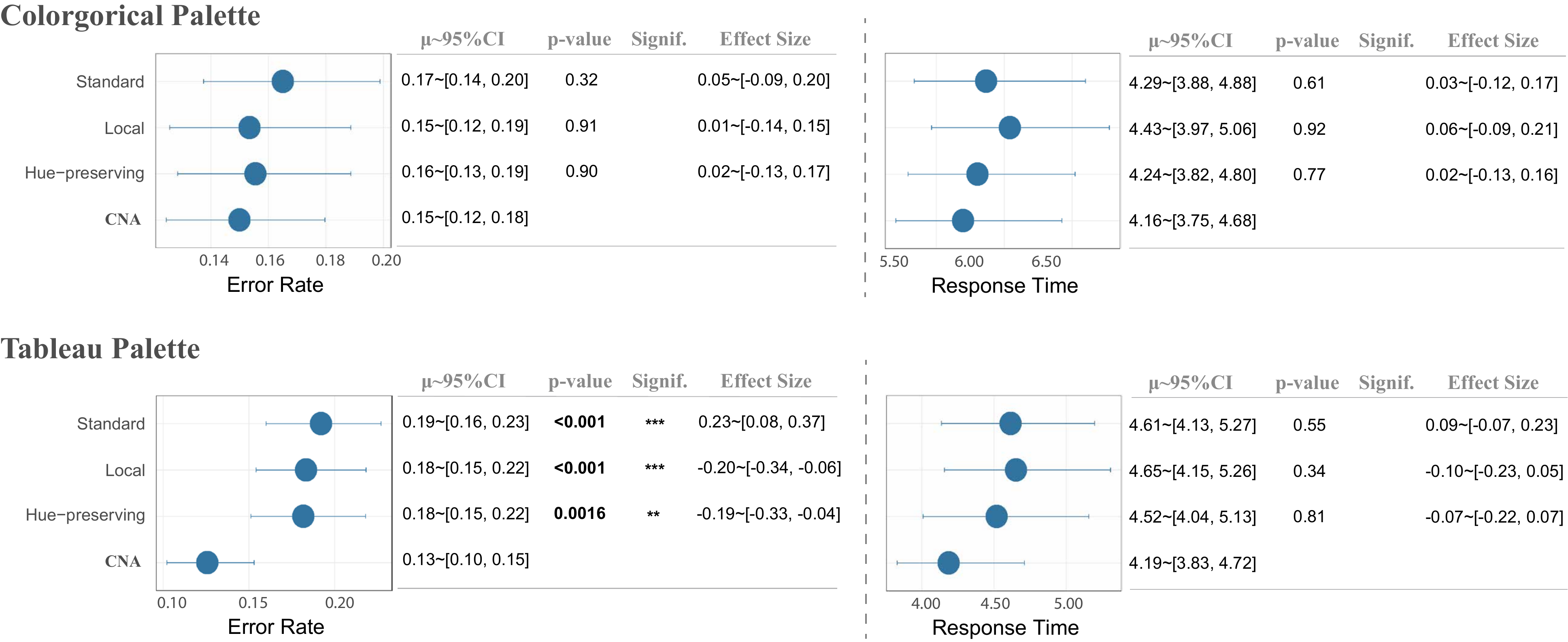}
	\caption{Results for the {\emph{Class Discrimination Task}}
	}
	\label{fig:palette_counting}
\end{figure}

The results are depicted in \autoref{fig:palette_counting}. In comparison with colors sampled from the Colorgorical palette, the error rate with our optimization method consistently remained lower than that of the fixed assignment strategy. 

It also presents the results for this task. Our proposed method surpasses the benchmarks that utilize colors sampled from the Tableau palette, characterized by fixed color and opacity assignment, in both error rate and response time. Notably, our algorithm significantly outperforms all benchmark conditions (\emph{Standard}, \emph{Local}, \emph{Hue-preserving}). This demonstrates the robustness of our approach in class discrimination tasks, particularly when compared with designer-crafted palettes like Tableau.

\subsection{Results for User Preference} 
\begin{figure}[!ht]
	\centering
	\includegraphics[width=0.98\linewidth]{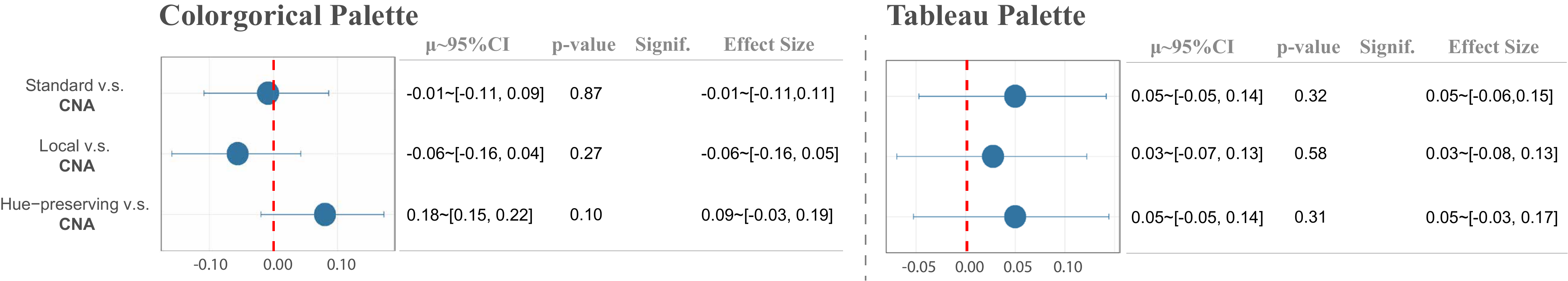}
	\caption{Results for the {\emph{User Preference Task}}. A preference score greater than 0 indicates that participants prefer our method over the respective benchmark.
	}
	\label{fig:preference}
\end{figure}

For each comparison, we assigned a score of 1 if the participant preferred our approach and -1 for preferring one of the other benchmarks. Zero was assigned for a neutral choice when the participant indicated no clear preference. The results are depicted in \autoref{fig:preference}, with a positive score indicating a preference for our technique.

Compared with colors sampled from the Colorgorical palette, our method appears preferable to \emph{Hue-preserving}, and comparable to the other two benchmarks (\emph{Standard}, \emph{Local}). As for the Tableau palette, our optimization has a slight advantage over all three benchmarks (\emph{Standard}, \emph{Local}, \emph{Hue-preserving}). However, none of these comparisons reached statistical significance. This indicates that in terms of aesthetics, our approach is likely similar to the benchmarks, suggesting that our method maintains visual appeal while enhancing other aspects of visualization.

\CreateLink{caseHist}{Visualizing Iris Dataset with semi-transparent overlapping histograms}
\section{Visualizing Iris Dataset with semi-transparent overlapping histograms}
\begin{figure}[!ht]
\centering
\includegraphics[width=0.98\linewidth]{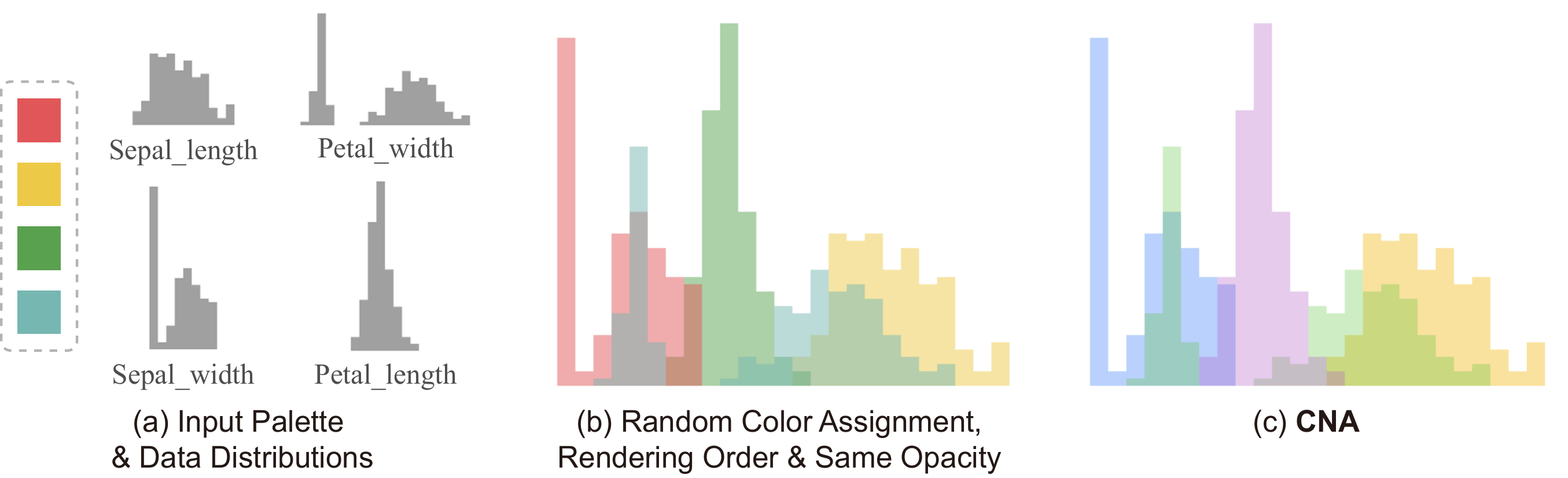}
\caption{
Visualizing the Iris dataset~\cite{misc_iris_53} with the overlapping histogram. With the input palette and data distributions (a), apply the default color assignment and the same opacity to the histogram (b); using our method, we automatically generate optimal color, opacity, and rendering order (c). Our optimization result achieves better readability of the overlapping histograms.
}
\label{fig:caseStudy}
\end{figure}

To further explore the potential of our proposed method, we conducted this case study on the well-known Iris dataset~\cite{misc_iris_53}, to compare the distribution difference of the four variables: sepal width, sepal length, petal width, and petal length. The histogram is first calculated by 25 bins (see \autoref{fig:caseStudy}-a) and then colorized with the Tableau 10 palette. Following the online course for overlapping histograms with Matplotlib in Python~\cite{matplotlib}, we set the opacity of each color to $0.5$. As shown in \autoref{fig:caseStudy}-b, with the setting of the same opacity and default color assignment, the distribution of sepal width(the green class) is hard to discriminate, due to the similar appearance to the mixed color from petal length(blue class) and sepal length(yellow class). The gray color from petal width(red color) and petal length(blue class) are very distinctive from the surrounding colors. 

Our optimization result avoids these problems, as shown in \autoref{fig:caseStudy}-c, we can see that our optimization removes the false colors by optimizing the color, hence the purple class, i.e., the petal length class, can be easily distinguished. Our algorithm also has a clear distribution for each class.

\CreateLink{caseHull}{Visualizing CIFAR100 Dataset with Cluster-encapsulating Hulls}
\section{Visualizing CIFAR100 Dataset with Cluster-encapsulating Hulls}
\begin{figure}[!ht]
\centering
\includegraphics[width=0.98\linewidth]{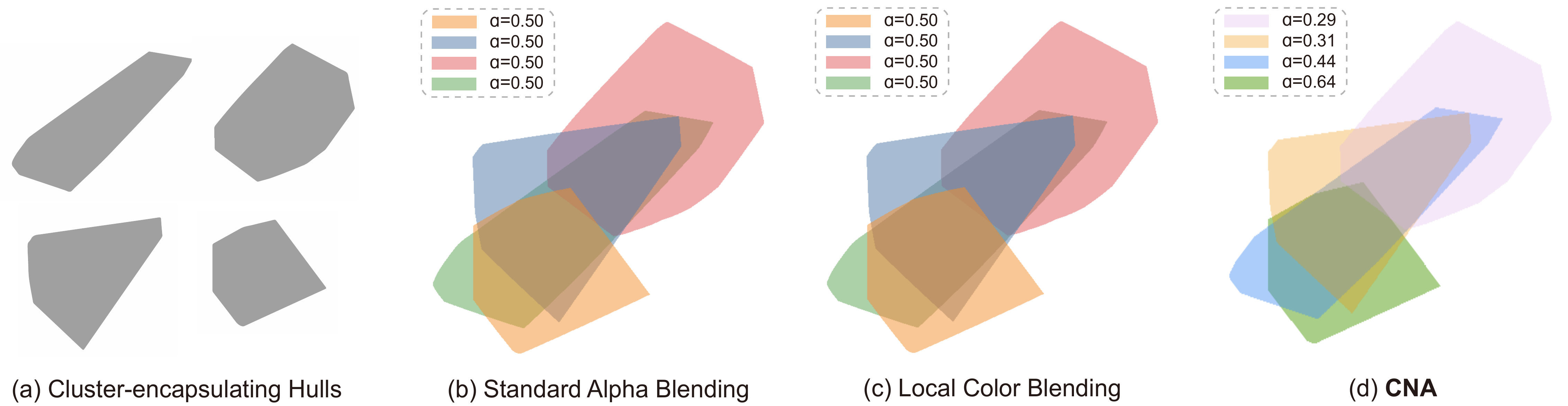}
 \caption{
 Visualizing the CIFAR100 dataset with cluster-encapsulating hulls. (a) Input cluster-encapsulating hulls for each class; (b) applying random color assignment and uniform opacity(0.5) with standard alpha blending model to the hulls, leading to many false colors; (c) using local color blending model does not reduce false colors; (d) our optimal colors, opacity settings, and rendering order, make the shape distribution of each class easy to figure out, while also ensuring good color separability for all segments.
 }
 \vspace*{-3mm}
 \label{fig:caseStudy-hull}
 \end{figure}

 Cluster-encapsulating hulls are often used as an alternative for scatterplot visualizations~\cite{elmqvist2008rolling, schreck2008butterfly}, as they allow a better perception of the distribution of whole clusters. However, due to the potential overlapping of complex shapes, this kind of visualization may lead to ambiguous colors~\cite{luboschik2010new}. To illustrate the applicability of our method in this domain, we generated cluster-encapsulating hulls from four CIFAR100 object classes~\cite{krizhevsky2009learning}. As shown in \autoref{fig:caseStudy-hull}-a, we first colorized the hulls using the Tableau-10 palette with random color assignment. Using a fixed opacity of $\alpha=0.5$ (see \autoref{fig:caseStudy-hull}-b, the distribution of the green class becomes ambiguous, especially for parts overlapping with the red class. 
 Applying a local color blending model (\autoref{fig:caseStudy-hull}-c) improves the depth order perception, but still does not reduce false colors. Using our optimization, the optimal color and opacity assignments reduce false colors. For example, the different parts of the orange cluster are easier to discriminate, as compared with a randomly assigned green color in \autoref{fig:caseStudy-hull}-b~\&~c. The extent of clusters is now more clear with our method (\autoref{fig:caseStudy-hull}-d), and the visualization overall can be read with minimal ambiguity. 

\CreateLink{stimuli}{Palettes and histograms used in the evaluation}
\section{Palettes and histograms used in the evaluation}
\begin{figure}[!ht]
\centering
\includegraphics[width=0.98\linewidth]{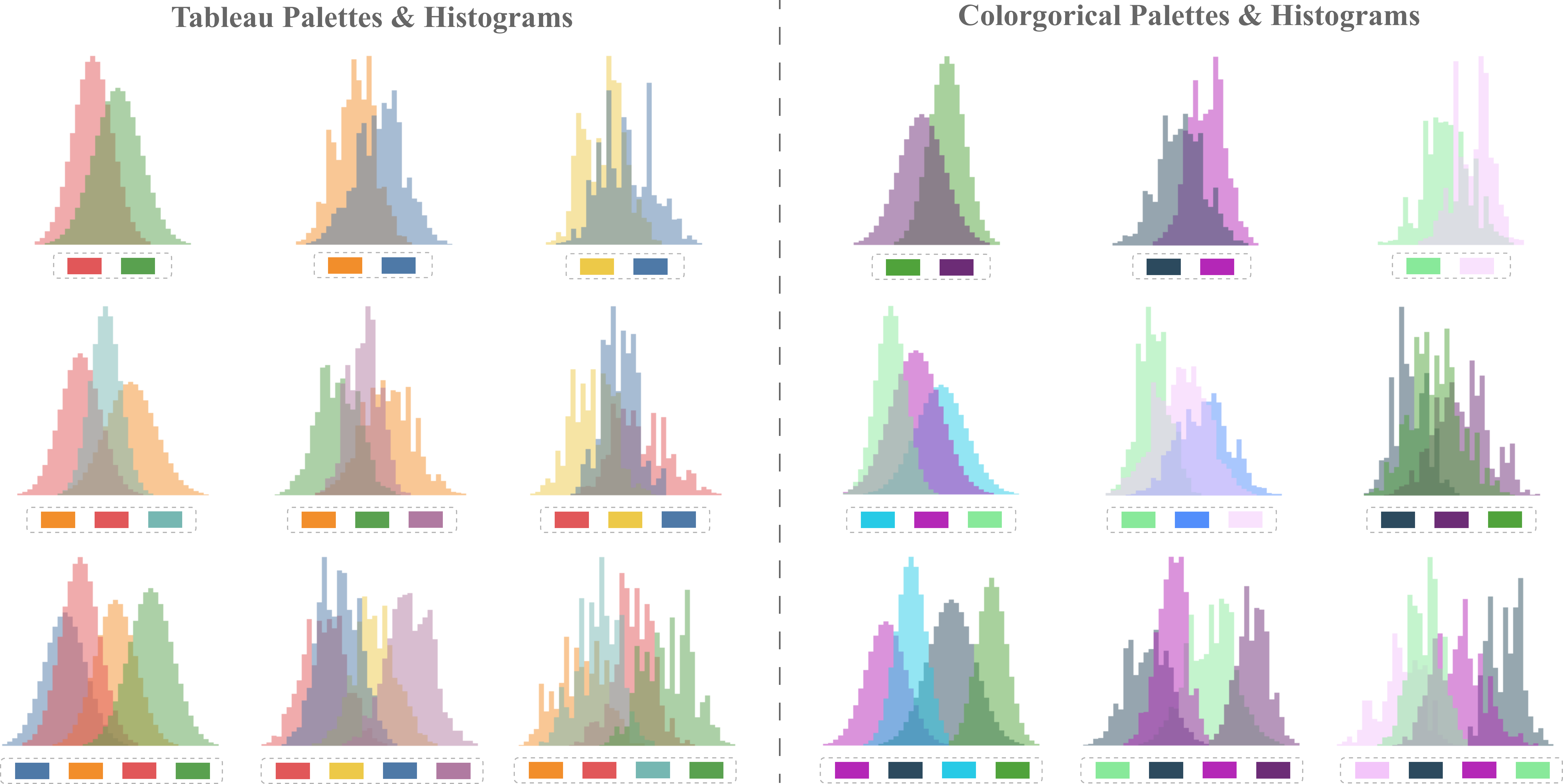}
\caption{
Stimuli used in the evaluation section, generated with both Tableau and Colorgorical palettes, applying a uniform opacity of 0.5 and consistent rendering order, based on the \emph{standard alpha blending model}.
}
\label{fig:stimuli-standard}
\end{figure}

\begin{figure}[!ht]
\centering
\includegraphics[width=0.98\linewidth]{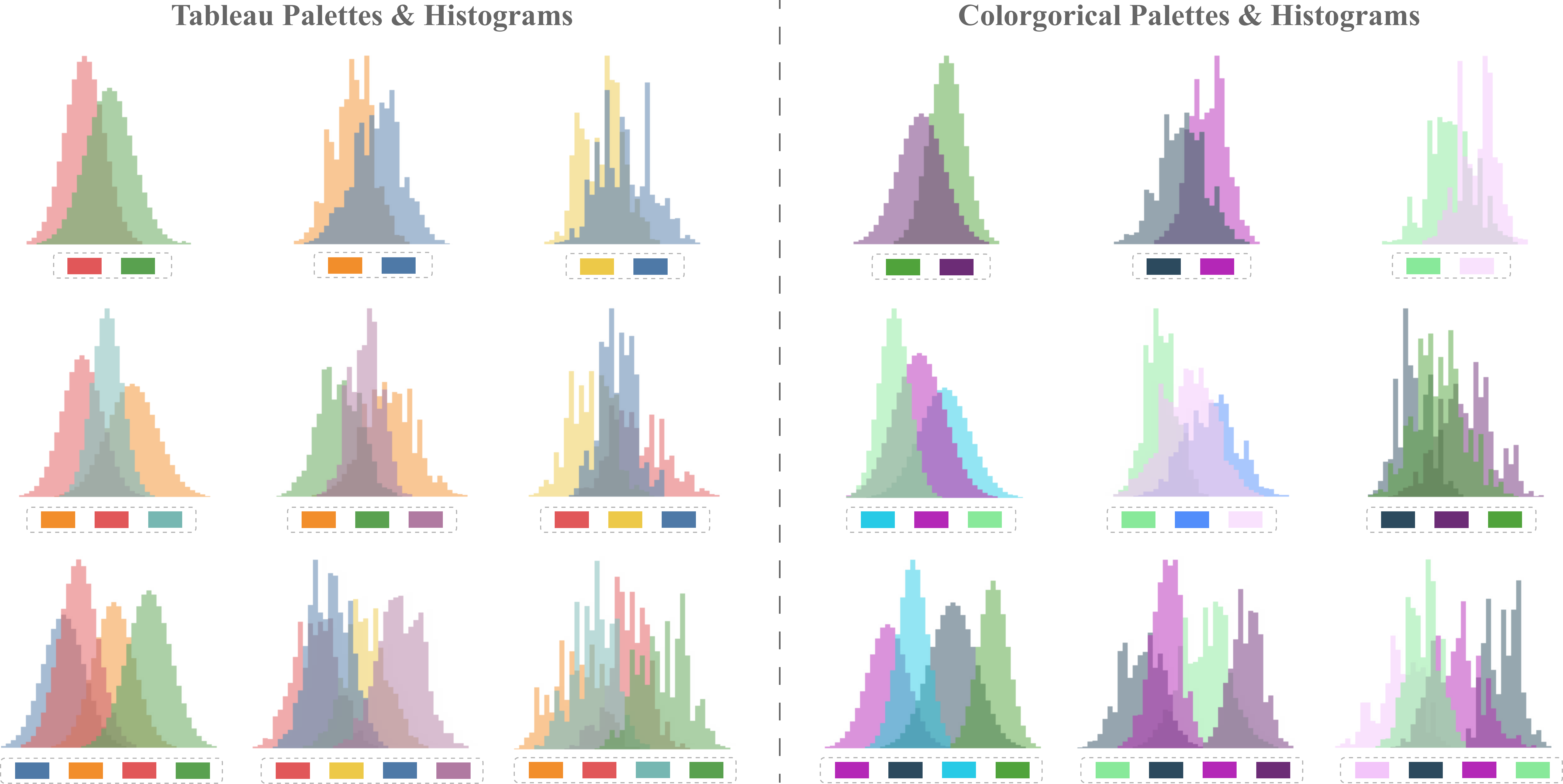}
\caption{
Stimuli used in the evaluation section, generated with both Tableau and Colorgorical palettes, applying a uniform opacity of 0.5 and consistent rendering order, based on the \emph{local color blending model}.
}
\label{fig:stimuli-local}
\end{figure}

\begin{figure}[!ht]
\centering
\includegraphics[width=0.98\linewidth]{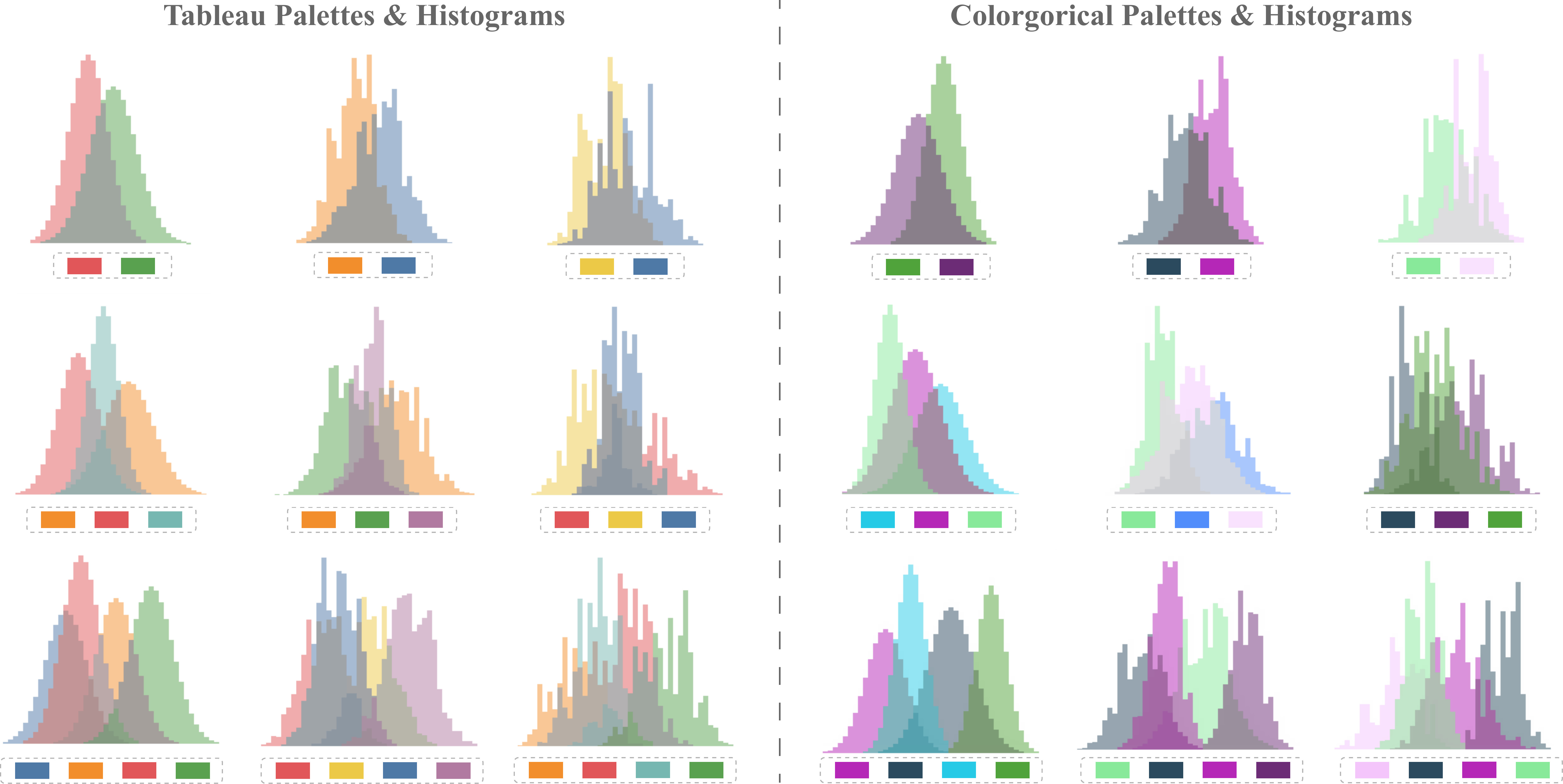}
\caption{
Stimuli used in the evaluation section, generated with both Tableau and Colorgorical palettes, applying a uniform opacity of 0.5 and consistent rendering order, based on the \emph{hue-preserving color blending model}.
}
\label{fig:stimuli-hue}
\end{figure}

\begin{figure}[!ht]
\centering
\includegraphics[width=0.98\linewidth]{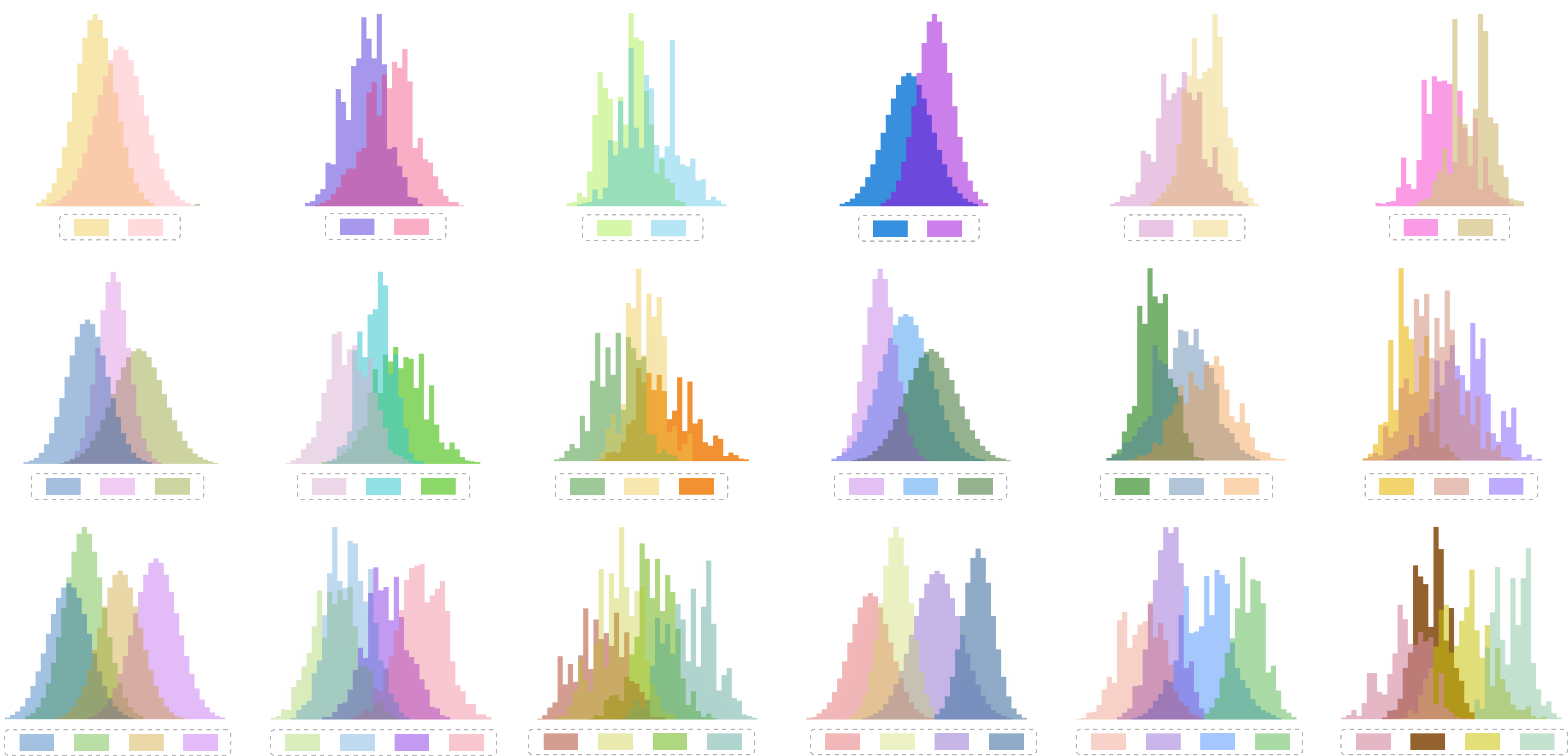}
\caption{
The stimuli used in the evaluation section, generated by our color-name aware optimization (\emph{CNA}). This approach automatically determines optimal settings for color, transparency, and rendering order to produce the final visualizations, ensuring clear discriminability of all segments while enhancing the perception of the whole from its constituent parts. The palettes displayed beneath each histogram are rendered using the auto-generated colors along with their corresponding opacity settings.
}
\label{fig:stimuli-cna}
\end{figure}

We developed a collection of 18 multi-class overlapping histograms, each consisting of two, three, or four classes. As shown in \autoref{fig:stimuli-standard}, these histograms exhibited a diverse array of characteristics, encompassing varying degrees of distribution smoothness and overlap. To color-code these histograms, we employ color palettes obtained equally from Tableau\cite{tableau} and Colorgorical\cite{Gramazio17}, ensuring a balanced application between the two sources. We maintain a uniform opacity of 0.5 and a consistent rendering order to simulate typical usage scenarios. Depending on the number of classes in the histogram, we randomly select sub-palettes comprising two, three, or four colors. The stimuli corresponding to the four comparison methods—standard alpha blending, local color blending, hue-preserving color blending, and our color-name aware optimization (CNA)—are presented in \autoref{fig:stimuli-standard}, \autoref{fig:stimuli-local}, \autoref{fig:stimuli-hue}, and \autoref{fig:stimuli-cna}, respectively. The first three benchmark methods use the same palette, uniform opacity of 0.5 and consistent rendering order, compared with our optimized results (CNA), allowing a direct comparison of how each blended method handles the color coding of overlapping histograms.

\clearpage  

\bibliographystyle{abbrv-doi-hyperref}
\bibliography{opacity}